\pdfoutput=1
\documentclass[twocolumn]{aastex63}
\usepackage{mathtools}
\usepackage{dsfont}

\received{}
\revised{}
\accepted{}

\graphicspath{{./}{figures/}}

\shorttitle{Star Formation in Nuclear Rings}
\shortauthors{Moon et al.}

\begin{document}

\title{Star Formation in Nuclear Rings with the TIGRESS Framework}

\author[0000-0002-6302-0485]{Sanghyuk Moon}
\affiliation{Department of Physics \& Astronomy, Seoul National University, 1 Gwanak-ro, Gwanak-gu, Seoul 08826, Republic of Korea}
\affiliation{SNU Astronomy Research Center, Seoul National University, 1 Gwanak-ro, Gwanak-gu, Seoul 08826, Republic of Korea}
\author[0000-0003-4625-229X]{Woong-Tae Kim}
\affiliation{Department of Physics \& Astronomy, Seoul National University, 1 Gwanak-ro, Gwanak-gu, Seoul 08826, Republic of Korea}
\affiliation{SNU Astronomy Research Center, Seoul National University, 1 Gwanak-ro, Gwanak-gu, Seoul 08826, Republic of Korea}
\author[0000-0003-2896-3725]{Chang-Goo Kim}
\affiliation{Department of Astrophysical Sciences, Princeton University,
  Princeton, NJ 08544, USA}
\author[0000-0002-0509-9113]{Eve C.\ Ostriker}
\affiliation{Department of Astrophysical Sciences, Princeton University,
  Princeton, NJ 08544, USA}

\email{moon@astro.snu.ac.kr, wkim@astro.snu.ac.kr, cgkim@astro.princeton.edu, eco@astro.princeton.edu}

\begin{abstract}
Nuclear rings are sites of intense star formation at the centers of barred galaxies.
To understand what determines the structure and star formation rate (SFR; $\dot{M}_{\rm SF}$) of nuclear rings, we run semi-global, hydrodynamic simulations of nuclear rings subject to constant mass inflow rates $\dot{M}_{\rm in}$.
We adopt the TIGRESS framework of Kim \& Ostriker to handle radiative heating and cooling, star formation, and related supernova (SN) feedback.
We find that the SN feedback is never strong enough to destroy the ring or quench star formation everywhere in the ring. Under the constant $\dot{M}_{\rm in}$, the ring star formation is very steady and persistent, with the SFR exhibiting only mild temporal fluctuations. The ring SFR is tightly correlated with the inflow rate as $\dot{M}_{\rm SF}\approx 0.8\dot{M}_{\rm in}$, for a range of $\dot{M}_{\rm in}=0.125-8\,M_\odot\,{\rm yr}^{-1}$.
Within the ring, vertical dynamical equilibrium is maintained, with the midplane pressure (powered by SN feedback) balancing the weight of the overlying gas.
The SFR surface density is correlated nearly linearly with the midplane pressure, as predicted by the pressure-regulated, feedback-modulated star formation theory.
Based on our results, we argue that the ring SFR is causally controlled by $\dot{M}_\text{in}$, while the ring gas mass adapts to the SFR to maintain the vertical dynamical equilibrium under the gravitational field arising from both gas and stars.
\end{abstract}

\section{Introduction}\label{s:intro}

Nuclear rings were first identified in photographic plates as multiple ``hot spots'' near galaxy centers \citep{morgan58,sp65}, which have turned out to be manifestation of compact yet vigorous star-forming regions \citep[see also][and references therein]{kennicutt94,kk04}. They
are thought to form as a result of gas redistribution due to a bar potential (e.g., \citealt{comb85,buta86,shl90,gar91,bc96}). Indeed,
numerical simulations have consistently shown that the non-axsiymmetric torque exerted by a stellar bar in disk galaxies causes gas to move radially inward along dust lanes and form a ring near the centers \citep[e.g.,][]{athanassoula92,piner95,eng97,pat00,kim12a,kimstone12,li15}. It is uncertain what determines the ring locations, but theoretical proposals suggest it may be determined by radial extent of periodic orbits caused by a bar potential \citep{bgsbu,rt03}, balance between centrifugal force and external gravity \citep{kim12b}, or shear reversal \citep{sormani18}.
While bars are by far the most efficient agent driving mass inflow in galactic disks, other non-axisymmetric features such as spiral arms, elongated bulges, and ovals can also drive mass inflow to fuel starburst activity \citep[e.g.,][]{ath94,combes01,kimkim14,seo14,kim18}.

Observations indicate that the star formation rate (SFR) in the rings of normal barred galaxies spans a wide range $\sim 0.1-10\,M_\odot\,{\rm yr}^{-1}$ \citep{mazzuca08,ma18}, while the total gas mass range in the rings
is more limited, $\sim(1-6)\times 10^8\,M_\odot$ \citep{sheth05}.  The central molecular zone (CMZ), which is believed to be a nuclear ring in the Milky Way, contains gas mass of $\sim (3-7)\times 10^7\,M_\odot$ \citep{price00,mol11,tok19} and is forming stars at a rate $\sim 0.02-0.1\,M_\odot\,{\rm yr}^{-1}$, measured by counting numbers of young stellar objects or estimating the ionizing photon luminosity that trace recent star formation activity in a time period $\lesssim 10\,{\rm Myr}$ \citep{yusef09,immer12,longmore13,koepferl15}. It has been noted that the observed SFR in the CMZ is a factor of $\sim 10$ smaller than what is expected for its gas mass or column density \citep{longmore13,kruijssen14}. Using numerical simulations, \citet{sk13} and \citet{seo19} showed that the ring SFR is closely related to the mass inflow rate to the ring rather than the ring mass, suggesting that the low current SFR of the CMZ is due to small mass inflow rates in the near past. Since this result can in principle depend on the treatment of star formation and feedback adopted in those works, it needs to be confirmed using new simulations with more realistic treatment of relevant physics in higher resolution. We note that there is observational evidence that the mass inflow rate to the CMZ varies considerably with time and currently has a very small value \citep{sb19}.

As an alternative scenario, \citet{kruijssen14} proposed that the ring SFR undergoes quasi-periodic variations between starburst and quiescent phases, and the CMZ is currently in the quiescent phase \citep[see also,][]{elmegreen94,kk15,krumholz17,torrey17,armillotta19}. In this scenario, the inflowing gas gradually piles up until the ring becomes gravitationally unstable and undergoes intense star formation. Associated strong stellar feedback terminates the starburst phase rapidly, causing the ring to become quiescent until its mass grows sufficient to trigger another burst.
\citet{krumholz17} ran numerical simulations based on vertically-integrated, axisymmetric, one-dimensional (1D) models and found that the ring SFR exhibits quasi-periodic oscillations with period $\sim 20\,{\rm Myr}$, even when the mass inflow rate is held constant.

However, it is questionable whether the quasi-periodic behavior of the ring SFR seen in the 1D models of \citet{krumholz17} exists in three-dimensional (3D) simulations for the CMZ \citep{armillotta19,tress20,sormani20}, or realized in observed galaxies. For instance, \citet{armillotta19} performed a global simulation of the Milky Way using {\sc gizmo} \citep{gizmo}. They modeled a stellar bar by a rigidly-rotating gravitational potential, and found that the ring SFR  varies by more than an order of magnitude over a timespan of $\sim 500\,{\rm Myr}$, even though the gas mass in the CMZ stays relatively constant. While the short-period ($\sim50\,$Myr) cycle in their ring SFR is likely modulated by feedback, the long-period ($\sim200\,$Myr) cycle that  dominates the SFR might be compromised by the orbital motions of a large molecular cloud which is unresolved and somehow survives in their simulation. Similar global simulations of \cite{sormani20} using {\sc arepo} \citep{arepo,weinberger20} found that the gas depletion time in the CMZ is quite steady and that the SFR is directly proportional to the time-varying CMZ mass (see also \citealt{tress20}).  The temporal changes of the CMZ mass and SFR in the models of \citet{sormani20} might be driven by the time variation in the mass inflow rate, as suggested by \citet{seo19}.

Diverse results from the 3D simulations mentioned above imply that there is no consensus as to what controls the ring SFR, gas mass evolution, and detailed dynamical properties. In fully global simulations, the mass inflow rate is naturally time-variable since the gas density near the bar ends and along the dust lanes is highly inhomogeneous \citep[e.g,][]{seo19,armillotta19,tress20,sormani20}. The time-dependent mass inflow rate causes the ring size, shape, and mass to vary significantly with time, making it difficult to isolate key factors that determine the ring SFR.
For a more controlled study, in this paper we construct semi-global models that focus on a nuclear ring and nearby regions, without explicitly including a stellar bar in the simulations.
Instead, our model have a stream of gas with prescribed properties entering though the domain boundaries, mimicking gas inflows along the dust lanes in global simulations \citep[e.g.,][]{athanassoula92,kim11,kim12a,sbm15,shin17}.  We handle star formation and associated supernova (SN) feedback by adopting the Three-phase Interstellar medium in Galaxies Resolving Evolution with Star formation and Supernova
feedback (TIGRESS) algorithms developed by \citet{tigress}. While our models cannot capture the triggering by the bar potential  of large-scale gas inflows, they allow us to investigate the ring region itself with high resolution, and to explore the behavior of the ring SFR when the mass inflow rate is kept constant in time (at chosen levels).

Our semi-global models are also useful for investigating the details of star formation regulation in nuclear rings. \citet{oml10} and \citet{os11} developed analytical equilibrium models for the self-regulation of SFR in normal and starburst regions of galactic disks, in which the equilibrium SFR is set by the balance between the weight of the interstellar medium (ISM) and the midplane pressure, with the required pressure provided primarily by SN feedback and far ultraviolet (FUV) heating. This equilibrium model has been validated through a series of local shearing-box simulations \citep{kko11,kok13,kim15}, including spiral arms \citep{kko20}, and for more extreme star-forming regions \citep{so12}. In this work, we explore whether the self-regulation theory is also applicable to the semi-global model of the galactic centers characterized by high SFRs and short dynamical timescales.

The remainder of this paper is organized as follows. In Section \ref{s:method}, we describe our numerical methods and our treatment of gas streams through the domain boundaries, and briefly summarize the TIGRESS framework for star formation and SN feedback. In Section \ref{s:result}, we present the temporal and morphological evolution of our models as well as the star formation histories. In Section \ref{s:stats}, we present various physical quantities characterizing nuclear rings and explore their correlations, testing the self-regulation theory of star formation. Finally, we summarize and discuss our results in Section \ref{s:discussion}.

\section{Numerical Methods}\label{s:method}

In this paper, we use the TIGRESS framework to study star formation and SN feedback in a nuclear ring located near a galaxy center. Ring formation is driven by stellar bars, which  cause gas to flow radially inward while still retaining enough angular momentum to circularize at some distance from the galactic nucleus.  In the present work we do not model the bar explicitly, instead imposing the gas inflows via boundary conditions (see below).

\subsection{Basic Equations}\label{s:equations}

Our simulation domain is a Cartesian cube with side length $L$, encompassing a nuclear ring.
The simulation domain rotates at an angular frequency ${\bf \Omega}_p=\Omega_p\hat{\bf z}$, where this represents the pattern speed of a bar (on larger scale than we are simulating).
The equations of hydrodynamics in the rotating frame read
\begin{equation}\label{eq:cont}
    \frac{\partial\rho}{\partial t} + \boldsymbol\nabla\cdot\left(\rho {\bf v}\right) = 0,
\end{equation}
\begin{equation}
\begin{split}
    \frac{\partial(\rho {\bf v})}{\partial t} + \boldsymbol\nabla\cdot\left(\rho {\bf v}{\bf v} + P\mathds{I}\right) = - 2\rho{\bf \Omega}_p\times{\bf v}-\rho\boldsymbol\nabla\Phi_{\rm tot},
\end{split}
\end{equation}
\begin{equation}\label{eq:energy}
\begin{split}
    \frac{\partial}{\partial t}\left(\frac{1}{2}\rho v^2 + \frac{P}{\gamma-1}\right) + \boldsymbol\nabla\cdot\left[\left(\frac{1}{2}\rho v^2 + \frac{\gamma P}{\gamma - 1}\right){\bf v}\right]\\= -\rho{\bf v}\cdot\boldsymbol\nabla\Phi_{\rm tot} -\rho{\cal L},
\end{split}
\end{equation}
\begin{equation}\label{eq:Poisson}
    \boldsymbol\nabla^2\Phi_\text{self} = 4\pi G(\rho + \rho_{\rm sp}),
\end{equation}
where ${\bf v}$ is the gas velocity in the rotating frame, $P$ is the gas pressure, $\mathds{I}$ is the identity matrix, $\rho\mathcal{L}$ is the net cooling rate per unit volume, and $\Phi_{\rm tot}=\Phi_{\rm cen} + \Phi_{\rm ext} + \Phi_{\rm self}$ is the total gravitational potential, consisting of the centrifugal potential $\Phi_{\rm cen}=-\frac{1}{2}\Omega_p^2(x^2+y^2)$, the external gravitational potential $\Phi_{\rm ext}$ giving rise to the background rotation curve, and the  self-gravitational potential $\Phi_{\rm self}$  of gas with density $\rho$ and newly-formed star particles with density $\rho_{\rm sp}$.

\begin{figure}[t]
  \centering
  \includegraphics[width=\linewidth]{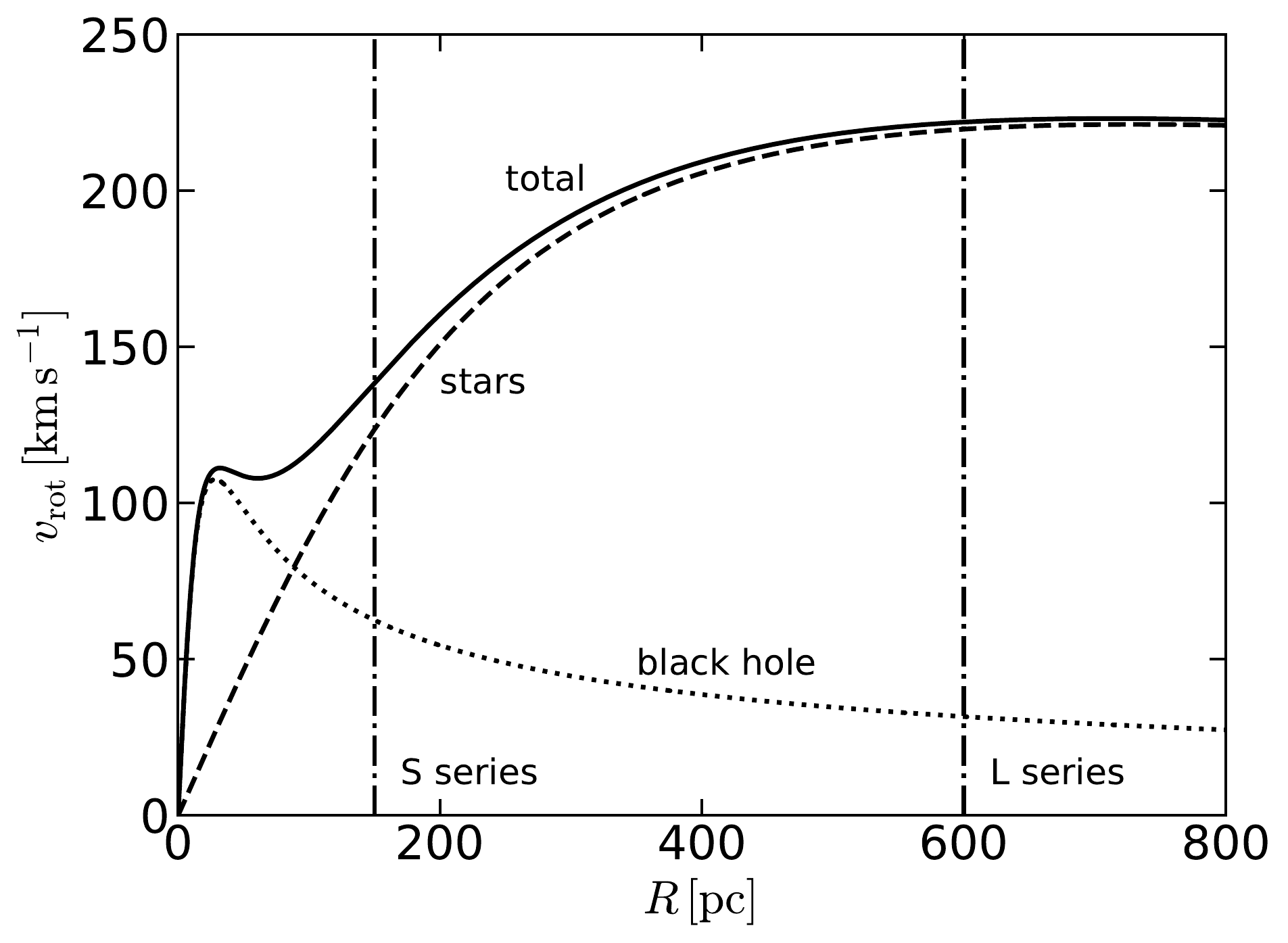}
  \caption{Rotation curve of our galaxy model.
    The solid line draws the circular velocity $v_\text{rot}=(Rd\Phi_b/dR + Rd\Phi_\text{BH}/dR)^{1/2}$, while the dashed and dotted lines show the individual contributions of $\Phi_b$ and $\Phi_\text{BH}$ to $v_\text{rot}$.
    Two vertical lines at $R_{\rm ring}=150\,{\rm pc}$ and $R_{\rm ring}=600\,{\rm pc}$ mark the ring positions in our {\tt S} and {\tt L}-series models, respectively.
  }%
  \label{fig:rotcurve}
\end{figure}

We adopt a model for the external potential based on the archetypal barred-spiral galaxy NGC 1097, which posses a star-forming nuclear ring with radius of $R_{\rm ring}\sim 700\,{\rm pc}$ \citep{hsieh11}.
\citet{onishi15} found that the observed gas kinematics near the galaxy center is consistent with the velocity field derived from the combined gravitational potential of stellar mass distribution and the supermassive black hole with mass $M_{\rm BH}=1.4\times 10^8\,M_\odot$, which we represent with a Plummer potential
\begin{equation}
    \Phi_{\rm BH} = -\frac{GM_{\rm BH}}{\sqrt{r^2+r_{\rm BH}^2}},
\end{equation}
where $r_{\rm BH}=20\,{\rm pc}$ is the softening radius.
To represent stellar mass distribution, we use a modified Hubble profile with the stellar volume density
\begin{equation}
    \rho_b = \frac{\rho_{b0}}{(1+r^2/r_b^2)^{3/2}},
\end{equation}
and corresponding gravitational potential
\begin{equation}\label{eq:bulge_potential}
    \Phi_b = -\frac{4\pi G\rho_{b0} r_b^3}{r}\ln\left(\frac{r}{r_b}+\sqrt{1+\frac{r^2}{r_b^2}}\right),
\end{equation}
with $\rho_{b0}=50\,M_\odot\,{\rm pc^{-3}}$ and $r_b=250\,{\rm pc}$ chosen so that the circular velocity at $r \sim 1 \,\text{kpc}$ and shape at smaller scale are similar to the rotation curve for NGC 1097 from \citet{onishi15}.
Figure \ref{fig:rotcurve} shows the resulting rotation curve derived from $\Phi_{\rm ext}=\Phi_b + \Phi_{\rm BH}$.

The cooling rate in general depends on the chemical composition, ionization state, and temperature, while the heating rate depends on composition, electron abundance, and the radiation and cosmic ray energy densities.
For the present work, in which we focus on dynamics rather than the exact thermal state,  we adopt a simplified  net cooling function $\rho{\cal L}$ in Equation \eqref{eq:energy} that consists of three parts:
\begin{equation}
    \rho{\cal L} = n_{\rm H}^2\Lambda - n_{\rm H}\Gamma_{\rm PE} -n_{\rm H}\Gamma_{\rm CR}.
\end{equation}
Here $n_{\rm H}^2\Lambda$ is the volumetric cooling rate, where $n_{\rm H}=\rho/(1.4271 m_{\rm H})$ is the hydrogen number density assuming the solar abundances.
We shall assume that $\Lambda$ depends only on the gas temperature $T$,
and use the fitting formula of \citet[see also \citealt{kko08}]{ki02} for $T < 10^{4.2}\,{\rm K}$, and the cooling function from \citet{sutherland93} for $T>10^{4.2}\,{\rm K}$, where the latter is based on collisional ionization equilibrium at solar metalicity.
The heating terms are $n_{\rm H}\Gamma_{\rm PE}$ representing the photoelectric heating rate by FUV radiation on dust grains, and $n_{\rm H}\Gamma_{\rm CR}$ representing the heating rate by cosmic ray (CR) ionization.
For the equation of state $P=\rho k_{\rm B} T / (\mu m_{\rm H})$ with the Boltzmann constant $k_{\rm B}$,  we allow the mean molecular weight $\mu(T)$ to vary with $T$ from $\mu_{\rm ato}=1.295$ for atomic gas to $\mu_{\rm ion}=0.618$ for ionized gas \citep[see][]{tigress}.

\begin{figure}[t]
  \centering
  \includegraphics[width=\linewidth]{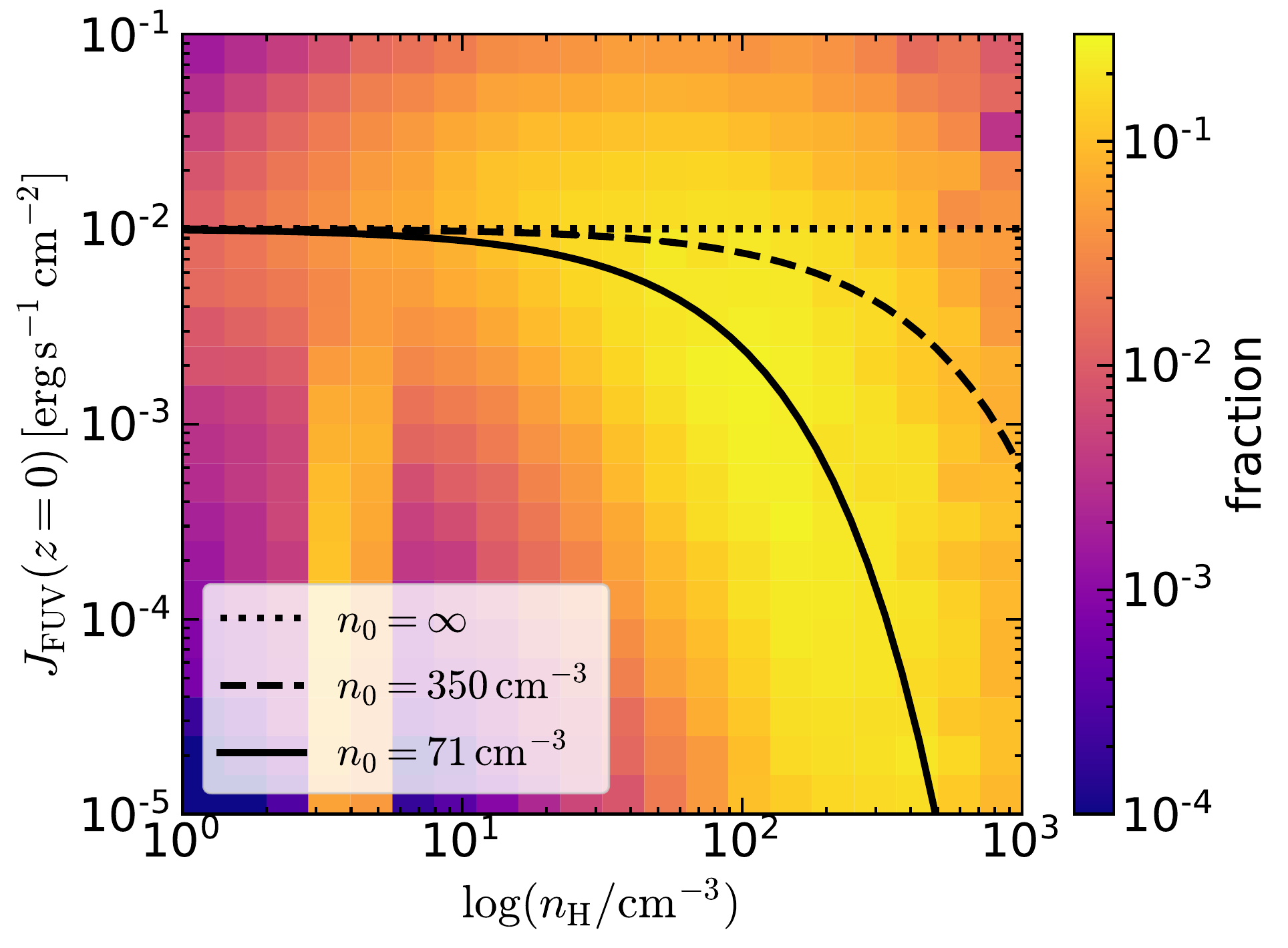}
  \caption{2D histogram of $J_{\rm FUV}$ and $n_{\rm H}$ in the $z=0$ plane, using our fiducial model.
    $J_{\rm FUV}$ is obtained by post-processing snapshots at $t=100-110\,{\rm Myr}$, using the gas density distribution and locations and luminosities of all star particles, and employing the adaptive ray-tracing algorithm of \citet{jgkim17}.
    The dotted, dashed, and solid lines represent Equation \eqref{eq:JFUV} with $n_0=\infty$, $350\,{\rm cm}^{-3}$, and $71\,{\rm cm}^{-3}$, respectively.
    Note that the  vertical feature near $n_{\rm H}=4\rm\,cm^{-3}$ is due to the gas along the inflowing streams.
  }%
  \label{fig:radps}
\end{figure}

The main source of the FUV radiation is young massive stars, which in our simulations are a constant fraction of the mass of star cluster particles formed when gas collapses.
In addition, we also allow for metagalactic FUV radiation.
We therefore take the PE heating rate per hydrogen $\Gamma_{\rm PE}$ as
\begin{equation}\label{eq:GammaPE}
  \Gamma_{\rm PE} = \Gamma_{\rm PE,0}\left(\frac{\mu(T)-\mu_{\rm ion}}{\mu_{\rm ato}-\mu_{\rm ion}}\right)\left(\frac{J_{\rm FUV}}{J_{\rm FUV,0}}+0.0024\right),
\end{equation}
where we take $\Gamma_{\rm PE,0} = 2\times 10^{-26}\,{\rm erg\,s^{-1}}$ \citep{ki02} and $J_{\rm FUV,0} = 2.1\times 10^{-4}\,{\rm erg\,s^{-1}\,cm^{-2}\,sr^{-1}}$ \citep{draine78} as normalizing factors based on solar neighborhood conditions.
The term in the first parentheses in Equation \eqref{eq:GammaPE} is to make the photoelectric heating completely shut off in the fully ionized gas.
The small additional factor in the last parentheses in Equation \eqref{eq:GammaPE} comes from the metagalactic radiation \citep{sternberg02}.

The FUV intensity would have large values in the regions near star particles and small values in deep inside clouds away from star particles due to dust attenuation. While it would be desirable to apply full radiative transfer to compute the FUV intensity $J_{\rm FUV}({\bf x},t)$ throughout the domain, time-dependent ray-tracing from every star particle
would be prohibitively expensive given the large ($\gtrsim 200$) number of sources in our simulations.
Instead, we adopt a simpler and less computationally expensive approach.  We calculate the total FUV luminosity ${\cal L}_{\rm FUV}$ of all star particles in the simulation domain (see Section~\ref{s:starpar_feedback}) and use this to set the local  $J_{\rm FUV}$ in a cell with density $n_H$ according to
\begin{equation}\label{eq:JFUV}
    J_{\rm FUV} = \frac{{\cal L}_{\rm FUV}}{4\pi L^2}\left(\frac{1-E_2(\tau_\perp/2)}{\tau_\perp}\right)e^{-n_{\rm H}/n_0}.
\end{equation}
Here $\tau_\perp=\kappa_d\Sigma$ with $\kappa_d=10^3\,\text{cm}^2\,\text{g}^{-1}$ is the vertical optical depth for $\Sigma\equiv M_{\rm gas}/L^2$ the average gas surface density, $E_2$ is the second exponential integral, and $n_0$ is a turnover density. $M_{\rm gas}$ is the total gas mass in the computational domain.
The first two factors in Equation \eqref{eq:JFUV} corresponds to the solution of the radiation transfer equation in a  plane-parallel geometry (see, e.g., \citealt{oml10}), while the exponential term takes into account the local shielding of FUV radiation inside dense clumps with $n_{\rm H} \gtrsim n_0$.

To motivate Equation \eqref{eq:JFUV} and determine an appropriate value of $n_0$ for each of our models, we first run simulations by taking $n_0=\infty$ (i.e, without FUV shielding) and select eleven snapshots during $100\,{\rm Myr} \leq t\leq 110\,{\rm Myr}$ after a nuclear ring already formed (see Section \ref{s:evolution}).
We then post-process the snapshots by applying the adaptive ray-tracing algorithm developed by \citet{jgkim17} to directly measure $J_{\rm FUV}$ produced by all star particles.
Figure \ref{fig:radps} plots the normalized two-dimensional (2D) histogram of $J_{\rm FUV}$ and $n_{\rm H}$ in the midplane of our fiducial model (see below) at $t=100$--$110$ Myr.
The dotted, dashed, and solid lines draw Equation \eqref{eq:JFUV} with $n_0=\infty$, $350\,{\rm cm}^{-3}$, and $71\,{\rm cm}^{-3}$, respectively, the last of which best describes $J_{\rm FUV}(z=0)$ resulting from the ray-tracing method.  For each model we use the same procedure to compute $n_0$;  Table~\ref{tb:models} lists the adopted values obtained in this way.
Overall, models with higher inflow rate and smaller ring size have higher gas density and thus larger $n_0$.

In addition to photoelectric heating, we include CR heating which is responsible for heating the cold and dense gas for which photoelectric heating almost shuts off due to the exponential factor in Equation \eqref{eq:JFUV}.  We assume the CR heating rate is proportional to the SFR surface density $\Sigma_{\rm SFR}$,
%We assume that the CRs are attenuated
also allowing for attenuation
by a factor of $\Sigma_0/\Sigma$ above a critical gas surface density $\Sigma_0 = 10.7\,M_\odot\,{\rm pc}^{-2}$ \citep{nw17}. We normalize by the CR heating rate and the SFR surface density in the solar neighborhood, $\Gamma_{\rm CR,0} = 3.2\times 10^{-27}\,{\rm erg\,s^{-1}}$ and $\Sigma_{\rm SFR,0}=3\times 10^{-3}\,M_\odot\,{\rm yr^{-1}\,kpc^{-2}}$ \citep{gow17,nw17}. We thus have
\begin{equation}\label{eq:cr}
  \Gamma_{\rm CR} = \Gamma_{\rm CR,0} \left( \frac{\mu(T)-\mu_{\rm ion}}{\mu_{\rm ato}-\mu_{\rm ion}} \right) \frac{\Sigma_{\rm SFR}}{\Sigma_{\rm SFR,0}} \min\left\{1, \frac{\Sigma_0}{\Sigma} \right\},
\end{equation}
where the factor in the parentheses shuts off CR heating by ionization in fully-ionized gas.\footnote{In this work, we do not consider CR heating by scattering off free electrons in fully-ionized gas (e.g., \citealt{draine11}).}

\subsection{Gas Inflow Streams}\label{s:streams}

\begin{figure}[t]  \centering
  \includegraphics[width=\linewidth]{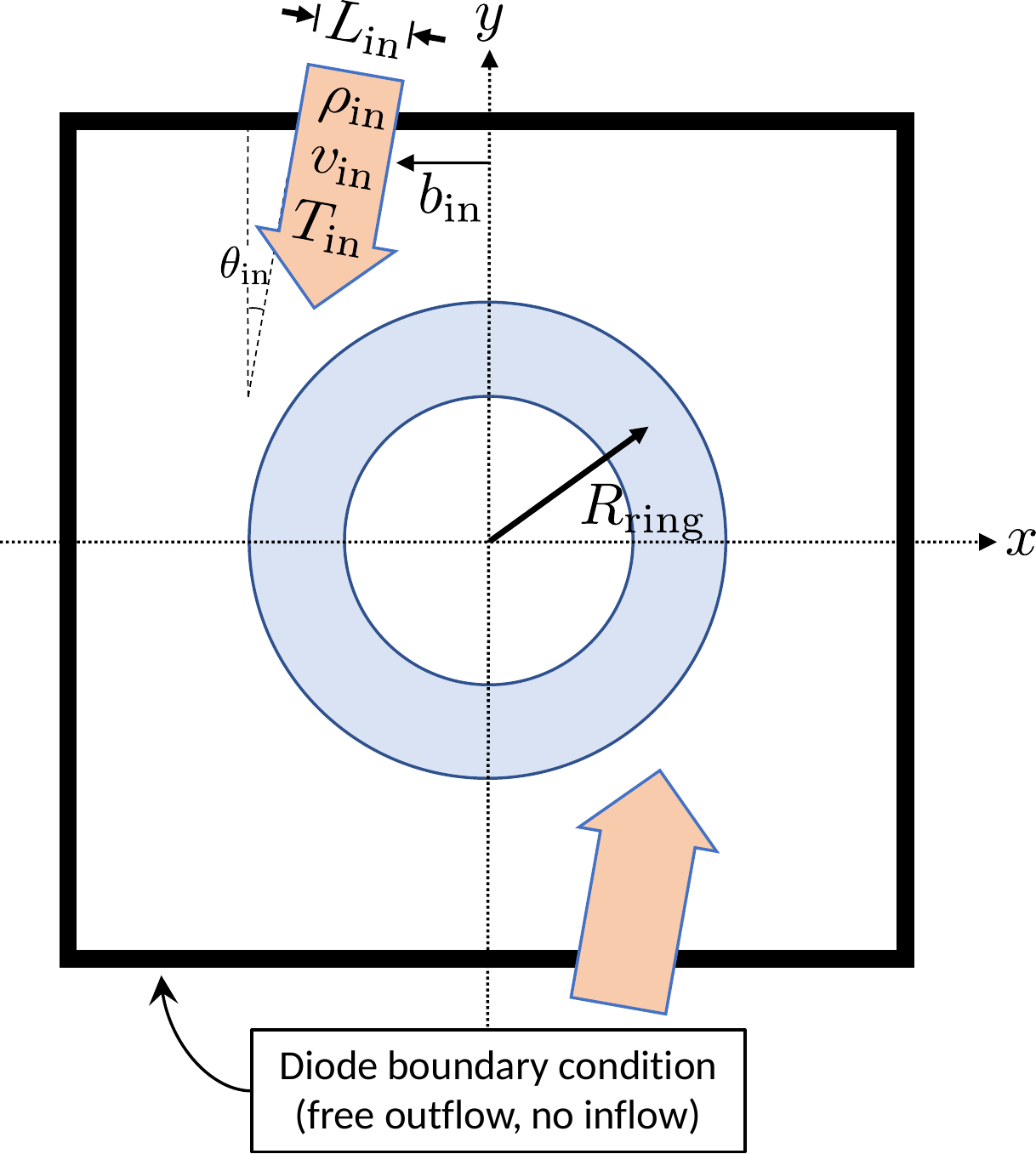}
  \caption{Schematic of the imposed gas inflows in our simulations, mimicking bar-driven inflows.
    Two gas streams flow into the computational domain through square-shaped nozzles with side $L_\text{in}$, located at the positive and negative $y$-boundaries (large arrows).
    $\theta_\text{in}=10^\circ$ and $b_\text{in}$ refer to the inclination angle and the impact parameter of the inflow stream relative to the $y$-axis.
    The blue shaded annulus represents the expected location for the ring formation. See text for details.
  }
  \label{fig:galcen_schematic}
\end{figure}

In our simulations, the hydrodynamic effect of a non-axisymmetric bar is implemented via idealized gas streams originating on the $y$-boundaries,
as depicted in Figure \ref{fig:galcen_schematic}. We create the inflows via a pair of square-shaped nozzles each of size $L_{\rm in}^2$, offset by impact parameter $b_{\rm in}$ relative to the $y$-axis. The direction of the inflow velocity is inclined to the $y$-axis by angle $\theta_{\rm in}$ to reflect that dust lanes are usually inclined relative to the bar semi-major axis (e.g., \citealt{comeron09,kim12a}).\footnote{We find that gas streams escape the computational domain without forming a ring when $\theta_{\rm in}$ is too small.}

We vary the density $\rho_{\rm in}$ and speed $v_{\rm in}$ of the streams in order to control both the mass inflow rate $\dot{M}_{\rm in}$ and the ring radius $R_{\rm ring}$ that forms.
From the condition that the specific angular momentum of the stream \emph{in the inertial frame} is equal to that of circular ring consistent with the background rotation at radius $R=R_{\rm ring}$, we find
\begin{equation}\label{eq:vin}
  v_{\rm in}(x,y) = \frac{R_{\rm ring}v_{\rm rot}(R_{\rm ring})-R^2\Omega_p}{|x\cos\theta_{\rm in}-y\sin\theta_{\rm in}|},
\end{equation}
where $R=\sqrt{x^2+y^2}$ is the galactocentric radius at the location of the nozzle.
Note that $v_{\rm in}$ depends on $x$ and $y$, indicating the inflow velocity varies across the nozzles; on the boundary $|y|=L/2$.
The mass inflow rate is then given by
\begin{equation}\label{eq:Mdot}
  \dot{M}_{\rm in} = 2\int_{-L_{\rm in}/2}^{L_{\rm in}/2}\int_{b_{\rm in}}^{b_{\rm in}+L_{\rm in}} \rho_{\rm in}v_{\rm in}\cos\theta_{\rm in}\,dxdz.
\end{equation}
For inflowing gas, we set the temperature to $T_{\rm in}=2\times 10^4\,{\rm K}$, which is typical of the warm neutral medium in our models. The choice of $T_{\rm in}$ is immaterial, however, because the temperature of the inflowing gas is quickly adjusted according to the heating and cooling rates once the stream enters the computational domain.

We caution the reader that if the simulation domain is large enough that $R^2\Omega_p > R_{\rm ring}v_{\rm rot}(R_{\rm ring})$ at the nozzles,
it is not possible to choose a value for the inflow velocity consistent with a
%the angular momentum of the inflowing gas is always greater than that of the
circular orbit at $R=R_{\rm ring}$.
%for any positive $v_{\rm in}$.
%In such a case, our semi-global model does not apply;
In this case, simple consideration of angular momentum conservation for ring formation is inadequate. Instead, it would be necessary to allow for a bar torque such that the inflowing gas can lose enough angular momentum to settle on a circular orbit at $R_{\rm ring}$. For our idealized semi-global setup, the box size is therefore limited.

\subsection{Star Particles and SN Feedback}\label{s:starpar_feedback}

A complete description for the creation and evolution of star particles and the prescription for treating supernovae can be found in \citet{tigress}.
Here, we briefly summarize the key components.  A cell undergoing gravitational collapse spawns a sink/star particle if the following three conditions hold simultaneously:
(1) the gas density in the cell exceeds the Larson-Penston density threshold $\rho_{\rm LP} = 8.86c_s^2/(G\Delta x^2)$, for grid spacing $\Delta x$; (2) the cell lies at a local potential minimum; and (3) the velocity is converging in all three directions.  A portion of the mass from a volume, $(3\Delta x)^3$, is removed from the grid and assigned to the sink particle upon creation.
Each particle represents a star cluster that fully samples the \citet{kroupa01} initial mass function (IMF).
Particles act as sinks, accreting gas from their neighboring cells, until the onset of first SN explosion occurring at $t\sim 4\,{\rm Myr}$.
The FUV luminosities of sink particles are assigned based on their mass and age using  {\tt STARBURST99} \citep{starburst99} assuming a fully sampled Kroupa IMF.

For every sink particle with mass $m_{\rm sp}$ and mass-weighted mean age $t_m$, we estimate the expected number of supernova, ${\cal N}_{\rm SN} = m_{\rm sp}\xi_{\rm SN}(t_m)\Delta t$, during hydrodynamic time step $\Delta t$, where $\xi_{\rm SN}(t_m)$ is the SN rate tabulated in {\tt STARBURST99}.
With our spatial resolution $\Delta x = 2\,{\rm pc}$ or $4\,{\rm pc}$, ${\cal N}_{\rm SN}$ is smaller than unity.
We therefore turn on SN feedback only if ${\cal N}_{\rm SN} > {\cal U}_{\rm SN}$, where ${\cal U}_{\rm SN} \in [0,1)$ denotes a uniform random number.

Each supernova returns mass $M_{\rm ej}= 10\,M_\odot$, momentum, and energy to the neighboring cells.
In the TIGRESS framework, the amount of momentum or energy injected depends on the local density and the resolution.
If the ambient density is too high for the Sedov-Taylor stage to be resolved, we assume the remnant has already entered the snowplow phase and thus inject the expected final radial momentum $p_*=2.8\times 10^5\,M_\odot\,{\rm km}\,{\rm s}^{-1}(n_{\rm H}/{\rm cm}^{-3})^{-0.17}$ \citep{ko15}.
If the density is sufficiently low such that the Sedov-Taylor stage is expected to be at least partially resolved, $72\%$ of the total SN energy $E_{\rm SN}=10^{51}\,{\rm erg}$ is injected as thermal energy, while the remaining $28\%$ is injected as kinetic energy associated with the radial momentum.
In our simulations, $80$--$90\%$ of all SNe are resolved.

We follow the motion of sink particles by solving the equations of motion
\begin{equation}\label{eq:sp_eom}
    \ddot{\bf x} = -\boldsymbol\nabla\Phi_{\rm tot} - 2{\bf \Omega}_p\times\dot{\bf x}
\end{equation}
in the rotating frame.
The original integrator used in \citet{tigress} is for the equations of motion in a local shearing box and therefore inapplicable for our purpose.
The usual explicit leap-frog integrator is also inappropriate, because it loses its symplectic nature in the presence of the velocity-dependent Coriolis force.
Noting that the right hand side of Equation \eqref{eq:sp_eom} has the same form as the Lorentz force in electromagnetism, we integrate Equation \eqref{eq:sp_eom} using the Boris algorithm \citep{boris70}, which is frequently adopted in kinetic codes for advancing charged particles under electromagnetic fields. In Appendix \ref{s:boris}, we describe our implementation of the Boris algorithm and present a test result.

\begin{deluxetable*}{lcccccccc}
    \tablecaption{Model parameters\label{tb:models}}
    \tablehead{
\colhead{Model} &
\colhead{$R_{\rm ring}$} &
\colhead{$\rho_b(R_{\rm ring})$} &
\colhead{$n_{\rm H,in}$} &
\colhead{$\dot{M}_{\rm in}$} &
\colhead{$\bar{v}_{\rm in}$} &
\colhead{$v_{\rm cir}$} &
\colhead{$t_{\rm orb}$} &
\colhead{$n_0$}\\
\colhead{} &
\colhead{$({\rm pc})$} &
\colhead{$(M_\odot\,{\rm pc}^{-3})$} &
\colhead{$({\rm cm^{-3}})$} &
\colhead{$(M_\odot\,{\rm yr^{-1}})$} &
\colhead{$({\rm km\,s^{-1}})$} &
\colhead{$({\rm km\,s^{-1}})$} &
\colhead{$({\rm Myr})$} &
\colhead{$({\rm cm}^{-3})$}\\
\colhead{(1)} &
\colhead{(2)} &
\colhead{(3)} &
\colhead{(4)} &
\colhead{(5)} &
\colhead{(6)} &
\colhead{(7)} &
\colhead{(8)} &
\colhead{(9)}
    }
    \startdata
{\tt\ L0} & 600 & 2.84 & 0.286 & 0.125 & 154 & 200 & 18.4 & 18\\
{\tt\ L1} & 600 & 2.84 & 1.15 & 0.5   & 154 & 200 & 18.4 & 35\\
{\tt\ L2}\tablenotemark{$*$} & 600 & 2.84 & 4.58 & 2 & 154 & 200 & 18.4 & 71\\
{\tt\ L3} & 600 & 2.84 & 18.3 & 8     & 154 & 200 & 18.4 & 142\\\hline
{\tt\ S0} & 150 & 31.5 & 5.71 & 0.125 & 123 & 133 & 6.92 & 71\\
{\tt\ S1} & 150 & 31.5 & 22.8 & 0.5   & 123 & 133 & 6.92 & 142\\
{\tt\ S2} & 150 & 31.5 & 91.3  & 2     & 123 & 133 & 6.92 & 283\\
{\tt\ S3} & 150 & 31.5 & 365  & 8     & 123 & 133 & 6.92 & 567
    \enddata
    \tablenotetext{$*$}{Fiducial model.}
\end{deluxetable*}

\subsection{Models}\label{s:models}

We consider two series of models that differ in $R_{\rm ring}$, the ``target'' size of the ring.
The large ring models ({\tt L} series) have $R_{\rm ring}=600\,{\rm pc}$, nozzle impact parameter $b_{\rm in}=320\,{\rm pc}$, and nozzle width $L_{\rm in}=200\,{\rm pc}$ (see Figure~\ref{fig:galcen_schematic}).  The {\tt L} series models have domain size $L=2048\,{\rm pc}$ and number of cells per dimension $N=512$, yielding grid spacing $\Delta x = L/N=4\,{\rm pc}$.
We construct small ring models ({\tt S} series) by scaling down the {\tt L} series by a factor of four, such that $R_{\rm ring}=150\,{\rm pc}$, $b_{\rm in}=80\,{\rm pc}$, $L_{\rm in}=50\,{\rm pc}$, and $L=512\,{\rm pc}$.
We take $N=256$, corresponding to $\Delta x=2\,{\rm pc}$, in order to mitigate a time step constraint arising from smaller cell size in the {\tt S} series.
For all models we adopt $\theta_{\rm in}=10^\circ$.
For both series, we consider four different values for the inflow rate\footnote{As long as $\dot{M}_{\rm in}$ is fixed, different combinations of $\rho_{\rm in}$ and $L_{\rm in}$ does not lead to any noticeable differences on the ring properties.}, $\dot{M}_{\rm in} = 1/8$, $1/2$, $2$, and $8\,M_\odot\,{\rm yr}^{-1}$.
For the angular velocity of our computational domain, we take $\Omega_p = 36\,{\rm km\,s^{-1}\,kpc^{-1}}$, equal to the bar pattern speed in NGC 1097 \citep{pinol14}.

Table \ref{tb:models} lists the parameters of all models.
Column (1) gives the model name.
Column (2) and (3) give $R_\text{ring}$ and the bulge stellar density $\rho_b$ at $R_{\rm ring}$, respectively.
Columns (4) and (5) give $n_{\rm H,in}=\rho_{\rm in}/(1.4271m_{\rm H})$, and $\dot{M}_\text{in}$, respectively.
Column (6) gives the mean inflow velocity $\bar{v}_{\rm in}\equiv\dot{M}_{\rm in}/(2\rho_{\rm in}L_{\rm in}^2\cos\theta_{\rm in})$.
Columns (7) and (8) list the circular velocity $v_{\rm cir}=v_\text{rot}(R_\text{ring})-R_\text{ring}\Omega_p$ and the orbital period $t_{\rm orb}=2\pi R_{\rm ring}/v_{\rm cir}$ of the ring in the rotating frame, respectively. The circular velocity in the inertial frame is $222\,{\rm km\,s^{-1}}$ and $138\,{\rm km\,s^{-1}}$ for {\tt L} and {\tt S} series, respectively.
Column (9) gives the value of $n_0$ we take for the dust attenuation (see Equation \ref{eq:JFUV}). We take model {\tt L2} with $R_\text{ring}=600\,$pc and $\dot{M}_\text{in}=2\,M_\odot\,{\rm yr^{-1}}$ as our fiducial model.

The initial condition of our models is near-vacuum, filled with rarefied gas with $n_{\rm H} = 10^{-5}\,\exp\left[{-|z|/(50\,{\rm pc})}\right]\,{\rm cm^{-3}}$ and $T=2\times 10^4\,{\rm K}$, rotating at ${\bf v} = \sqrt{ R (\partial\Phi_{\rm tot}/(\partial R) }\hat{\boldsymbol\phi}$.
The subsequent evolution is governed entirely by the mass inflow from the boundaries.

We integrate Equations \eqref{eq:cont}--\eqref{eq:Poisson} using a modified version of the {\tt Athena} code \citep{athena}, which solves the equations of hydrodyanmics or  magnetohydrodynamics using finite-volume Godunov methods. %based on  Riemann solver, while automatically preserving the divergence-free constraint on the magnetic fields by using the integral form of the induction equation averaged over the cell area rather than the volume \citep{gardiner05,gardiner08}.
In the present work, we do not include magnetic fields.
Our simulations use the van Leer integrator \citep{stone09}, Roe's Riemann solver with H-correction \citep{sanders98}, and second-order spatial reconstruction.   When needed, we apply first-order flux correction \citep{lemaster09}.
%to attain maximal stability.
We solve the Poisson equation via FFT convolution with open boundary conditions \citep{skinner15}; the density of star particles is included using a particle-mesh approach as in \citet{tigress}.

Within the nozzle region on the boundaries, we apply inflow boundary conditions as described in Section \ref{s:streams}.
For the rest of the boundaries, we take diode boundary conditions: we extrapolate the hydrodynamic variables from the last two active zones to the ghost cells, and  set the normal velocity to zero if the gas is inflowing.
This allows gas to freely escape from the computational domain, while ensuring no inflow occurs except through the nozzles.

\section{Evolution}\label{s:result}

In this section, we describe overall temporal and morphological evolution of our fiducial model, focusing on ring formation, star formation histories, and distributions of gas and star particles. Steady-state physical quantities averaged over the nuclear ring and their correlations will be presented in Section \ref{s:stats}.

\subsection{Overall Evolution of the Fiducial Model}\label{s:evolution}

\begin{figure*}[htpb]
  \centering
  \includegraphics[width=\linewidth]{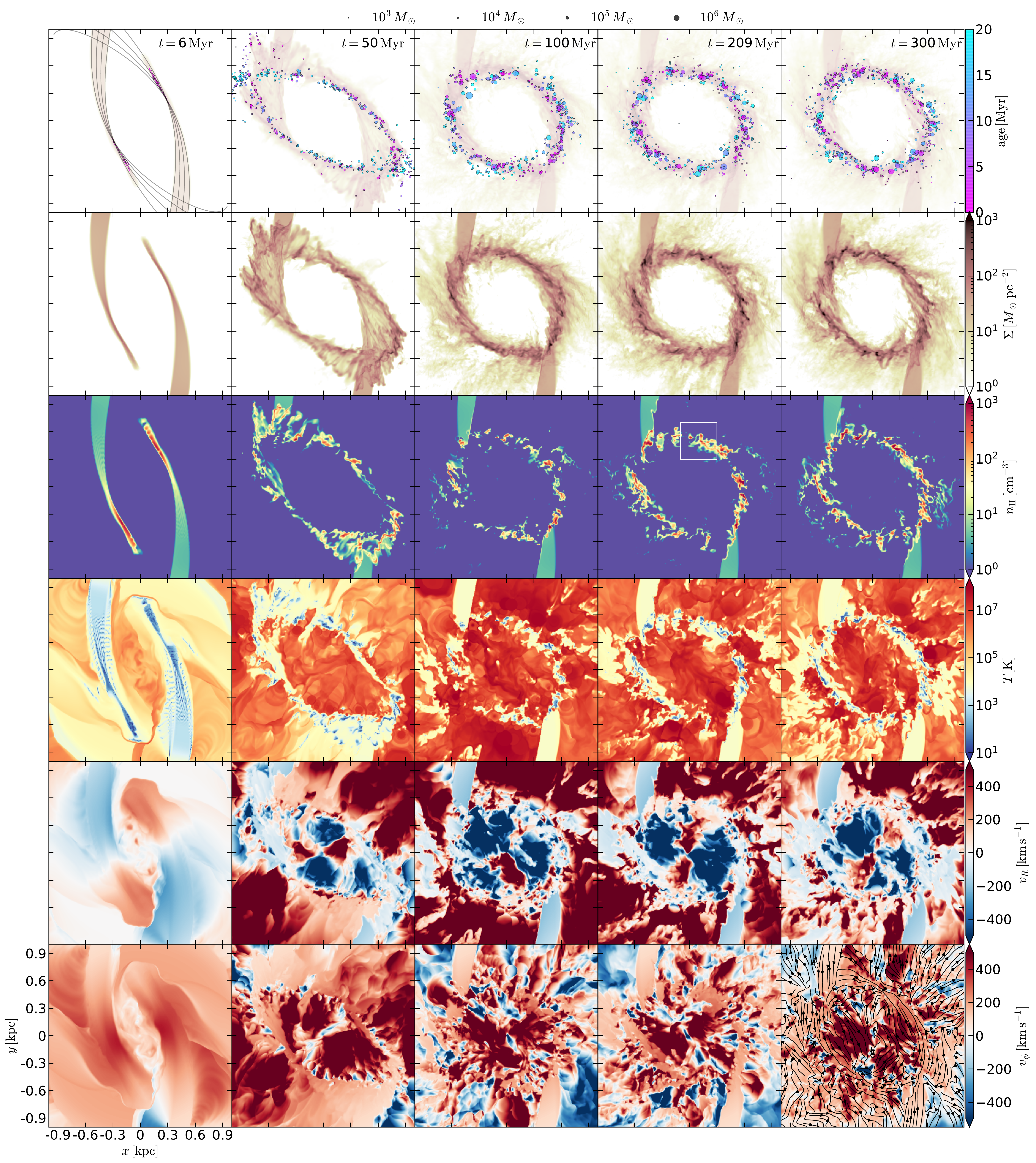}
  \caption{
    Face-on views of model {\tt L2} at $t=6, 50, 100, 209, 300\,{\rm Myr}$ (columns from left to right).
    From top to bottom, rows display: the projected positions  of sink particles, integrated gas surface density $\Sigma$, hydrogen number density $n_\text{H}$, temperature $T$, radial velocity $v_R$, and azimuthal velocity $v_\phi$ in the rotating frame.
    The last four rows show slices through $z=0$.
    In the top row, $\Sigma$ is also shown in the background for reference.
    The gray solid lines in the leftmost top panel draw the expected orbits of pressureless particles injected at the nozzles.
    The white square box in the fourth panel of the third row marks the zoom-in region shown in Figure \ref{fig:closeup}.
    The black solid lines with arrows in the rightmost bottom panel plot the gas streamlines at the midplane.
    Gas streams injected from the nozzles initially follow ballistic orbits, colliding with the stream from the opposite side at $t\sim 8.5\,{\rm Myr}$.
    The collision drives strong shocks which gradually remove the kinetic energy from the gas streams.
    Eventually, gas settles down to a roughly circular orbit and forms a nuclear ring with radius $R\approx 600\,{\rm pc}$.
  }%
  \label{fig:evolution}
\end{figure*}

\begin{figure*}[htpb]
  \centering
  \includegraphics[width=\linewidth]{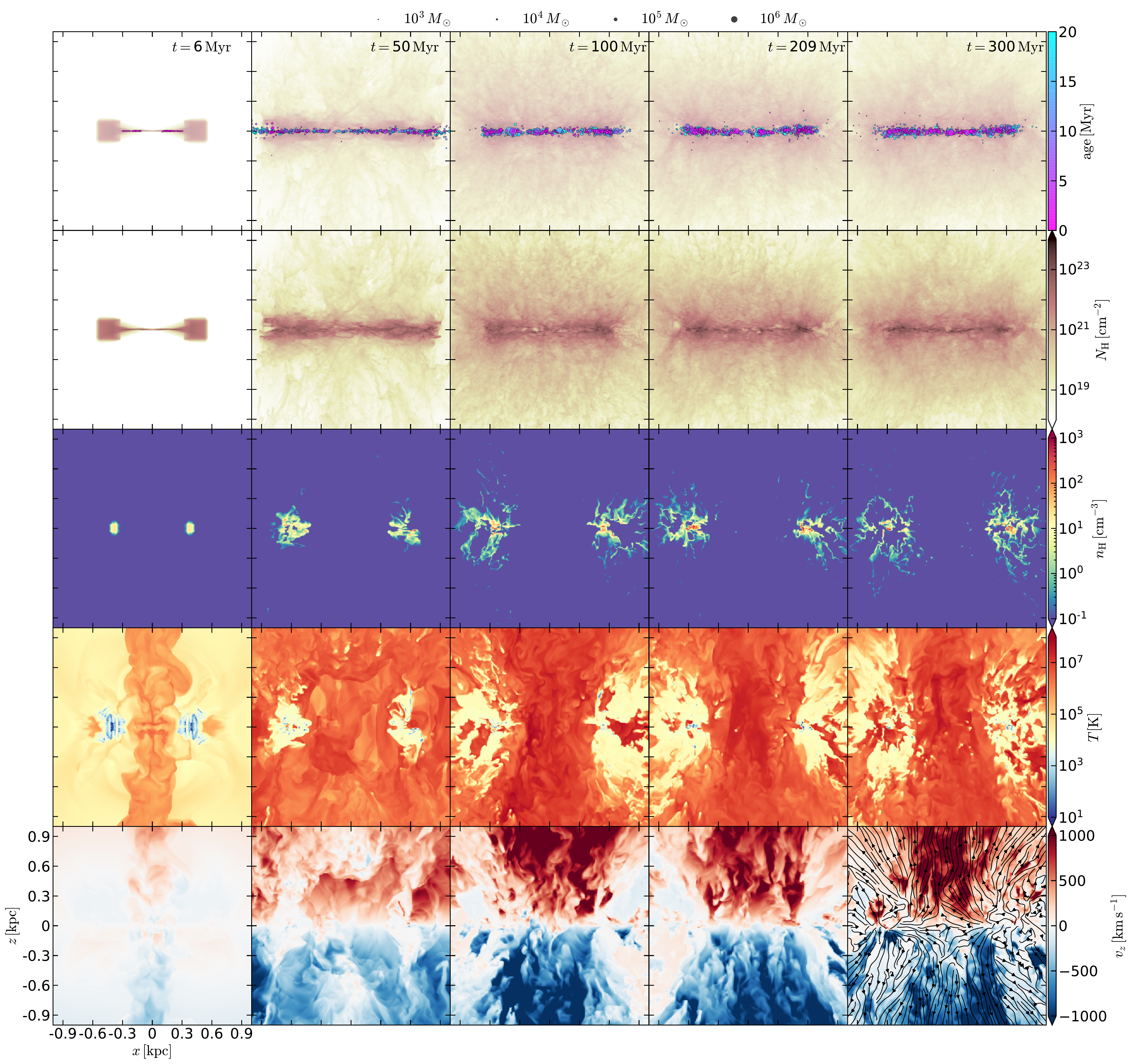}
  \caption{Edge-on views of model {\tt L2} at $t=6, 50, 100, 209, 300\,{\rm Myr}$ (columns from left to right).
    From top to bottom, rows display: the projected positions of the sink particles, hydrogen column density $N_{\rm H} = \int n_{\rm H}\,dy$, hydrogen volume density $n_\text{H}$, temperature $T$, and vertical velocity $v_z$.
    The last three rows show slices through $y=0$.
    The black solid lines with arrows in the rightmost bottom panel plot the gas streamlines.
    Superbubbles produced by interactions of multiple SN explosions lead to persistent winds.
  }%
  \label{fig:edgeon}
\end{figure*}

Figures \ref{fig:evolution} and \ref{fig:edgeon} provide a visual impression of overall time evolution of model {\tt L2}, in the $x$-$y$ plane and $x$-$z$ plane, respectively.

At early time, the gas streams injected through the nozzles closely follow the ballistic orbits shown as gray solid lines in the leftmost top panel.
Orbit crowding, as manifested by convergence of the ballistic orbits near $(x, y)=(\pm 0.3, \pm 0.2)\,{\rm kpc}$, triggers the first star formation in the inflowing streams at $t=5.1\,\text{Myr}$.

In about a half of the (rotating-frame) orbital time $t_{\rm orb}/2 = 9.2\,{\rm Myr}$, the two inflowing streams start to collide with each other and produce strong shocks with Mach number of $\sim 15$.
Star formation then begins
at the contact points where the inflowing streams collide.
The sink particles produced at very early time have highly eccentric orbits close to the ballistic orbits of the streams and thus most of them leave the computational domain. SN feedback from these particles (prior to their escape) produces hot gas that fills most of the volume, inside and outside the gas streams.

\begin{figure}[t]
  \centering
  \includegraphics[width=\linewidth]{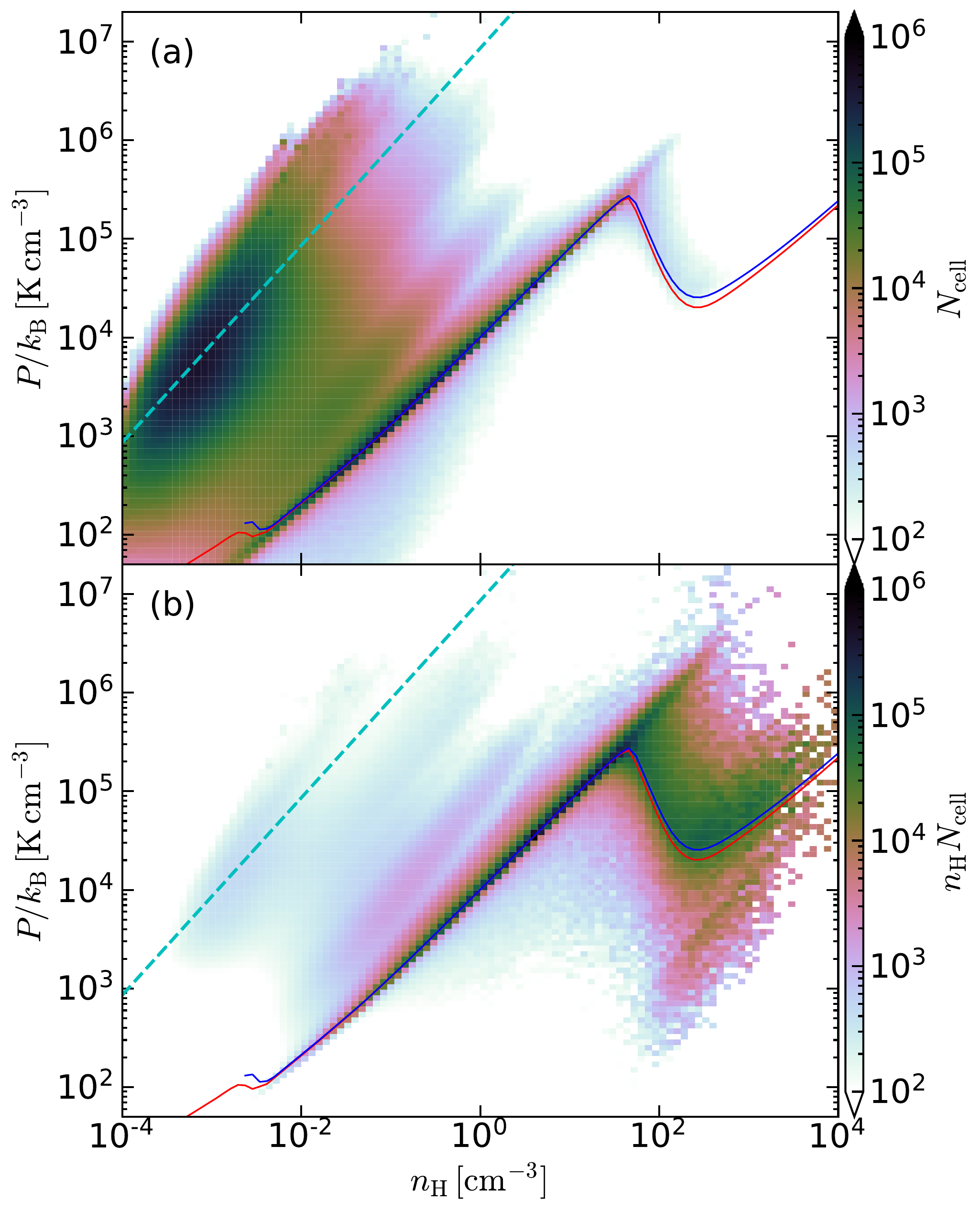}
  \caption{(a) Volume-  and (b) density-weighted joint PDF in $P/k_{\rm B}$ and $n_{\rm H}$ of model {\tt L2}  at $t=250\,{\rm Myr}$.
   The red line is the thermal equilibrium curve under the instantaneous radiative heating rate at that epoch, while the blue line corresponds to
   an equilibrium curve with additional turbulent heating rate  $\Gamma_\text{turb}$ (see text), which yields slightly better agreement with the mean PDFs.
    Most of the warm and cold medium follows the blue line, with some scatter due to intermittent shock heating.
    The dashed line indicates the mean temperature $T=3.7\times 10^6\,{\rm K}$ of the hot medium.
   }%
  \label{fig:phase}
\end{figure}

Unlike the early sink particles, however, gas loses a significant amount of kinetic energy at every passage of the contact points and orbits become less eccentric, eventually creating a ring-like shape. While star-forming regions are concentrated near the contact points at early time, they soon become widely distributed  as strong SN feedback makes the ring highly clumpy everywhere, enabling local collapse. At $t\sim 100\,{\rm Myr}$, the system reaches a quasi-steady state in which the ring morphology and various statistical quantities including the ring gas mass and SFR do not change appreciably over time.
The radius of the nuclear ring is $R\sim 600\,\text{pc}$ in model {\tt L2}, consistent with the expectation from the angular momentum conservation (Equation \ref{eq:vin}). For the fiducial model, the in-plane width of the gas ring is $\sim 200\,{\rm pc}$.
After steady state is reached, the star formation in model {\tt L2} (and other models) is not
concentrated in preferred regions of the ring
but occurs randomly in space and time. The star formation/feedback cycle does not have strong bursts (from either spatial or temporal correlations in gas collapse). As a result, the steady inflow rate adopted in our models gives rise to a steady SFR with small temporal fluctuations for model {\tt L2} (and other models). SN feedback never destroys the ring completely in model {\tt L2} as shown in Figure \ref{fig:evolution}. We find that our models generally show the steady star formation and persistence of the ring.

SN feedback as well as gravitational and thermal instabilities produce cold and dense cloudlets distributed around the nuclear ring. Figure \ref{fig:phase}  display the volume- and mass-weighted probability distribution functions (PDFs) of $n_{\rm H}$ and $P/k_{\rm B}$ at $t=250\,{\rm Myr}$. The volume fractions of the hot ($T>2\times 10^4\,{\rm K}$) and the cold-warm ($T<2\times 10^4\,{\rm K}$) phases are $84\%$ and $16\%$, while their mass fractions are $2\%$ and $98\%$, respectively.

The mean thermal pressure of the cold-unstable medium with $T<5050\,{\rm K}$ is somewhat enhanced above the equilibrium curve with the instantaneous total heating rate at that epoch, shown as the red line. This enhancement is observed in all epochs, suggesting that it is not caused by chance due to fluctuation of heating rate. Instead, this enhanced pressure (or temperature) may be attributed to dissipation of (turbulent) kinetic energy in the ring. To get a rough estimate of the kinetic energy dissipation rate due to (marginally-resolved) cloud-scale turbulence, we calculate the mass-weighted vertical velocity dispersion $\sigma_z\sim 12\,{\rm km\,s^{-1}}$ and the scale height $H \sim 26\,{\rm pc}$ of the cold-unstable medium, and estimate the turbulent heating rate (per hydrogen) as $\Gamma_{\rm turb} = \tfrac{3}{2}\mu_{\rm H}m_{\rm H}\sigma_z^3/H$. The blue line shows the equilibrium curve when $\Gamma_{\rm turb}$ is additionally included, showing a slightly better agreement with the PDFs for the cold-unstable medium. At $n_{\rm H}=10^{3}\,{\rm cm}^{-3}$, the heating rates due to FUV, CR, and (assumed) turbulent dissipation are $\Gamma_{\rm PE} = 5.8\times 10^{-5}\,\Gamma_{\rm PE,0}$, $\Gamma_{\rm CR} = 6.2\,\Gamma_{\rm PE,0}$, and $\Gamma_{\rm turb}=3.9\,\Gamma_{\rm PE,0}$, respectively. We note that actual kinetic energy dissipation rate might be even larger than $\Gamma_{\rm turb}$ considering numerical dissipation. Although our treatment of the FUV and CR heating is rather simplified, the above result suggests that the turbulent heating could be a major heating source for the dense gas where the radiation is heavily shielded \citep[e.g.,][]{ginsburg16}.

\begin{figure*}[t]
  \centering
  \includegraphics[width=\linewidth]{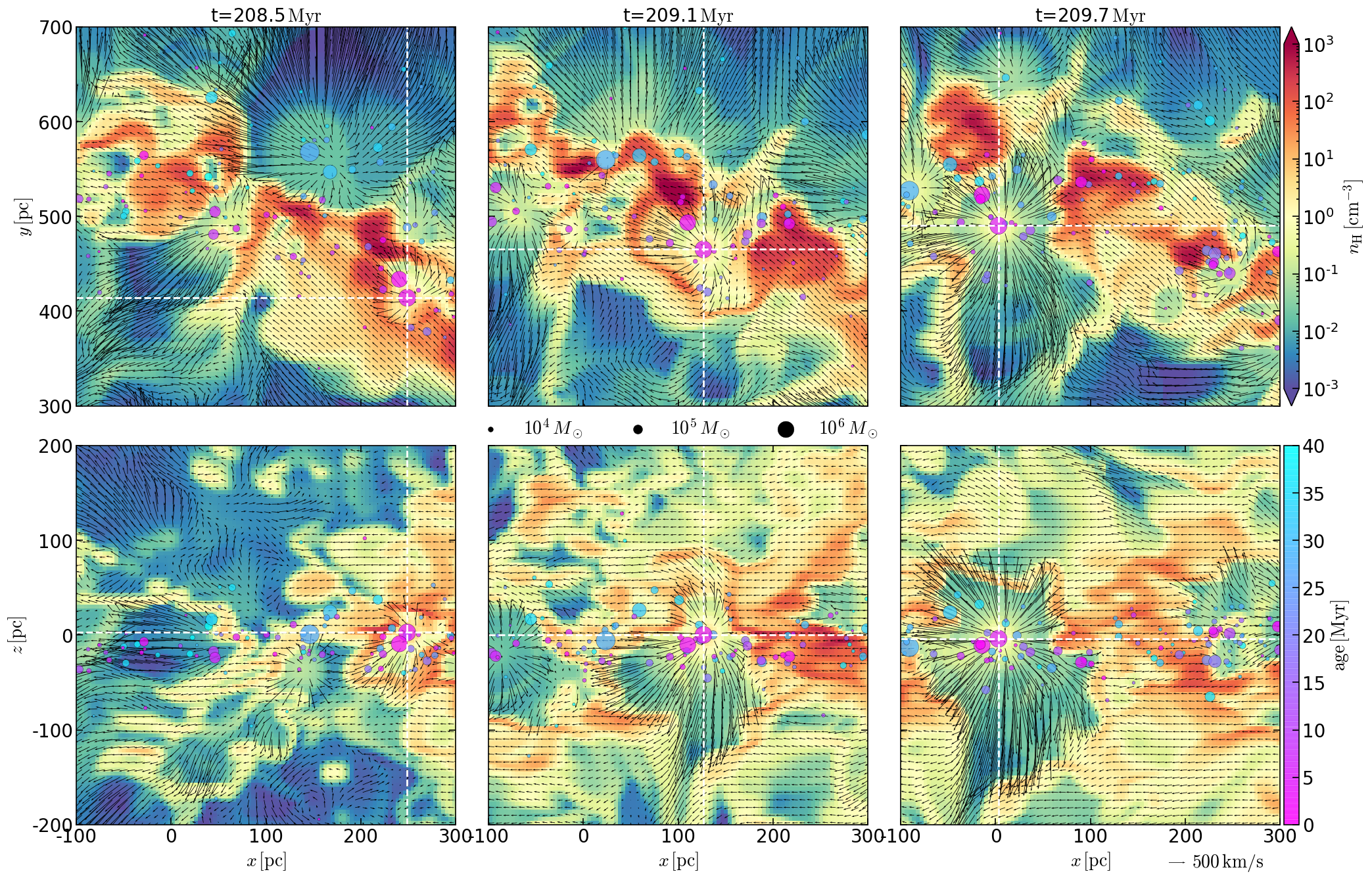}
  \caption{
  Close-up view of the ring: three consecutive snapshots of model {\tt L2} at $t=208.5, 209.1, 209.7\,{\rm Myr}$ from left to right.  The top panels show the density and velocities $(v_x, v_y)$ in the $z=0$ plane at $-0.1{\rm\,kpc}\leq x \leq 0.3{\rm\,kpc}$ and $0.3{\rm\,kpc}\leq y \leq 0.7{\rm\,kpc}$.  Also shown are sink particles within $|z|\leq 0.2 {\rm\,kpc}$.
  Vertical  and horizontal dashed lines trace the location of a selected sink particle, which moves over a few 100 pc within a few Myr, due to its orbital motion (including epicyclic oscillations).
    The bottom panels show the density and velocities $(v_x, v_z)$ at $y=0.41, 0.47, 0.49\,$kpc, corresponding to the $y$ coordinates of the selected sink particle in the top panels.  Other sink particles from the top panels are also shown.
    Arrows show the velocity field, where the arrow length is proportional to the velocity magnitude in $x$-$y$ (top panels) and $x$-$z$ (bottom panels) plane.
  }%
  \label{fig:closeup}
\end{figure*}

Figure \ref{fig:edgeon} shows that most of the star formation takes place in the high-density gas near the midplane, while the distribution of lower-density gas extends vertically up to $|z| \sim 500\,{\rm pc}$ due to SN feedback.  For example, at $t=300\,{\rm Myr}$ (last column) there are three large bubbles centered at $(x, z)\sim (-700\,{\rm pc}, 150\,{\rm pc})$, $\sim (-500\,{\rm pc}, 100\,{\rm pc})$, and $\sim (-800\,{\rm pc}, -150\,{\rm pc})$ in the negative-$x$ portions of the ring, that lift up the gas to high latitude. Heated by the SN shocks, gas inside the bubbles reaches $\sim 10^7\,{\rm K}$.
Sometimes the hot gas inside the superbubbles breaks out through the cold-warm medium, such as the bubble at $(x,z)\sim (-800\,{\rm pc}, -150\,{\rm pc})$.  Away from the midplane the hot gas dominates.

Superbubbles created by repeated feedback from relatively young cluster particles typically have a diameter of $\sim (100-200)\,$pc, comparable to the ring width of $\sim 200\,$pc, so that a fraction of the feedback energy and momentum escape from the ring through the holes like champagne flows. This is illustrated in Figure \ref{fig:closeup} which displays the density and velocity fields as well as star particles with age less than 40 Myr in a zoomed-in region within the ring for model {\tt L2} at $t=208.5, 209.1, 209.6\,$Myr.
A superbubble surrounding the star particle with $(x,y)=(250, 410)\,{\rm pc}$ at $t=208.5\,{\rm Myr}$ expands and blows out, dispersing a part of the ring and compressing the gas nearby, as the particles moves to $(x,y) = (3, 466)\,$pc at $t=209.7\,{\rm Myr}$.
Since orbits of star particles deviate from that of the gas, relatively old particles can explode in the regions outside the ring.
For instance, the clusters near $(x,y)=(150,570)\,{\rm pc}$ at $t=208.5\,{\rm Myr}$ are located near the outer edge of the ring.
In this case, only a small fraction of the feedback energy contributes to turbulence in the ring material.
Because the ring gas is quite spatially confined in radial direction and the gas and stellar orbits differ, much of the feedback energy is transferred to the gas outside the star-forming ring, resulting in lower feedback yield to cold-warm gas than in previous simulations with more uniform distribution of gas in the horizontal direction (see discussion in Section \ref{s:self_regulation}).

\begin{figure*}[t]
  \centering
  \includegraphics[width=\linewidth]{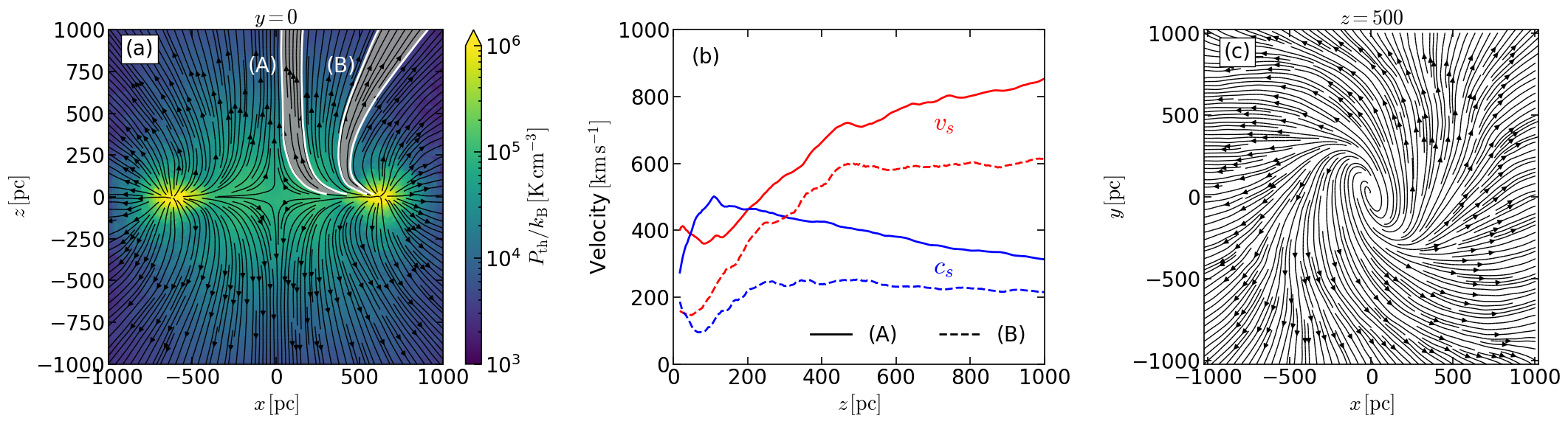}
  \caption{(a) Gas streamlines in model {\tt L2} overlaid over the thermal pressure in the $x$-$z$ plane averaged over $t=200$ to $300\,{\rm Myr}$.
    The ring appears as bull's-eyes at $|x|\sim 600\,{\rm pc}$.
    Streamlines emerge from the ring due to the SN feedback and lead to large-scale hot superwinds.
    (b) Vertical profile of the poloidal velocity $\left<v_s\right> \equiv (\left<v_R\right>^2 + \left<v_z\right>^2)^{1/2}$ (red) and the sound speed $\left<c_s\right>$ (blue) along the selected streamlines (A) and (B) shown in (a).
    (c) Time-averaged streamlines in the $x$-$y$ plane at $z=500\,$pc, showing the helical wind.
  }%
  \label{fig:wind}
\end{figure*}

Hot gas created by individual SN shocks merges together to launch large, coherent outflows resembling galactic winds.  These quasi-conical outflows are clearly visible in the bottom row of Figure~\ref{fig:edgeon}.
For the same model, Figure \ref{fig:wind}(a) plots in the $x$-$z$ plane gas streamlines overlaid over the thermal pressure.
For these streamlines, we average over $t=200$--$300\,{\rm Myr}$ and include all of the gas in the $y=0$ slice. The ring shows up like bull's-eyes at $|x|\sim 600\,{\rm pc}$ in the pressure map, from which the streamlines emerge.
Figure \ref{fig:wind}(b) plots the vertical profiles of the time-averaged poloidal velocities and the sound speed of the gas along the selected streamlines shown in Figure \ref{fig:wind}(a). It is apparent that the wind is accelerated beyond the sonic point, readily reaching $\approx(600-900)\,{\rm\,km\,s^{-1}}$ at $z=1\,$kpc. Figure \ref{fig:wind}(c) plots the time-averaged streamlines at $z=500\,$pc, showing that the winds are helical, with the rotational velocity decreasing as gas moves outward radially conserving angular momentum.

We note that the subsonic to supersonic transition of the winds was absent in the previous local simulations in which the streamlines cannot open up due to the %eometrical limitation
combination of a relatively small box and SNe throughout the midplane region
\citep[e.g.,][]{martizzi16,ko18}.
Here, the relatively small size of the star-forming ring within the domain allows streamlines to open up, leading to the characteristic bi-conical shape and supersonic transition often seen in both observations \citep{strickland04, yukita12} and simulations \citep{wps09, fielding17, srt18}.  We note that in contrast to previous simulations of central starburst-driven winds where the locations of SN feedback were imposed by hand, here the SN location distribution arises naturally from star formation within the ring.

\subsection{Star Formation}\label{s:sfh}

\begin{figure}[htpb]
  \centering
  \includegraphics[width=\linewidth]{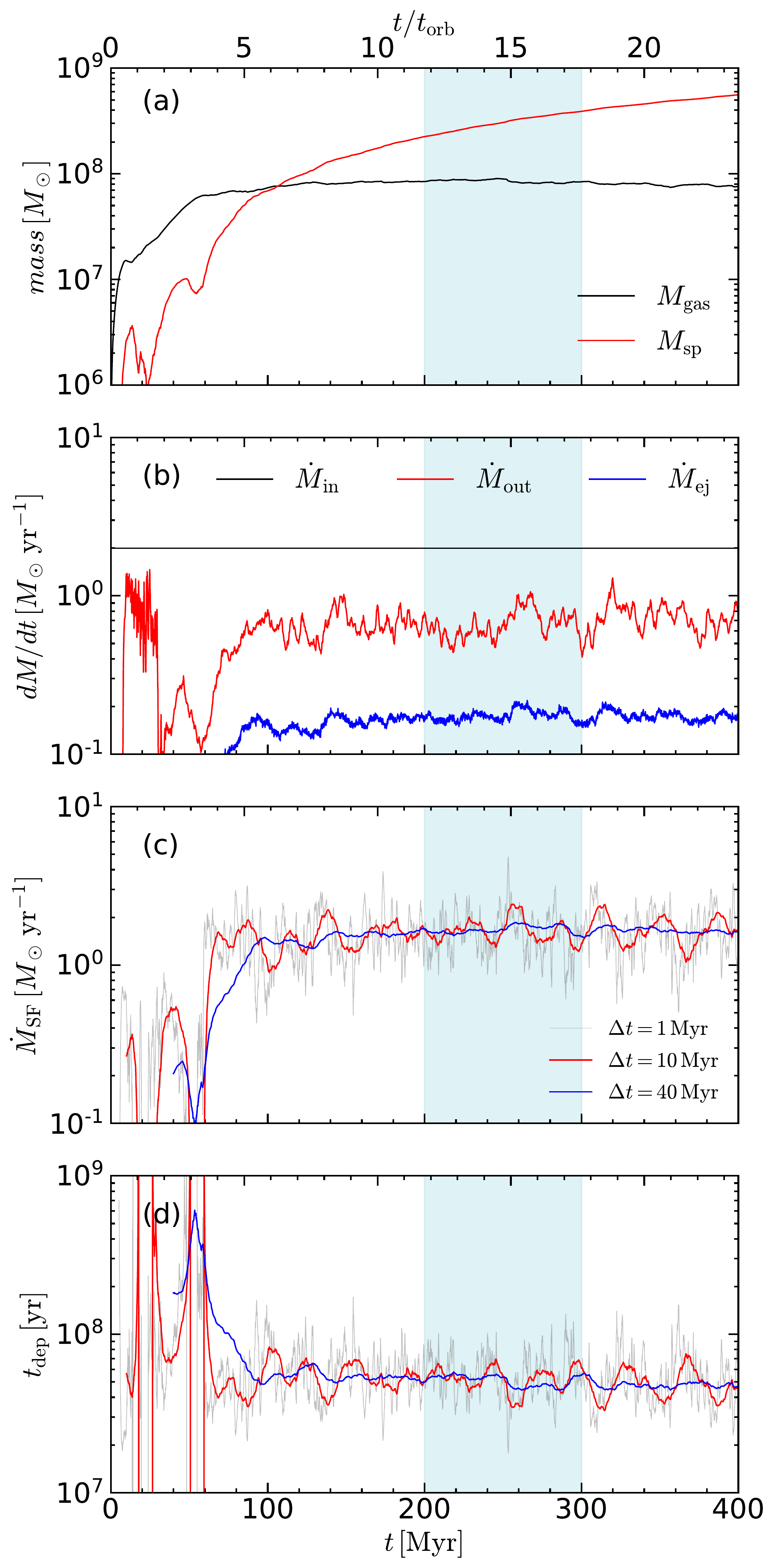}
  \caption{Temporal history in model {\tt L2} of (a)  the total gas mass ($M_{\rm gas}$; black), the total mass of the sink particles representing star clusters ($M_{\rm sp}$; red), (b) the mass inflow rate ($\dot{M}_{\rm in}$; gray) through the nozzles, the mass outflow rate ($\dot{M}_{\rm out}$; red) through the domain boundaries, and the rate of the mass return from SN feedback ($\dot{M}_{\rm ej}$; blue), (c) the SFR averaged over $\Delta t=1\,{\rm Myr}$ (gray),  $\Delta t=10\,{\rm Myr}$ (red), and $40\,{\rm Myr}$ (blue), and
    (d) the gas depletion time $t_{\rm dep} = M_{\rm gas}/\dot{M}_{\rm SF}$ within the domain.
    The mean gas mass, SFR, and the depletion time in the shaded region ($200\,{\rm Myr} \leq t \leq 300\,{\rm Myr}$) are $8.5\times 10^7\,M_\odot$, $1.7\,M_\odot\,{\rm yr}^{-1}$, and $51.2\,{\rm Myr}$, respectively.
  }%
  \label{fig:mass_sfr_history}
\end{figure}

Figure \ref{fig:mass_sfr_history}(a) plots the temporal evolution of  the total mass in the gas $M_\text{gas}$ and in the sink particles $M_\text{sp}$ in model {\tt L2}. The total gas mass saturates to $M_{\rm gas}\sim 8.5\times 10^7\,M_\odot$ within $\sim3t_\text{orb}$, while $M_\text{sp}$  steadily increases with time, except for the initial phase when stars formed in the inflowing streams escape from the computational box.

Figure \ref{fig:mass_sfr_history}(b) plots the mass inflow rate $\dot{M}_\text{in}$ to the box, outflow rate $\dot{M}_\text{out}$ from the box, and deposition rate $\dot{M}_\text{ej}$ from SN ejecta in model {\tt L2}.
The mass inflow rate is fixed to $\dot{M}_{\rm in}=2\,M_\odot\,\text{yr}^{-1}$ at the nozzles, and $\dot{M}_\text{ej}\sim 0.2\,M_\odot\,\text{yr}^{-1}$ is roughly constant after $t=100\,$Myr when the SFR reaches a quasi-steady state.
The mass outflow rate through the domain boundaries is quite large at early time ($t\lesssim 30\,{\rm Myr}$) due to the gas leaving the domain through the $y$-boundaries before gas orbits are circularized (Figure \ref{fig:evolution}). Except for these early transients, $\dot{M}_\text{out}$ is dominated by SN-driven outflows and saturates to $\sim 0.67\,M_\odot\,\text{yr}^{-1}$ with some fluctuations.

We calculate the SFR at time $t$ by
\begin{equation}
  \dot{M}_{\rm SF}(t, \Delta t) = \frac{M_{\rm sp}(t)-M_{\rm sp}(t-\Delta t)}{\Delta t},
\end{equation}
where $\Delta t$ is a chosen time window for averaging.
We take $\Delta t=1$, $10$, or $40\,{\rm Myr}$ to allow for different timescales pertinent to common observational tracers of the SFR (see, e.g., \citealt{kennicutt12}):
$\Delta t=1\,{\rm Myr}$ corresponds to the \emph{instantaneous} SFR, while $\Delta t=10$ and $40\,{\rm Myr}$ are appropriate for ${\rm H}\alpha$ and radio free-free/recombination lines, or to FUV/IR tracers, respectively.
Figure \ref{fig:mass_sfr_history}(c) and (d) plot $\dot{M}_{\rm SF}$ and the gas depletion time $t_{\rm dep}\equiv M_{\rm gas}/\dot{M}_{\rm SF}$, respectively.
At early time, the SFR is lower than the inflow rate and gas builds up in the ring, increasing $M_{\rm gas}$ eligible for star formation.
The SFR increases with time until the system enters a quasi-steady state.
In model {\tt L2}, the steady-state SFR is $\dot{M}_{\rm SF} \sim 1.66\,M_\odot\,{\rm yr^{-1}}$, corresponding to $83\%$ of $\dot{M}_{\rm in}$. The average depletion time after a steady-state is reached is $t_{\rm dep} = 51.2\,{\rm Myr}$.

While the mean values of the SFR is insensitive to $\Delta t$, temporal fluctuations decrease with $\Delta t$.  Fluctuation amplitudes are $0.55$, $0.28$, and $0.09\,M_\odot\,{\rm yr}^{-1}$ for $\Delta t=1, 10$, and $40\,{\rm Myr}$, respectively. Our simulations do not exhibit a  burst/quench cycle as seen in the model of \citet{krumholz17} and simulation of \citet{torrey17}, nor the long-term variation in the depletion time seen in \citet{armillotta19} (who found a range $t_{\rm dep}\sim 10^8-10^9\, {\rm yr})$.
More similar to our results were those of  \citet{sormani20}, who ran moving-mesh simulations with star formation and feedback targeting star formation in the CMZ, and found that the SFR and the depletion time are quite steady with time, with only modest (within factor two) variations.
Our results suggest that under a constant inflow rate, the SFR in nuclear rings would be quite steady.  This would also implies that bursty behavior in real systems is due to variations in the feeding rate from larger scales.
%is remarkably steady.
We shall compare the numerical results with the prediction of the self-regulation theory \citep{oml10,os11} in Section \ref{s:self_regulation}.

\subsection{Other Models}\label{s:other_models}

\begin{figure*}[htpb]
  \centering
  \includegraphics[width=.9\linewidth]{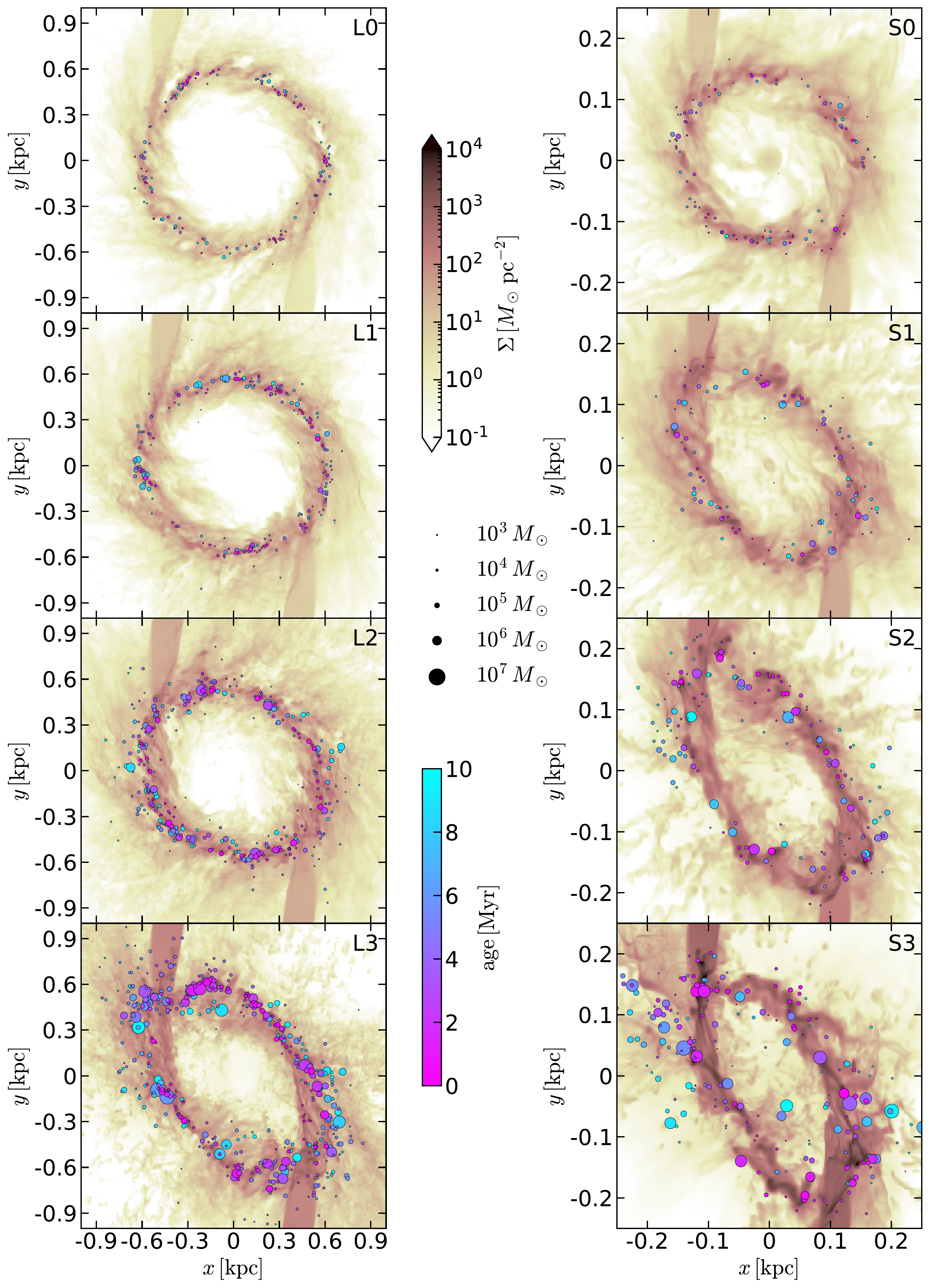}
  \caption{Distributions of the gas surface density $\Sigma$ and sink particles with age less than 10 Myr, for all models at $t=250\,{\rm Myr}$.
    The left and right columns are for the {\tt L} and {\tt S} series, respectively (note difference in box size). From top to bottom, rows correspond to the models with $\dot{M}_\text{in}=0.125, 0.5, 2$, and $8\,M_\odot\,{\rm yr^{-1}}$.
    All panels share the same color scale for $\Sigma$ and the same symbol size and color scale for star particle mass and age, given in the middle.
  }%
  \label{fig:model_comparison}
\end{figure*}

Evolution of the other models is qualitatively similar to that of the fiducial model, although the ring size and shape, SFR, etc. depend significantly on the model parameters $R_\text{ring}$ and $\dot{M}_\text{in}$.  Figure \ref{fig:model_comparison} compares distributions in the $x$-$y$ plane of gas and sink particles for all the models at $t=250\,$Myr, after a steady-state is reached.
The left and right columns correspond to the {\tt L} and {\tt S} series, respectively, with increasing $\dot{M}_\text{in}$ from top to bottom.
In both series, the mean surface density of the ring $\Sigma_\text{ring}$ and its ellipticity increase with $\dot{M}_\text{in}$.
The increase of $\Sigma_\text{ring}$ with $\dot{M}_\text{in}$ is because the associated higher SFR yields larger thermal and turbulent pressures via feedback that support the ring against stronger gravity (see Section \ref{s:self_regulation}).
At higher SFR, gas turns into stars before the inflowing streams are fully circularized, yielding a more eccentric ring.
Rings in the {\tt S} series are overall more eccentric compared to their counterparts in the {\tt L} series. This is because the ratio $t_{\rm dep}/t_\text{orb}$ is smaller in the {\tt S} series, implying that more gas is consumed by star formation before the orbit circularization.
We note that some barred galaxies including NGC 986, NGC 1365, NGC 3351, and NGC 5383 possess an eccentric nuclear ring at their centers, similarly to our models with a high inflow rate.

\begin{figure}[t]
  \centering
  \includegraphics[width=\linewidth]{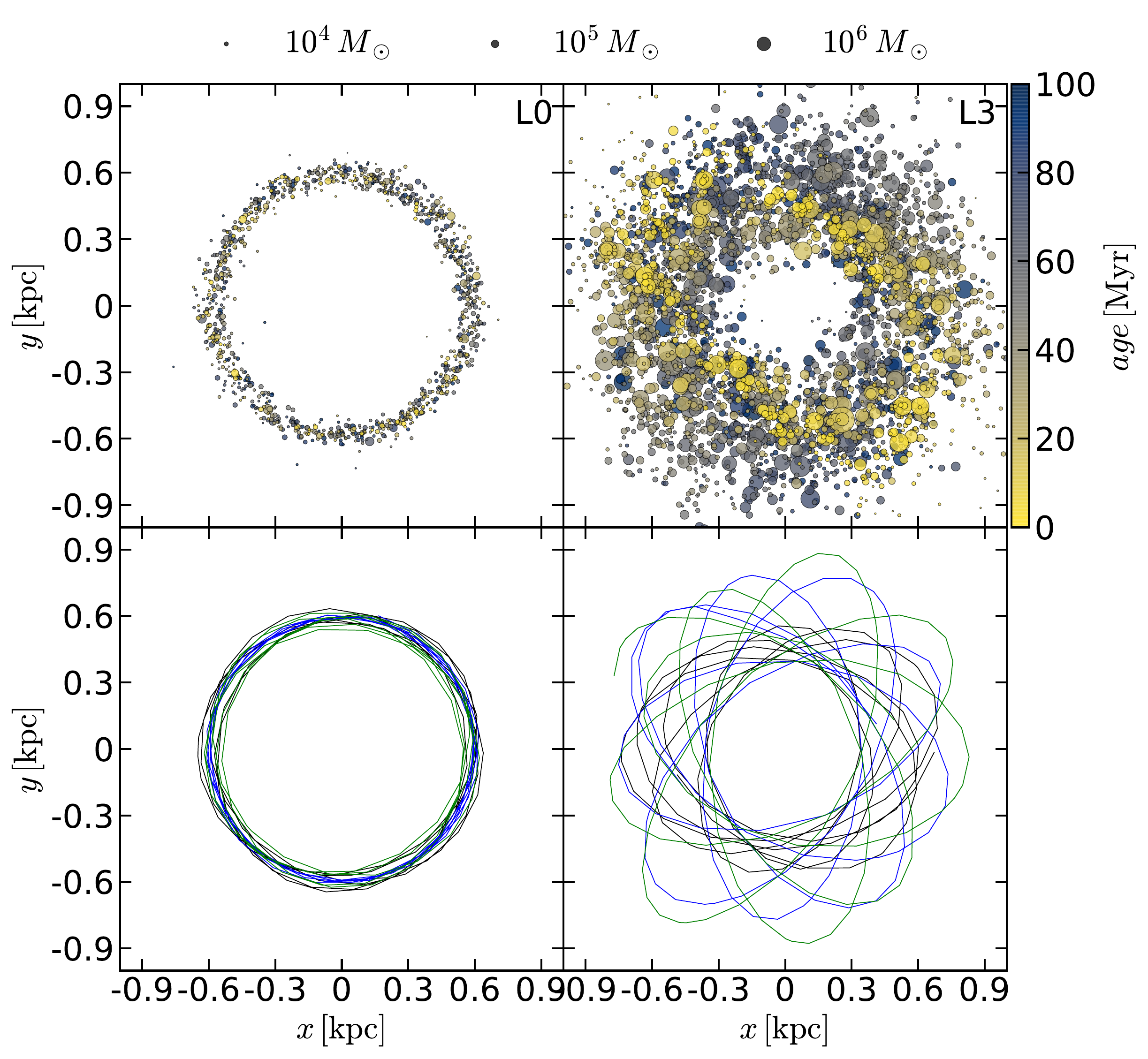}
  \caption{{\it Top}: Spatial distributions of sink particles younger than $100\,{\rm Myr}$ at $t=300\,{\rm Myr}$ for models {\tt L0} (left) and {\tt L3} (right).
    {\it Bottom}: Representative trajectories of some sink particles during $t=200$--$300\,{\rm Myr}$.
  }%
  \label{fig:starpar}
\end{figure}

Figure \ref{fig:model_comparison} shows that the masses of individual sink particles, on average, increase with $\dot{M}_\text{in}$ due to the increase of $\Sigma_\text{ring}$.
Because sink particles inherit the gas velocity from which they form, their initial orbits are eccentric, similarly to the gas ring.
Unlike the gas, however, the sink particles do not suffer direct collisions at the contact points and their orbits freely precess under the total gravitational potential.
As a result, the spatial distribution of the sink particles deviates from that of the gas.
The deviation is more prominent in models with larger $\dot{M}_\text{in}$ due to more eccentric orbits of the sink particles.

Figure \ref{fig:starpar} compares the projected distributions of sink particles and their orbits in the $x$-$y$ planes between models {\tt L0} and {\tt L3}.
Since the ring is almost circular in model {\tt L0}, the orbits of the sink particles are also nearly circular.
Consequently, both the gas and the sink particles form a narrow circular annulus near $R_{\rm ring}$.
In model {\tt L3}, however, the orbits of the sink particles have high eccentricities and precess.
As they age, they diffuse out of the gaseous ring and also precess, occupying a much wider range of radius. Young particles with $t_m \lesssim 10\,{\rm Myr}$ are still found quite close to the eccentric gaseous  ring for model {\tt L3}.
While our simple sink particles retain their individual identities, in reality the older massive clusters could be disrupted  \citep{grijs12, vaisanen14} and form a psuedobulge \citep{kk04}.

In our simulations, the cold-warm gas with $T<2\times 10^4\,{\rm K}$ comprises about $98\%$ of the total mass. To quantify the physical properties of the cold-warm gas in our simulations, we compute the midplane values of the mass-weighted turbulent velocity dispersion $\sigma_z$ and sound speed $c_s$, and also measure the scale height $H$, via
\begin{equation}\label{eq:sigmaz}
  \sigma_z = \left(\frac{\iiint_{z=-\Delta z}^{z=\Delta z} \rho v_z^2\Theta\,dxdydz}{\iiint_{z=-\Delta z}^{z=\Delta z} \rho\Theta\,dxdydz}\right)^{1/2},
\end{equation}
\begin{equation}
  c_s = \left(\frac{\iiint_{z=-\Delta z}^{z=\Delta z} P\Theta\,dxdydz}{\iiint_{z=-\Delta z}^{z=\Delta z} \rho\Theta\,dxdydz}\right)^{1/2},
\end{equation}
\begin{equation}\label{eq:scaleH}
  H = \left(\frac{\iiint_{z=-L/2}^{z=L/2} \rho z^2\Theta\,dxdydz}{\iiint_{z=-L/2}^{z=L/2} \rho\Theta\,dxdydz}\right)^{1/2},
\end{equation}
where $\Delta z=\Delta x$ is the grid spacing along the $z$-direction and the symbol $\Theta$ denotes the phase selector, such that $\Theta=1$ for the cold-warm medium ($T<2\times 10^4\,{\rm K}$) and $\Theta=0$ otherwise.

\begin{figure}[t]
  \centering
  \includegraphics[width=\linewidth]{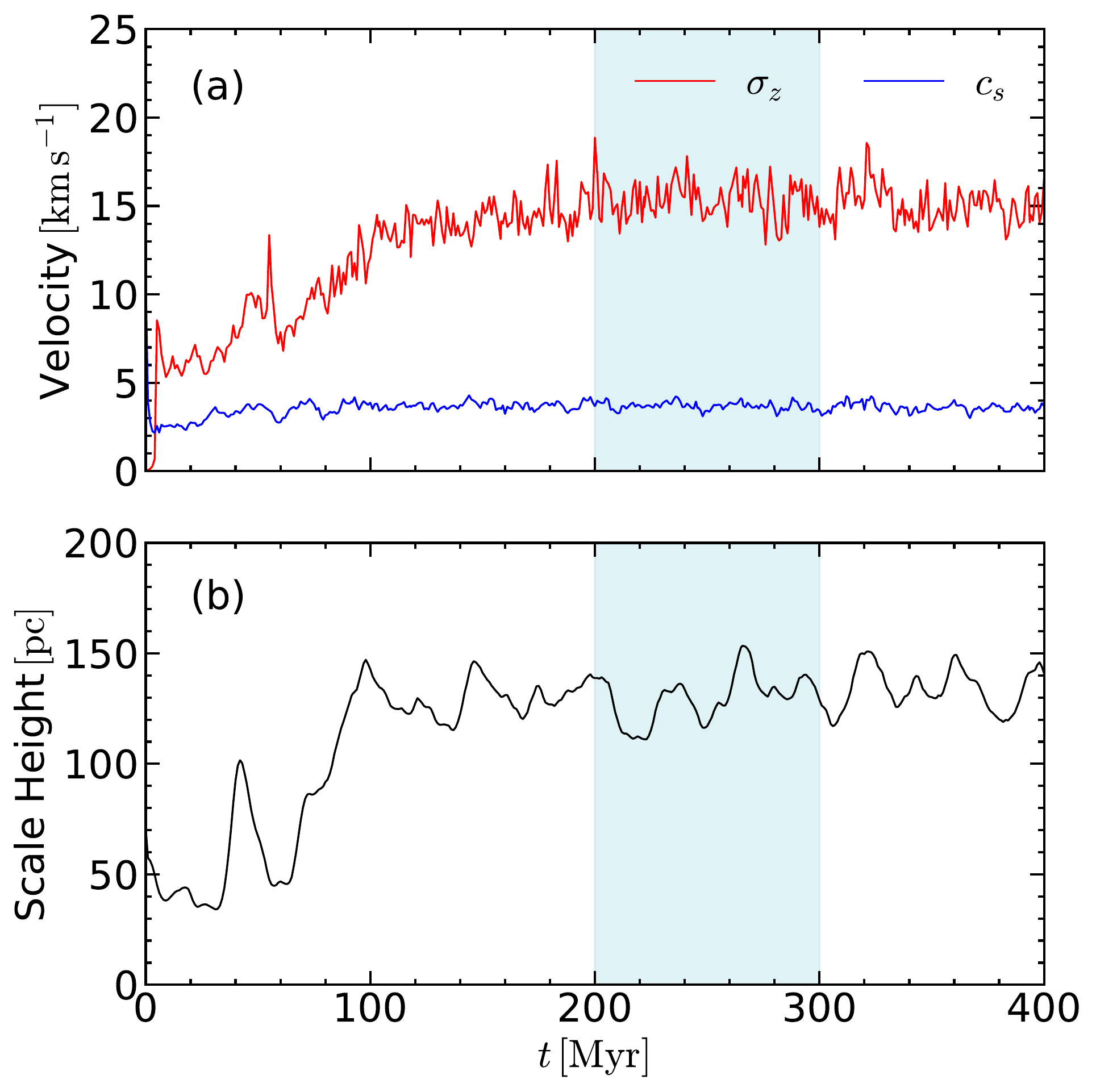}
  \caption{Temporal variations of (a) the vertical velocity dispersion ($\sigma_z$; red) and the sound speed ($c_s$; blue) and (b) the scale height $H$ of the cold-warm medium in model {\tt L2}. The shade indicates the time interval ($200$--$300\,{\rm Myr}$) over which time-averaged quantities are evaluated.}%
  \label{fig:veldisp_evolution}
\end{figure}

Figure \ref{fig:veldisp_evolution} plots the temporal history of $\sigma_z$, $c_s$, and $H$ for model {\tt L2}, showing that these quantities remain more or less constant after $t=100\,$Myr. Note that $\sigma_z\sim 4c_s$ in model {\tt L2}, indicating that the motions of the cold-warm gas are predominantly supersonic.
Table \ref{tb:warmcold} lists the mean values (with standard deviations) of $\sigma_z$, $c_s$, and $H$ averaged over a time span of $\Delta t=100\,{\rm Myr}$ after the system reaches a quasi-steady state.\footnote{\label{fn:tavg}We take an average over $t=300-400\,{\rm Myr}$ for models {\tt L0} and {\tt L1} and over $t=200-300\,{\rm Myr}$ for all the other models, as the former takes a longer time to reach a steady state.}

\begin{deluxetable}{cccc}[t]
  \tablecaption{Steady-state properties of the cold-warm gas
  \label{tb:warmcold}}
  \tablehead{
    \colhead{Model} & \colhead{$\sigma_z$}           & \colhead{$c_s$}        & \colhead{$H$}\\
    \colhead{(1)}   & \colhead{(2)}                  & \colhead{(3)}          & \colhead{(4)}\\
    \colhead{}      & \colhead{$({\rm km\,s^{-1}})$} & \colhead{$({\rm km\,s^{-1}})$} & \colhead{$({\rm pc})$}
  }
  \startdata
  L0  & $8.85\pm 0.66$ & $3.81\pm 0.20$ & $63.88\pm 4.87$ \\
  L1  & $12.6\pm 1.0$  & $3.83\pm 0.20$ & $95.0\pm 5.7$ \\
  L2  & $15.4\pm 1.1$  & $3.69\pm 0.22$ & $130\pm 11$   \\
  L3  & $17.4\pm 1.6$  & $3.30\pm 0.24$ & $151\pm 8$   \\\hline
  S0  & $11.9\pm 1.3$  & $3.66\pm 0.31$ & $33.8\pm 1.9$ \\
  S1  & $13.1\pm 1.3$  & $3.31\pm 0.32$ & $42.8\pm 2.9$ \\
  S2  & $14.8\pm 1.6$  & $2.89\pm 0.28$ & $42.5\pm 2.8$ \\
  S3  & $23.6\pm 3.9$  & $2.89\pm 0.35$ & $21.5\pm 2.9$
  \enddata
  \tablecomments{(1) Model name. (2) Vertical turbulent velocity dispersion at the midplane. (3) Isothermal sound speed at the midplane. (4) Scale height of the gas.}
\end{deluxetable}

\section{Correlations of Statistical Quantities}\label{s:stats}

In this section, we present our measurements of various physical properties of the rings after quasi-steady state is reached, and explore correlations among these properties.

\subsection{Ring Properties}\label{s:ringprops}

\begin{figure}[t]
  \centering
  \includegraphics[width=\linewidth]{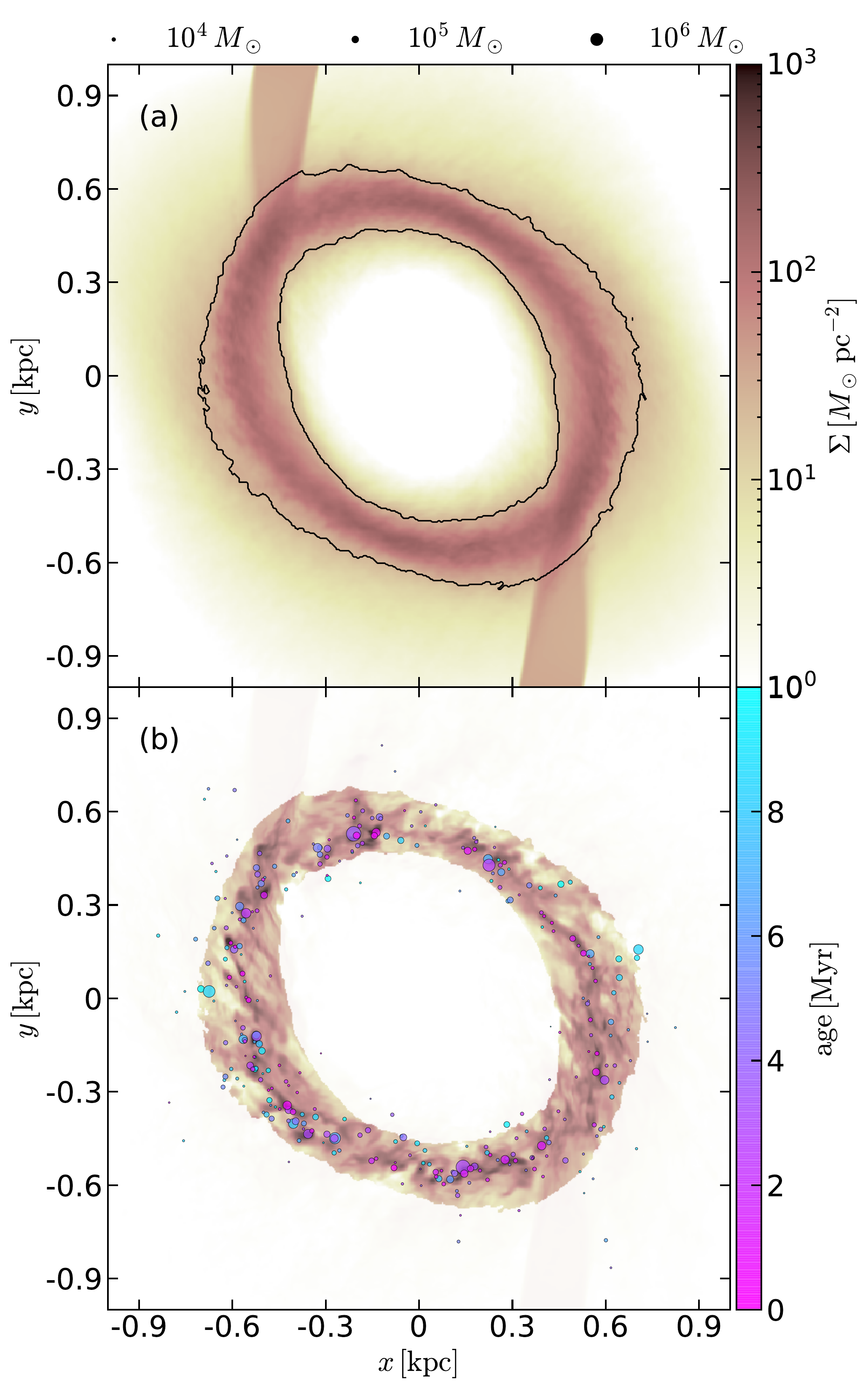}
  \caption{(a) Time-averaged surface density map of model {\tt L2}.
    The solid lines delineate the boundaries of the \emph{ring regions}.
    (b) Example of applying the ring mask to a snapshot of model {\tt L2} at $t=250\,{\rm Myr}$.
    Quantities such as $\Sigma$, $\Sigma_{\rm SFR}$, $\sigma$, $H$, etc. are measured inside the ring regions.
  }%
  \label{fig:ring}
\end{figure}

To properly characterize the average gas surface density $\Sigma_{\rm ring}$ of the ring, it is essential to measure the ring area $A_{\rm ring}$.
For this purpose, we place a circular aperture with radius $R_c$ to exclude the gas streams, and take a temporal average $\langle \Sigma\rangle$ of the gas surface density inside the aperture over a time span of $\Delta t=100\,{\rm Myr}$ after the system reaches a quasi-steady state.
We then identify the collection of cells with $\langle \Sigma\rangle > \Sigma_{\rm crit}$ as the ring region (or ring mask), where $\Sigma_{\rm crit}$ is determined such that the ring region contains  $90\%$ of the total gas mass within $R_c$.
We adjust $R_c$ until it matches the outer semi-major axis of the ring.
Figure \ref{fig:ring}(a) plots as black contours the ring boundaries constructed by this method, overlaid over the time-averaged surface density for model {\tt L2}.
We measure the ring area $A_{\rm ring}$ bounded by the contours, and apply the time-averaged mask to individual snapshots to calculate the ring surface density and the SFR surface density as
\begin{equation}
  \Sigma_{\rm ring} \equiv \frac{M_{\rm ring}}{A_{\rm ring}},
\end{equation}
\begin{equation}\label{eq:SSFR}
  \Sigma_{\rm SFR}  \equiv \frac{\dot{M}_{\rm SF}(t,\Delta t =10\,{\rm Myr})}{A_{\rm ring}},
\end{equation}
where $M_{\rm ring}$ is the gas mass contained in the ring regions.
Note that in Equation \eqref{eq:SSFR}, $\dot{M}_{\rm SF}$ counts sink particles not only inside the ring but also outside the ring since they all have formed inside the ring.
The gas depletion time inside the ring is then given by $t_\text{dep,ring} =\Sigma_{\rm ring}/\Sigma_{\rm SFR}$, which is not much different from $t_{\rm dep}$ over the whole domain defined in Section \ref{s:sfh} since most mass is contained in the ring.
Figure \ref{fig:ring}(b) overlays the ring mask on top of a surface density map at $t=250\,$Myr for model {\tt L2}, illustrating that most mass is enclosed within the mask.
Table \ref{tb:ringprops1} lists the above steady-state physical properties of the rings for all models.
%, measured by the above procedure.

\begin{deluxetable*}{ccccccccc}[t]
  \tablecaption{Steady-state ring properties\label{tb:ringprops1}}
  \tablehead{
    \colhead{Model}     & \colhead{$M_{\rm ring}$}          & \colhead{$\dot{M}_{\rm SF}$} & \colhead{$A_{\rm ring}$}
                        & \colhead{$\Sigma_{\rm ring}$}     & \colhead{$\Sigma_{\rm SFR}$}  & \colhead{$t_{\rm dep,ring}$}
                        & \colhead{$P_{\rm th}$}            & \colhead{$P_{\rm turb}$} \\
    \colhead{(1)}       & \colhead{(2)}                     & \colhead{(3)}                 & \colhead{(4)}
                        & \colhead{(5)}                     & \colhead{(6)}                 & \colhead{(7)}
                        & \colhead{(8)}                     & \colhead{(9)} \\
    \colhead{}          & \colhead{$(10^7\,M_\odot)$}       & \colhead{$(M_\odot\,{\rm yr}^{-1})$}      & \colhead{$({\rm kpc}^2)$}
                        & \colhead{$(\,M_\odot\,{\rm pc}^{-2})$} & \colhead{$(M_\odot\,{\rm yr}^{-1}\,{\rm kpc}^{-2})$}  & \colhead{$({\rm Myr})$}
                        & \colhead{$(10^6\,k_{\rm B}\,{\rm K\,cm^{-3}})$} & \colhead{$(10^6\,k_{\rm B}\,{\rm K\,cm^{-3}})$}
  }
  \startdata
 {\tt L0}                    & $1.25\pm 0.05$ & $0.0897\pm 0.0166$ & $0.607$ & $20.7\pm 0.8$      &      $0.148\pm 0.027$      &      $144\pm 24$ & $0.128\pm0.022$ & $0.137\pm0.022$ \\
 {\tt L1}                    & $2.89\pm 0.07$ & $0.425\pm 0.065$ & $0.656$ & $44.0\pm 1.0$      &      $0.648\pm 0.099$      &      $69.3\pm 9.5$ & $0.424\pm0.051$ & $0.512\pm0.079$\\
 {\tt L2}                    & $6.58\pm 0.32$ & $1.67\pm 0.29$ & $0.805$ & $81.8\pm 3.9$      &      $2.07\pm 0.37$        &      $40.6\pm 7.0$ & $0.966\pm0.106$ & $1.49\pm0.19$\\
 {\tt L3}                    & $14.4\pm 0.4$ & $6.98\pm 0.76$ & $0.879$ & $164\pm 5$         &      $7.94\pm 0.86$        &      $20.9\pm 2.6$ & $1.99\pm0.32$ & $4.49\pm0.78$\\\hline
 {\tt S0}                    & $0.263\pm 0.009$ & $0.0891\pm 0.0115$ & $0.0475$ & $55.3\pm 1.8$      &      $1.88\pm 0.24$        &      $30.0\pm 4.6$ & $1.01\pm0.30$ & $1.44\pm0.33$\\
 {\tt S1}                    & $0.619\pm 0.043$ & $0.378\pm 0.056$ & $0.0567$ & $109\pm 8$         &      $6.67\pm 0.98$        &      $16.8\pm 3.4$ & $2.32\pm0.42$ & $4.26\pm0.82$\\
 {\tt S2}                    & $1.43\pm 0.07$ & $1.64\pm 0.15$ & $0.0604$ & $236\pm 12$        &      $27.1\pm 2.5$         &      $8.81\pm 1.07$ & $3.82\pm0.78$ & $12.7\pm2.7$\\
 {\tt S3}                    & $3.66\pm 0.26$ & $6.68\pm 0.52$ & $0.0705$ & $518\pm 37$        &      $94.8\pm 7.4$         &      $5.52\pm 0.67$ & $9.51\pm2.06$ & $57.4\pm19.7$
  \enddata
  \tablecomments{(1) Model name. (2) Total gas mass inside the ring. (3) Total star formation rate. (4) Area of the ring. (5) Mean surface density of the ring. (6) Averaged SFR surface density of the ring. (7) Gas depletion time of the ring. (8) Midplane thermal pressure. (9) Midplane turbulent pressure.}
\end{deluxetable*}

\begin{figure}[t]
  \centering
  \includegraphics[width=\linewidth]{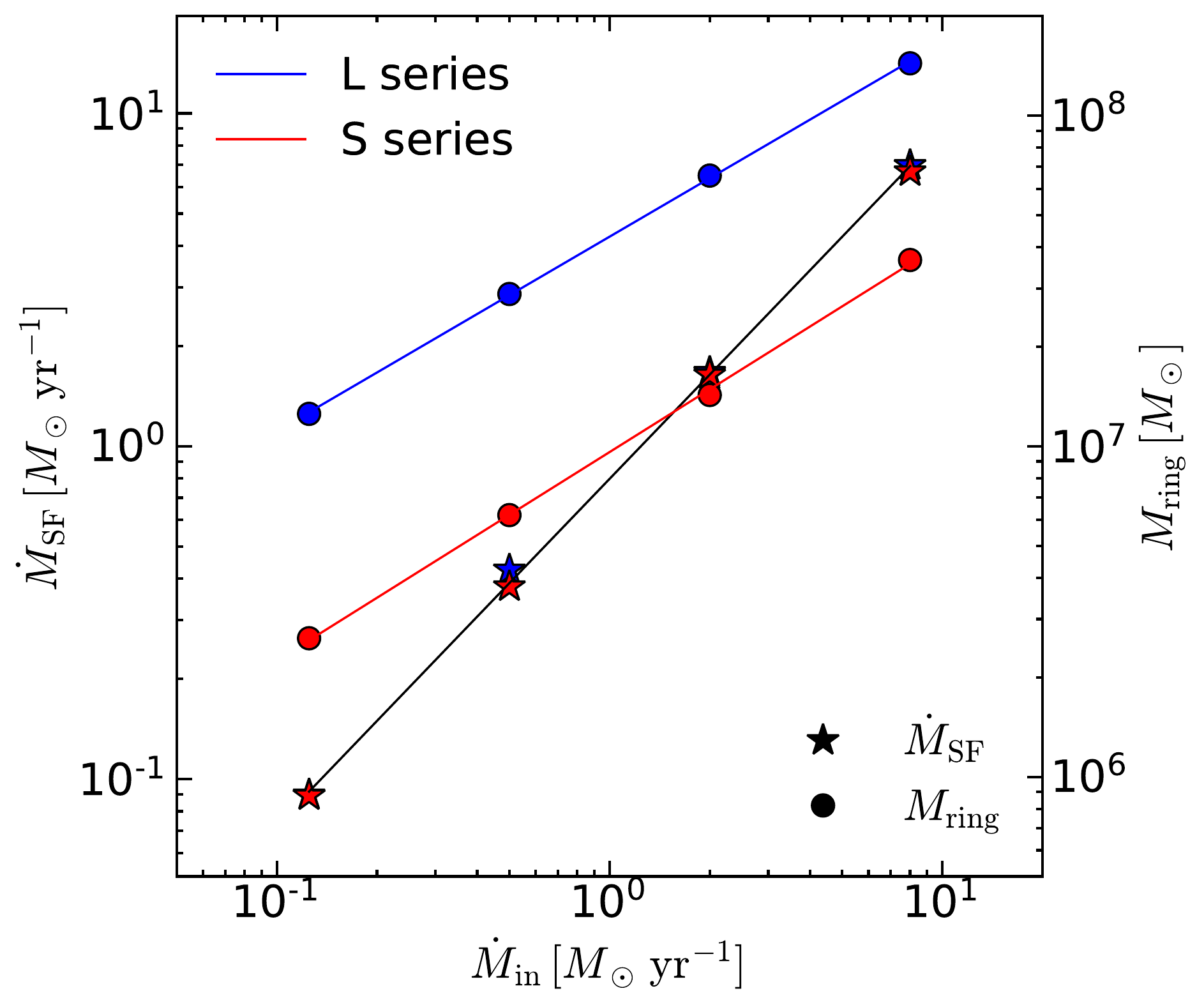}
  \caption{Star formation rate $\dot{M}_\text{SF}$ (star symbols, left scale) and the ring mass $M_\text{ring}$ (circles, right scale) against the inflow rate $\dot{M}_\text{in}$ for all models. The blue and red color correspond to the {\tt L} and {\tt S} series, respectively. The solid lines show the linear fits described in the text.}%
  \label{fig:Mdot_sfr_Mgas}
\end{figure}

Figure \ref{fig:Mdot_sfr_Mgas} plots the dependence of $\dot{M}_{\rm SF}$ and $M_\text{ring}$ on $\dot{M}_\text{in}$, showing strong correlations.  The ring mass in the {\tt L} series follows a relation $M_\text{ring}\approx 4\times 10^7\,M_\odot\,(\dot{M}_\text{in}/1\,M_\odot\,{\rm yr}^{-1})^{0.6}$, scaled up by a factor $\sim 4$ relative to the analogous relation for the  {\tt S} series, $M_\text{ring}\approx 1\times  10^7\,M_\odot\,(\dot{M}_\text{in}/1\,M_\odot\,{\rm yr}^{-1})^{0.6}$. In contrast, the SFR is practically the same, $\dot{M}_\text{SF}\approx 0.8\,\dot{M}_\text{in}$, for both series.
This demonstrates that in our models the SFR is determined by the mass inflow rate rather than the ring mass.
We note that in a given series, $M_{\rm ring}$ varies by a factor of $\sim 10$, while  $\dot{M}_\text{in}$ and  $\dot{M}_{\rm SF}$ vary by a factor of $\sim 80$.
This is because $\Sigma_{\rm SFR}$ is superlinearly proportional to $\Sigma_{\rm ring}$, which we will explore in Section \ref{s:self_regulation}.
This appears consistent with observations that ring masses do not vary much among galaxies, while SFRs vary widely \citep{sheth05, mazzuca08}.

\begin{figure}[htpb]
  \centering
  \includegraphics[width=\linewidth]{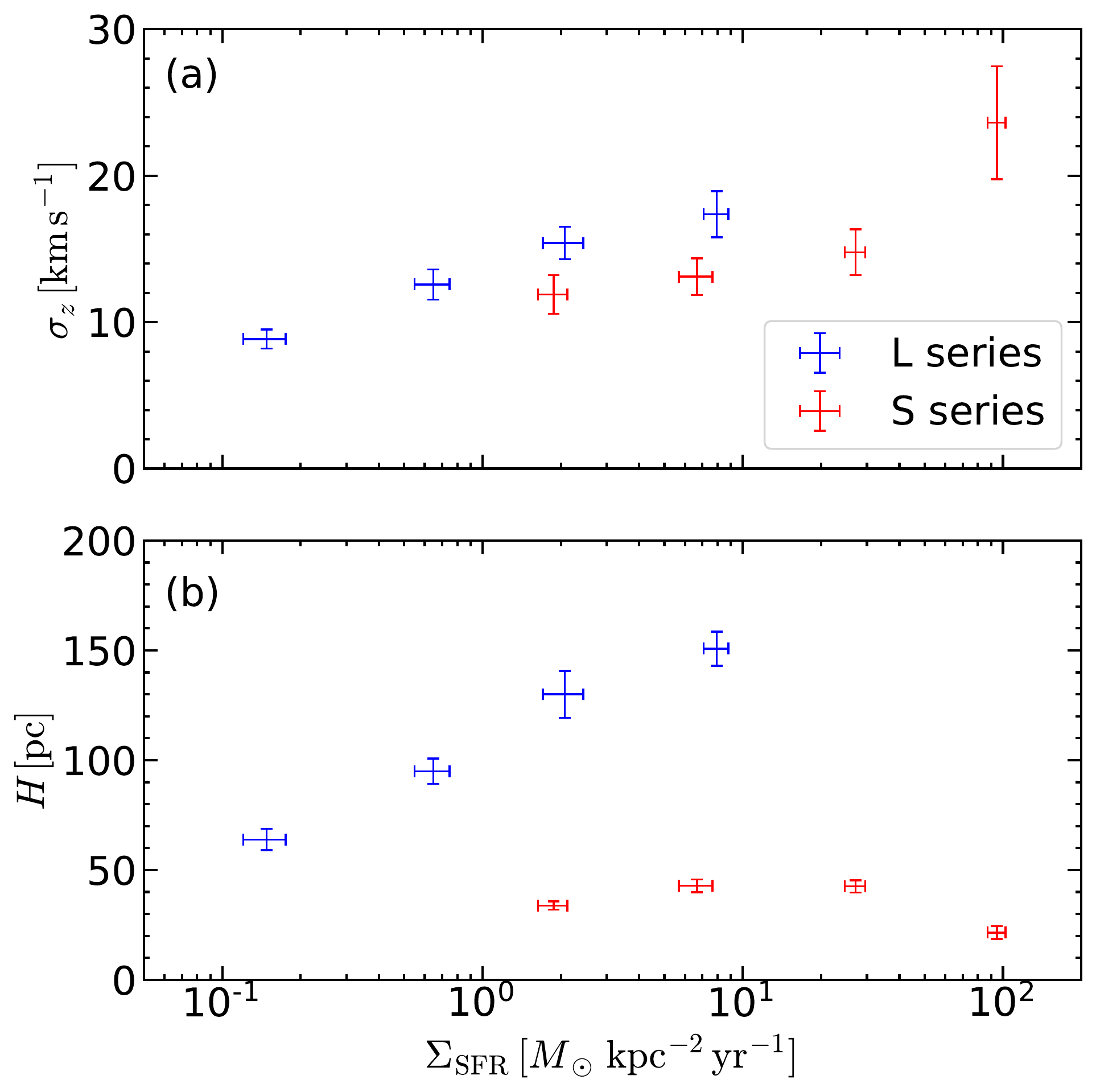}
  \caption{Dependence on the SFR surface density $\Sigma_\text{SFR}$ of (a) the time-averaged vertical turbulent velocity dispersion $\sigma_z$ and (b) the time-averaged vertical scale height $H$.}%
  \label{fig:veldisp_scaling}
\end{figure}

Figure \ref{fig:veldisp_scaling} plots the quasi-steady values of $\sigma_z$ and $H$, measured via Equations~\eqref{eq:sigmaz} and \eqref{eq:scaleH},
%in Section \ref{s:other_models}
against $\Sigma_{\rm SFR}$ for all models, with errorbars corresponding to the standard deviations.
Note that the turbulent velocity dispersion increases weakly with $\Sigma_{\rm SFR}$; the same increasing trend was also seen by \citet[but for a much smaller range of $\Sigma_{\rm SFR}$]{so12} and \citet{orr20}. At low $\Sigma_{\rm SFR}$, $\sigma_z$ is comparable to the solar neighborhood TIGRESS model of \citet{tigress}.  In analogous local-box TIGRESS models with higher gas and stellar density that yield $\Sigma_{\rm SFR}\sim 0.1-1\, M_\odot\,{\rm yr}^{-1}\,{\rm kpc}^{-2}$ (\citealt{kim20} and Ostriker \& Kim 2021, in preparation), the values of $\sigma_z$ are slightly higher, similar to the results shown here.

The scale height in the {\tt S} series is smaller than in the {\tt L} series because of the stronger external gravitational potential. In the {\tt L} series, $H$ increases with $\Sigma_\text{SFR}$ due to the increase in $\sigma_z$, while it in the {\tt S} series is almost constant or decreases with $\Sigma_\text{SFR}$ because of the increased gravity (see below).

\subsection{Vertical Dynamical Equilibrium and Star Formation Feedback}\label{s:self_regulation}

Because the nuclear rings in our simulations are not transient but persist over many orbital times, the weight of the ISM, $\mathcal{W}$, must be supported by the midplane pressure $P_{\rm mid}$ \citep[e.g.][]{bc90,elm94,wmht03}.
The pressure needed for vertical dynamical equilibrium can be maintained only if there are sources of energy and momentum, primarily from young, massive stars \citep{oml10,os11}: the thermal and turbulent pressures would otherwise decay due to radiative cooling and turbulent dissipation on short timescales.
In our simulations, the thermal and turbulent pressures are replenished by FUV and CR heating and SN feedback, with the latter being dominant.

\subsubsection{Vertical dynamical equilibrium}

\begin{figure}[t]
  \centering
  \includegraphics[width=\linewidth]{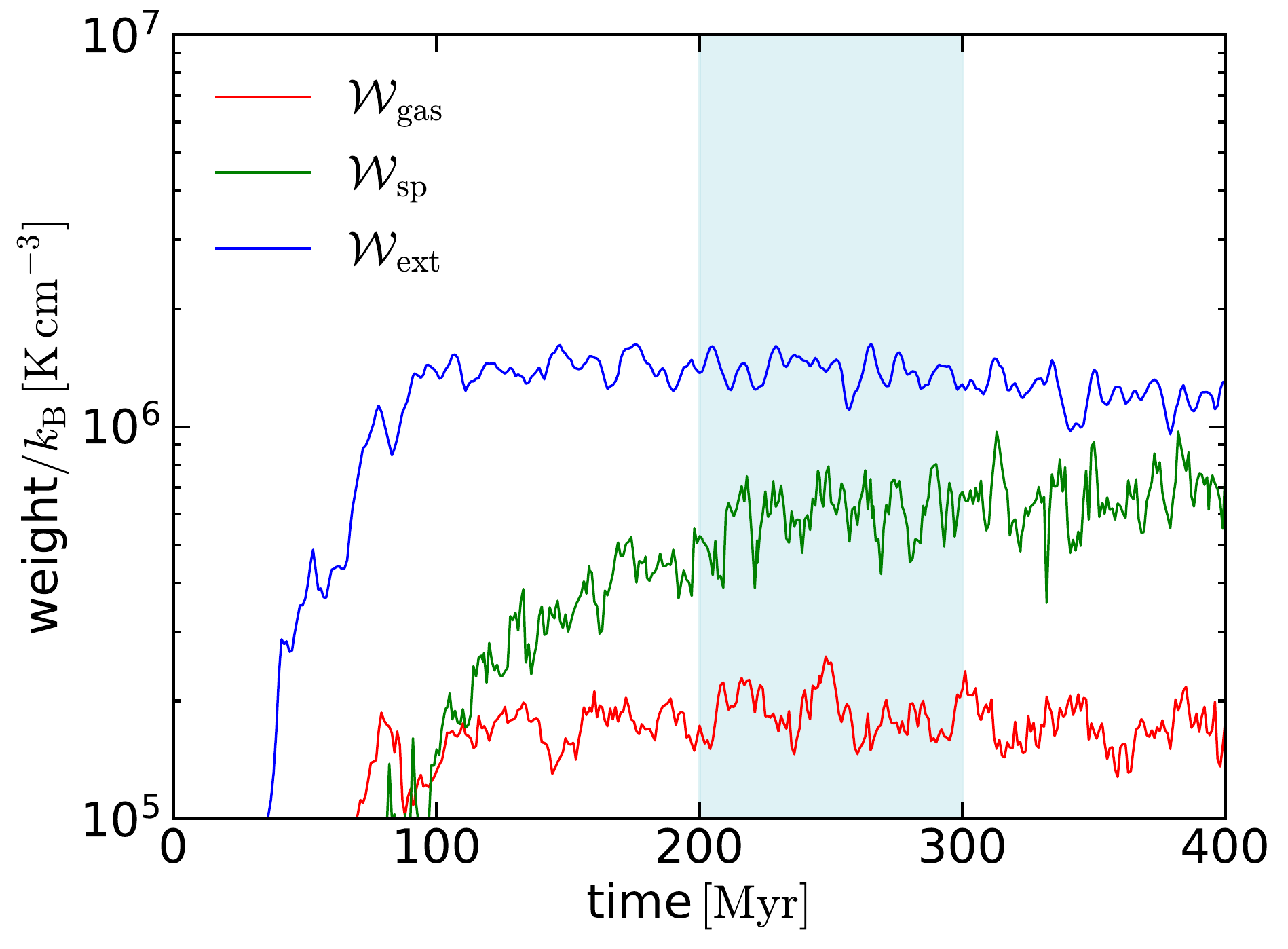}
  \caption{Time evolution of the gas weight contributions due to the gaseous self-gravity (${\mathcal W}_{\rm gas}$), the gravity of the sink particles (${\mathcal W}_{\rm sp}$), and the external potential (${\mathcal W}_{\rm ext}$) in model {\tt L2}. The shaded region indicates that time span over which the weights are averaged.
  }%
  \label{fig:weights_evolution}
\end{figure}

We measure the midplane thermal and turbulent pressures inside the ring by
\begin{align}\label{eq:Pth}
  P_{\rm th} & \equiv \frac{1}{2\Delta z A_{\rm ring}}  \int_{-\Delta z}^{\Delta z}\iint_{A_{\rm ring}} P\,dxdydz , \\
  \label{eq:Pturb}
  P_{\rm turb} &\equiv \frac{1}{2\Delta z A_{\rm ring}} \int_{-\Delta z}^{\Delta z}\iint_{A_{\rm ring}} \rho v_z^2\,dxdydz.
\end{align}
We separately measure the weight of the gas due to its own gravitational field, to the gravity of the sink particles, and to the external gravity from the stellar bulge as
\begin{equation}\label{eq:Wgas}
 {\mathcal W}_{\rm gas} \equiv \frac{1}{A_{\rm ring}} \iint_{A_{\rm ring}} \left(\int_0^{\infty} \rho \frac{\partial\Phi_{\rm gas}}{\partial z}\,dz\right)dxdy,
\end{equation}
\begin{equation}\label{eq:Wsp}
 {\mathcal W}_{\rm sp} \equiv \frac{1}{A_{\rm ring}} \iint_{A_{\rm ring}}\left(\int_0^{\infty} \rho \frac{\partial\Phi_{\rm sp}}{\partial z}\,dz\right)dxdy,
\end{equation}
\begin{equation}\label{eq:Wext}
 {\mathcal W}_{\rm ext} \equiv \frac{1}{A_{\rm ring}} \iint_{A_{\rm ring}}\left(\int_0^{\infty} \rho \frac{\partial\Phi_{\rm ext}}{\partial z}\,dz\right)dxdy,
\end{equation}
where $\Phi_{\rm gas}$ and $\Phi_{\rm sp}$ refer to the gravitational potential of the gas and sink particles, respectively, such that $\Phi_{\rm self} =\Phi_{\rm gas}+\Phi_{\rm sp}$.
Figure \ref{fig:weights_evolution} plots the temporal evolution of ${\mathcal W}_{\rm gas}$, ${\mathcal W}_{\rm sp}$, and ${\mathcal W}_{\rm ext}$ of model {\tt L2} as red, green, and blue lines, respectively.
After the system reaches a quasi-steady state ($t\gtrsim 100\,{\rm Myr}$), ${\mathcal W}_{\rm ext}$ and ${\mathcal W}_{\rm gas}$ do not vary much, while ${\mathcal W}_{\rm sp}$ keeps increasing due to the continuous creation of the sink particles.
The time-averaged weights over $t=200-300\,{\rm Myr}$ are ${\mathcal W}_{\rm gas}/k_B=1.9\times 10^5\,{\rm K\,cm^{-3}}$, ${\mathcal W}_{\rm sp}/k_B=6\times 10^5\,{\rm K\,cm^{-3}}$, and ${\mathcal W}_{\rm ext}/k_B=1.4\times 10^6\,{\rm K\,cm^{-3}}$, indicating that the gas weight in model {\tt L2} is mostly due to $\Phi_{\rm sp}$ and $\Phi_{\rm ext}$ rather than $\Phi_\text{gas}$.

We now compare the total midplane pressure $P_{\rm mid}=P_{\rm th}+P_{\rm turb}$ with the total weight ${\mathcal W}_{\rm tot}={\mathcal W}_{\rm gas}+{\mathcal W}_{\rm sp}+{\mathcal W}_{\rm ext}$ of the ISM, checking to what extent vertical dynamical equilibrium holds.
Vertical dynamical equilibrium requires
\begin{equation}\label{eq:Pdiff}
     P_{\rm mid} - P_{\rm top} = {\mathcal W}_{\rm tot},
\end{equation}
where $P_{\rm top}$ is the total pressure at the top boundary ($z=L/2$) defined as
\begin{equation}
P_{\rm top}\equiv \frac{1}{A_{\rm ring}} \iint (P+\rho v_z^2)_{z=L/2}\,dxdy.
\end{equation}
In normal situations, $P_{\rm top} \ll P_{\rm mid}$, leading to the usual equilibrium condition $P_{\rm mid}\approx {\mathcal W}_{\rm tot}$. If a system develops strong outflows with high ram pressure and the vertical extent is small, however, $P_{\rm top}$ may no longer be negligible compared to the midplane pressure.

\begin{figure}[t]
  \centering
  \includegraphics[width=\linewidth]{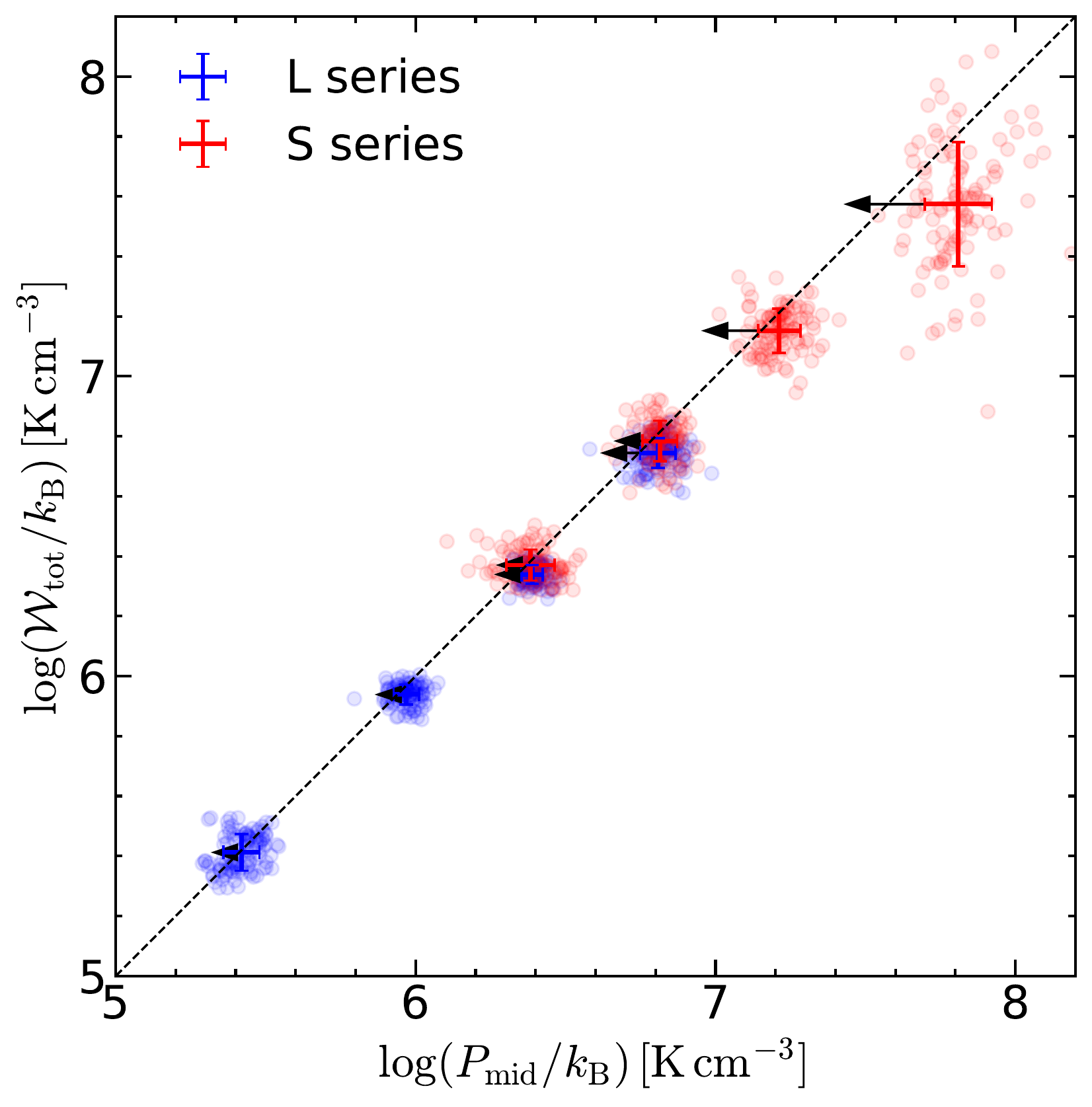}
  \caption{Total weight of the ISM against the total midplane pressure.
    Each circle corresponds to a different snapshot during $t=200-300\,{\rm Myr}$.
    Errorbars denote the temporal averages and standard deviations of each model.
    The dashed line corresponds to ${\mathcal W}_\text{tot}=P_{\rm mid}$.  Black arrows indicate the shifts of the mean values in the abscissa when $P_{\rm mid}$ is changed to $P_{\rm mid}-P_{\rm top}$.
  }%
  \label{fig:vertical_equilibrium}
\end{figure}

Figure \ref{fig:vertical_equilibrium} plots for individual snapshots ${\mathcal W}_{\rm tot}$ against $P_{\rm mid}$ over the interval $\Delta t=100\,{\rm Myr}$ after steady state is reached, using blue and red for the {\tt L} and {\tt S} series, respectively.  Additionally, points with errorbars indicate means and standard deviations for each model.
All models closely follow the ${\mathcal W}_{\rm tot}=P_{\rm mid}$ line, except for models {\tt S2} and  {\tt S3} which lie slightly below the line.  The black arrows denote the shifts of the mean values in the abscissa when $P_{\rm mid}$ is changed to $P_{\rm mid}-P_{\rm top}$, indicating that $P_{\rm top}$ is significant ($\approx 0.4P_{\rm mid}$) in models {\tt S2} and {\tt S3}.
After correcting for $P_{\rm top}$, all models satisfy Equation \eqref{eq:Pdiff}, demonstrating that vertical dynamical equilibrium is  maintained in an averaged sense.

\begin{figure}[t]
  \centering
  \includegraphics[width=\linewidth]{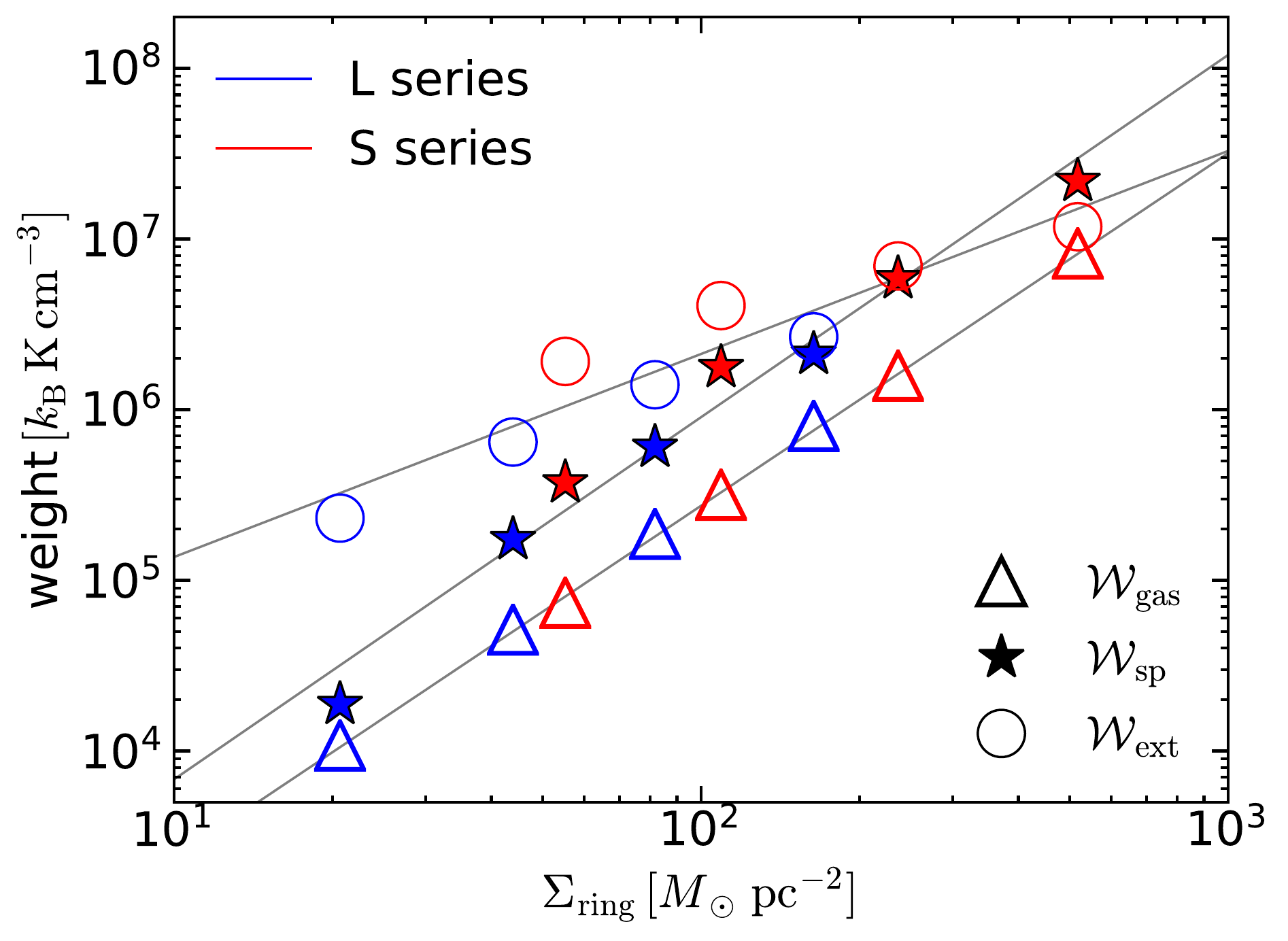}
  \caption{Dependence of the gas weights on the ring surface density. Blue and red symbols correspond to the {\tt L} and {\tt S} series, respectively. Triangles, stars, and circles represent the gas weight due to the gaseous self-gravity, the gravity of the sink particles, and the external potential, respectively. All quantities are averaged over $\Delta t=100\,{\rm Myr}$. Lines are linear fits to the numerical results with slopes of $2.07$, $2.12$, and $1.19$ for ${\mathcal W}_{\rm gas}$, ${\mathcal W}_{\rm sp}$, and ${\mathcal W}_{\rm ext}$, respectively.
  }%
  \label{fig:weights_scaling}
\end{figure}

\subsubsection{Scaling relations of the gas weights}
Figure \ref{fig:weights_scaling} plots the time-averaged gas weights of all models against $\Sigma_{\rm ring}$.
For a plane-parallel slab with total gas surface density $\Sigma_{\rm ring}$, ${\mathcal W}_{\rm gas} \approx \pi G\Sigma_{\rm ring}^2/2$; while this does not apply exactly  given the ring geometry, a quadratic scaling is still expected. If the scale height of the gas disk is larger than that of young stars but smaller than that of old stars, the corresponding weights for horizontally uniform disks would be ${\mathcal W}_{\rm sp}\approx \pi G\Sigma_{\rm ring}\Sigma_{\rm sp}$ and ${\mathcal W}_{\rm ext}\approx \Sigma_{\rm ring}\sigma_z(2G\rho_b/3)^{1/2}$, where $\Sigma_{\rm sp}$ is the surface density of the sink particles, $\rho_b$ is the volume density of old stars in the bulge at the midplane, and a Gaussian vertical profile is assumed \citep{os11}; again these can only  be approximate given the ring geometry, bulge stratification, etc.

Assuming that the weights are proportional to $\Sigma_{\rm ring}^p$ with a power-law index $p$, simple linear fits to our results yield $p=2.07$, $2.12$, and $1.19$ for ${\mathcal W}_{\rm gas}$, ${\mathcal W}_{\rm sp}$, and ${\mathcal W}_{\rm ext}$, respectively.
These are broadly consistent with the above prediction, under the condition that $\Sigma_{\rm sp}$ is roughly proportional to $\Sigma_{\rm ring}$ and that $\sigma_z$ depends on $\Sigma_{\rm ring}$ only weakly (from Figure \ref{fig:veldisp_scaling}).\footnote{Since $\rho_b$ is larger for the {\tt S} series than the {\tt L} series, ${\cal W}_{\rm ext}$ is offset upward for the former, as expected from ${\cal W}_{\rm ext} \propto \Sigma_{\rm ring} \sqrt{\rho_b}$.  With a constant SFR, $\Sigma_{\rm sp} \propto \Sigma_{\rm SFR} t$. We shall show that from self-regulation, $\Sigma_{\rm SFR}$ varies approximately linearly in ${\cal W}_{\rm tot}$, so for weight dominated by the external potential we expect $\Sigma_{\rm sp} \propto \Sigma_{\rm ring} t \sigma_z\sqrt{\rho_b} $, which would then yield ${\cal W}_{\rm sp} \propto \Sigma_{\rm ring}^2 t \sigma_z \sqrt{\rho_b}$. } Although the gas disks in our models neither have a plane-parallel geometry nor are well described by a Gaussian vertical profile, deviations of each weight contributions measured using Equations \eqref{eq:Wgas}--\eqref{eq:Wext} from analytical predictions are at most a factor of $\sim 3$.

The total weight of the gas is dominated by ${\mathcal W}_{\rm ext}$ at low $\Sigma_\text{ring}$, while ${\mathcal W}_{\rm sp}$ also becomes significant at high $\Sigma_\text{ring}$. For all models, self-gravity has a relatively minor contribution to the total weight, different from the models of \citet{os11} and \citet{so12}.
There, the weight in the external potential was (by design) smaller than the weight of the gas, and the weight from star particles was not considered because it was implicitly assumed that the starburst had a sufficiently short duration that significant stellar mass did not build up (the present models show that this indeed requires hundreds of Myr). In the local simulations of \citet{os11} and \citet{so12}, $M_{\rm gas}$  did not build up over time but was imposed from the initial conditions, so it could be large without also having $M_{\rm sp}$ large (unlike the case for the present models, per Figure~\ref{fig:Mdot_sfr_Mgas}).
In the present simulations, the value of $\rho_b$ in the ring region is relatively large, because nuclear rings form more easily in the presence of a strong central concentration \citep{athanassoula92,rt03,li15}.
%due to numerical requirements for achieving high resolution while simultaneously having ring formation without an explicit bar model.
We note that if the star formation efficiency within the sink particles is not 100\%, only a fraction of ${\mathcal W}_{\rm sp}$ would be regarded as being self-gravitational.
In creating a sink particle, we assume that all of the gas in a cell is immediately converted to  a star cluster.
In contrast, \citet{tress20} assumed only  $5\%$ of the sink particle mass actually represents the mass of the star cluster, while treating the remaining $95\%$ as gas `temporarily stored' in the sink particles which is later returned to the ambient ISM via SN feedback.  We have adopted the current approach for simplicity, but in future work it would be quite interesting to test whether a treatment of sink particles with mass loss would change the results for the various scaling relations studied in this work.

\begin{figure*}[htpb]
  \centering
  \includegraphics[width=\linewidth]{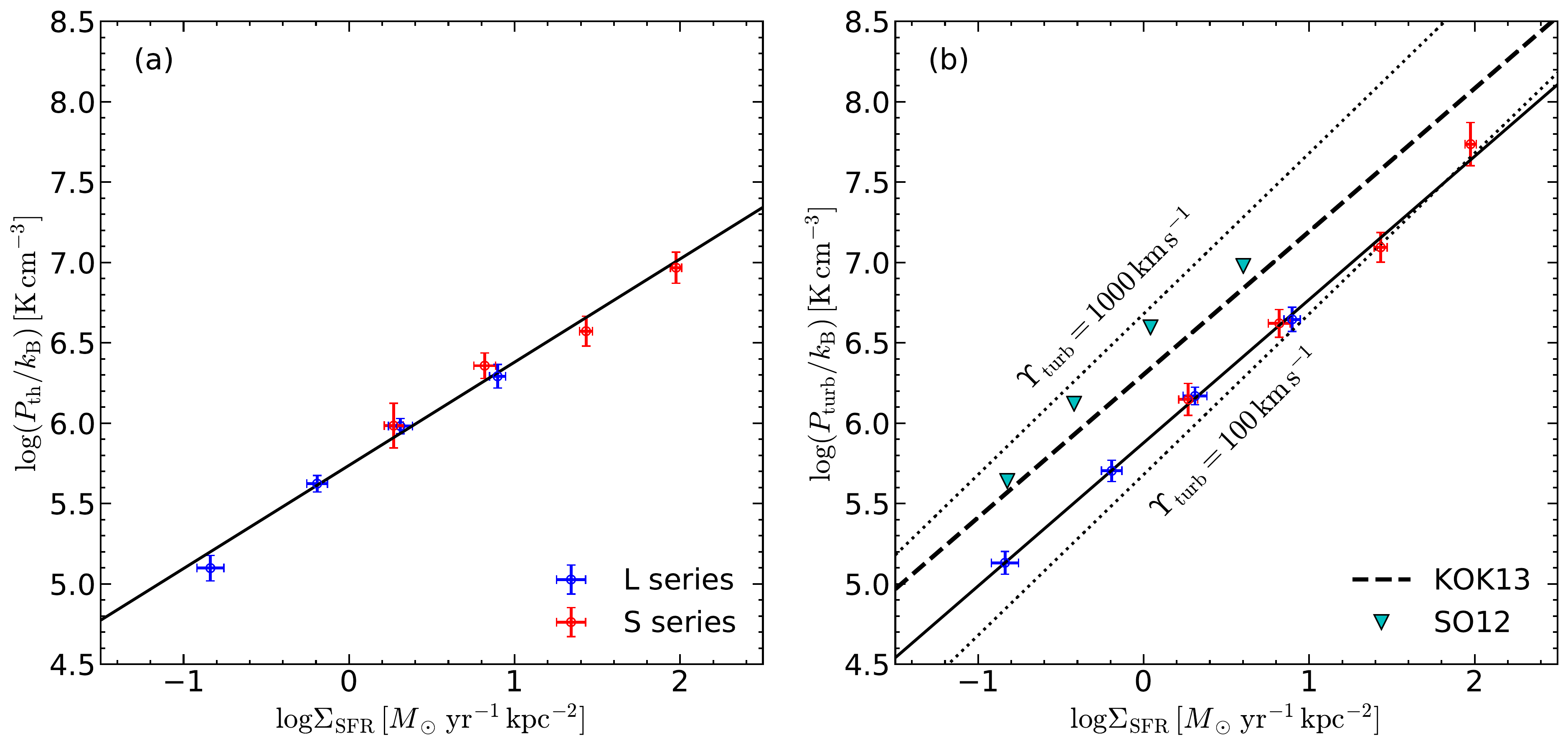}
  \caption{Mean (a) thermal and (b) turbulent pressure at the midplane, plotted against the mean SFR surface density within the ring, for each model (see also Table~\ref{tb:ringprops1}). The solid lines are the linear fits, Equations \eqref{eq:Pth_bestfit} and \eqref{eq:Pturb_bestfit}. In (b), cyan triangles are the results of \citet[Series S]{so12}, while the dashed line is the extrapolation of Equation (21) from \citet{kok13}. The two dotted lines correspond to $\Upsilon_{\rm turb}=100\,{\rm km\,s^{-1}}$ and $1000\,{\rm km\,s^{-1}}$ for reference.}%
  \label{fig:pturb_sfr}
\end{figure*}

\subsubsection{Pressure scaling relations and the feedback yields}\label{s:feedback_yields}
As expected, the pressures are directly correlated with the star formation rate per unit area.
Figure \ref{fig:pturb_sfr} plots $P_{\rm th}$ and $P_{\rm turb}$ against $\Sigma_{\rm SFR}$ as  points with errorbars,  corresponding to the mean values and standard deviations, respectively.
Our best fits, including both the {\tt S} and {\tt L} series, are
\begin{equation}\label{eq:Pth_bestfit}
  \frac{P_{\rm th}}{k_{\rm B}} = 5.47\times 10^5\,{\rm cm^{-3}\,K}\,\left( \frac{\Sigma_{\rm SFR}}{M_\odot\,{\rm kpc^{-2}\,yr^{-1}}} \right)^{0.64},
\end{equation}
\begin{equation}\label{eq:Pturb_bestfit}
  \frac{P_{\rm turb}}{k_{\rm B}} = 7.56\times 10^5\,{\rm cm^{-3}\,K}\,\left( \frac{\Sigma_{\rm SFR}}{M_\odot\,{\rm kpc^{-2}\,yr^{-1}}} \right)^{0.89},
\end{equation}
plotted as solid lines.
The thermal pressure increases with $\Sigma_{\rm SFR}$ sublinearly because of the adopted FUV shielding that is stronger for models with higher $\dot{M}_{\rm in}$.
The turbulent pressure driven by SN feedback is proportional almost linearly to $\Sigma_{\rm SFR}$.
As a result, the turbulent pressure dominates the thermal pressure for high $\Sigma_{\rm SFR}$ models, consistent with  \citet{os11}.  Figure~\ref{fig:pturb_sfr} also shows, for comparison, results from simulations in \citet{so12} modeling starburst regions, and the extrapolation of the relation
$
P_{\rm turb}/k_{\rm B} = 2.0\times 10^6\,{\rm cm^{-3}\,K}\,\left(\Sigma_{\rm SFR}/M_\odot\,{\rm kpc^{-2}\,yr^{-1}} \right)^{0.89}
$
found by \citet{kok13} (see their Equation 21) based on local simulations of normal star-forming galactic environments.

Equations \eqref{eq:Pth_bestfit} and \eqref{eq:Pturb_bestfit} can be rewritten as
\begin{align}
  P_{\rm th}   &= \Upsilon_{\rm th}\Sigma_{\rm SFR}, \\
  P_{\rm turb} &= \Upsilon_{\rm turb}\Sigma_{\rm SFR}.
\end{align}
where $\Upsilon_{\rm th}$ and $\Upsilon_{\rm turb}$ are the thermal\footnote{In our simulations, the thermal pressure mostly comes from hot bubbles created by SN feedback rather than FUV heating.} and the turbulent feedback yields, respectively, given by
\begin{align}
  \Upsilon_{\rm th} &= 114\,{\rm km\,s^{-1}}\,\left( \displaystyle \frac{\Sigma_{\rm SFR}}{M_\odot\,{\rm kpc^{-2}\,yr^{-1}}}  \right)^{-0.36},\label{eq:Upsth} \\
  \Upsilon_{\rm turb} &= 158\,{\rm km\,s^{-1}}\,\left( \displaystyle\frac{\Sigma_{\rm SFR}}{M_\odot\,{\rm kpc^{-2}\,yr^{-1}}}  \right)^{-0.11}.\label{eq:Upsturb}
\end{align}
The feedback ``yield'' for individual pressure terms was introduced by \citet{kko11} (see their Equations 11 and 12).  There, and also in \citet{kok13} and \citet{kim15}, the notation $\eta$ was adopted for the ratio between $P$ and $\Sigma_{\rm SFR}$, adopting common astronomical units of $10^3\,k_B\, {\rm cm^{-3}\,K}$ for the former and $10^{-3}M_\odot\,{\rm kpc^{-2}\,yr^{-1}}$ for the latter so that $\eta$ is dimensionless.  Since the ratio between $P$ and $\Sigma_{\rm SFR}$ is a naturally a velocity, we instead adopt units of ${\rm km\,s^{-1}}$ for yields $\Upsilon$.  The conversion is $\Upsilon /{\rm km\,s^{-1}} = 209 \eta$.
The turbulent yield in the present work is a factor $\sim 2-3$ smaller than that of \citet{so12} and the extrapolation of \citet{kok13}, where the latter is $\Upsilon_{\rm turb} = 420\,{\rm km\,s^{-1}}\,\left( \Sigma_{\rm SFR}/M_\odot\,{\rm kpc^{-2}\,yr^{-1}}  \right)^{-0.1}$ (converting from their Equation 23 to the present units). For the set of TIGRESS simulations described in \citet{kim20}, analysis to be presented in Ostriker \& Kim (2021, in prep.) also finds $\Upsilon_{\rm turb} \propto \Sigma_{\rm SFR}^{-0.1}$, with a coefficient $\sim 70\%$ higher than in Equation~\eqref{eq:Upsturb}. In Section \ref{s:discuss}, we will discuss possible causes for the lower $\Upsilon_{\rm turb}$ in the present models and star-forming rings more generally.

\subsubsection{Scaling relations of the star formation rate}
\begin{figure*}
    \plotone{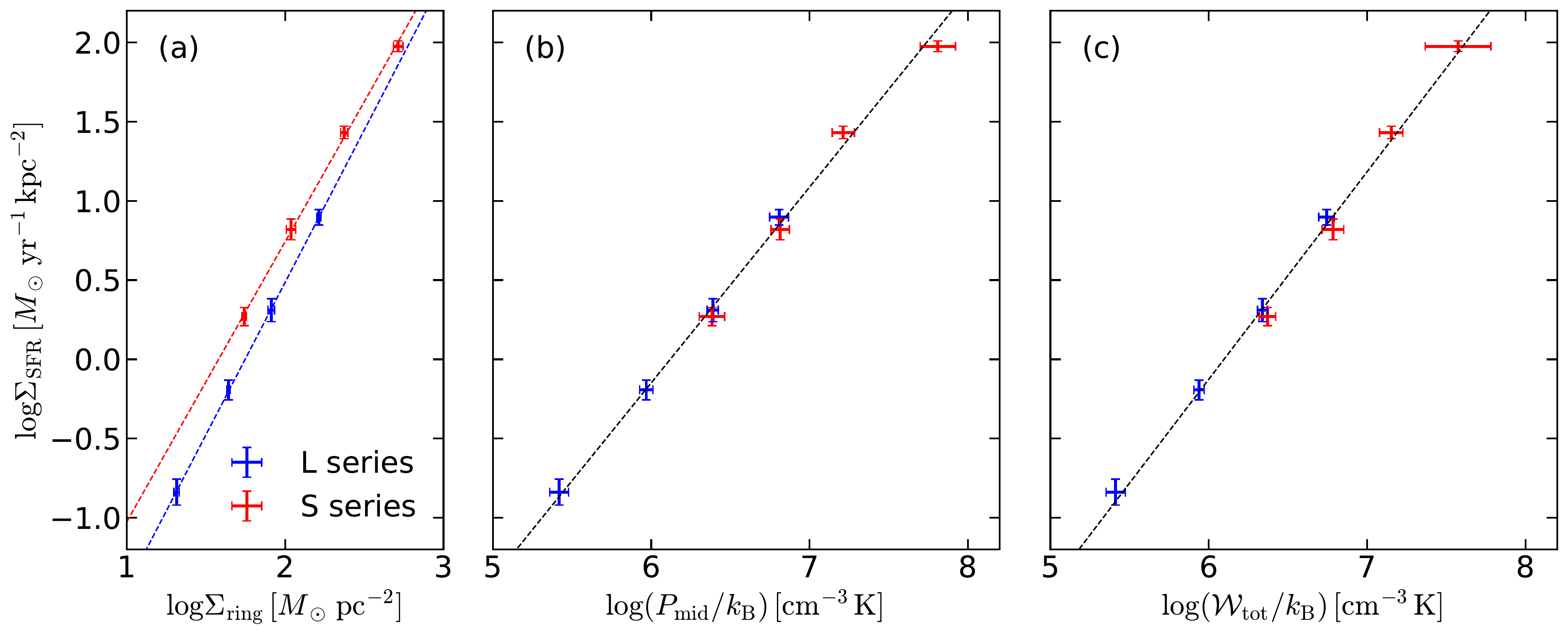}
    \caption{Dependence of the SFR surface density  $\Sigma_\text{SFR}$ on (a) the ring surface density $\Sigma_\text{ring}$, (b) the midplane pressure $P_\text{mid}$, and (c) the total gas weight ${\mathcal W}_{\rm tot}$. The {\tt L} and {\tt S} series lie on different lines, Equation \eqref{eq:KS}, in the $\Sigma_{\rm SFR}$--$\Sigma_{\rm ring}$ plane. In contrast, the two series follow the same relation, Equation \eqref{eq:SFRP}, over nearly three orders of magnitude in the $\Sigma_{\rm SFR}$--$P_{\rm mid}$ or $\Sigma_{\rm SFR}$--${\mathcal W}_{\rm tot}$ plane. The pressure (or weight) provides a better predictor for star formation than gas surface density alone because it allows for (varying) compression of the ISM by the stellar potential.}
    \label{fig:ks}
\end{figure*}

It is of much interest to determine what large-scale galactic properties provide the best prediction for the star formation rate. A simple correlation that has been extensively investigated empirically is the Kennicutt-Schmidt relation between gas and star formation surface densities \citep{schmidt59,kennicutt98}.
Figure \ref{fig:ks}(a) plots $\Sigma_{\rm SFR}$ against $\Sigma_\text{ring}$ for all of our models.
A linear fit to our models yields
\begin{subequations}\label{eq:KS}
 \begin{align}
   \Sigma_{\rm SFR} & = 3.0\,M_\odot\,{\rm yr}^{-1}\,{\rm kpc}^{-2}\,\left(\frac{\Sigma_\text{ring}}{10^2\,M_\odot\,{\rm pc}^{-2}}\right)^{1.9}, \, \text{for {\tt L} series},\label{eq:KS1} \\
   \Sigma_{\rm SFR} & = 5.5\,M_\odot\,{\rm yr}^{-1}\,{\rm kpc}^{-2}\,\left(\frac{\Sigma_\text{ring}}{10^2\,M_\odot\,{\rm pc}^{-2}}\right)^{1.8}, \, \text{for {\tt S} series}.\label{eq:KS2}
 \end{align}
 \end{subequations}
While these have similar scalings with $\Sigma_{\rm ring}$, the distinct offset between the relations for the two series makes plain that an additional parameter contributes in regulating $\Sigma_{\rm SFR}$: surface density is not by itself sufficient.

In the theory of feedback-modulated, self-regulated star formation, the key large-scale parameter is not the gas surface density by itself, but the combination of gas and stellar parameters that go into defining the ISM weight $\cal W$, as  described above.  The individual components of pressure scale as power laws in $\Sigma_{\rm SFR}$ (the source of heat and turbulence), as shown in Figure~\ref{fig:pturb_sfr}, and the pressure balances the gas weight, as shown in Figure~\ref{fig:vertical_equilibrium}.
Since turbulence dominates the pressure in galactic center environments and $\Upsilon_{\rm turb}$ varies only weakly, we expect a nearly linear scaling of $\Sigma_{\rm SFR}$ with $P_{\rm mid}$ or $\cal W$, and this is indeed evident in Figure \ref{fig:ks}(b),(c).

Quantitatively we find that both the {\tt L} and {\tt S} series follow a single relation
\begin{equation} \label{eq:SFRP}
  \Sigma_{\rm SFR} = 0.71\,M_\odot\,{\rm yr}^{-1}\,{\rm kpc}^{-2}\left(\frac{P_{\rm mid}/k_B}{10^6\,{\rm cm}^{-3}\,{\rm K}}\right)^{1.2},
\end{equation}
or
\begin{equation} \label{eq:SFRW}
  \Sigma_{\rm SFR} = 0.75\,M_\odot\,{\rm yr}^{-1}\,{\rm kpc}^{-2}\left(\frac{{\cal W}_{\rm tot}/k_B}{10^6\,{\rm cm}^{-3}\,{\rm K}}\right)^{1.3},
\end{equation}
which are plotted as dashed lines in Figure \ref{fig:ks}(b),(c).
For comparison, the extrapolation of Equation (26) from \citet{kok13} is $\Sigma_{\rm SFR} = 0.48\,M_\odot\,{\rm yr}^{-1}\,{\rm kpc}^{-2}\left(P_{\rm mid}/10^6\,{\rm cm}^{-3}\,k_B\,{\rm K}\right)^{1.18}$, lower by a factor $\sim 1.5$.
Similarly, the extrapolation of Equation (27) from \citet{kok13} is $\Sigma_{\rm SFR} = 0.33\,M_\odot\,{\rm yr}^{-1}\,{\rm kpc}^{-2}\left({\cal W}_\text{tot}/10^6\,{\rm cm}^{-3}\,k_B\,{\rm K}\right)^{1.13}$, lower by a factor $\sim 2.3$. In local-disk TIGRESS simulations, the coefficients in the relations corresponding to Equations~\eqref{eq:SFRP}-\eqref{eq:SFRW} are lower by factors $2.5-3$ (Ostriker \& Kim 2021, in prep.).
We discuss in Section~\ref{s:discuss} possible reasons for these offsets in $\Sigma_{\rm SFR}$-pressure relations.
The better agreement between {\tt S} and {\tt L} series in (b) and (c) than in (a) suggest that the $\Sigma_{\rm SFR}$--$P_{\rm mid}$ (or $\Sigma_{\rm SFR}$--$\cal W_\text{tot}$) relation is more fundamental than the $\Sigma_{\rm SFR}$--$\Sigma_\text{ring}$ relation for regulation of star formation in nuclear rings.

We can use the $\Sigma_{\rm SFR}$--pressure relation together with our previous results for scalings to interpret dependences of $\Sigma_{\rm SFR}$ on $\Sigma_{\rm ring}$.
At given $\Sigma_\text{ring}$, the {\tt S} series shows enhanced $\Sigma_{\rm SFR}$ compared to the {\tt L} series because of the stronger external gravity, which increases $\cal W_\text{tot}$.  This in turn requires higher $\Sigma_{\rm SFR}$ to maintain a higher $P_{\rm mid}$ through feedback.
However, the enhancement of $\Sigma_{\rm SFR}$ is only a factor of $\sim 2$, despite a factor of $11$ difference in $\rho_b(R_{\rm ring})$.
This is because stronger gravity and slightly lower $\sigma_z$ makes the disk thinner in the {\tt S} series (with $H\propto \sigma_z/\sqrt{\rho_b}$), leading to only a modest increase in the weight ${\cal W}_{\rm ext} \propto \rho_b H \propto \sqrt{\rho_b} \sigma_z$ of the ISM for given surface density.
The difference in the external stellar potential is the main reason that the $\Sigma_{\rm SFR}$--$\Sigma_\text{ring}$ relation is different between the {\tt L} and {\tt S} series.

As noted above, the three weight contributions are roughly given by ${\mathcal W}_{\rm gas} \sim \pi G\Sigma_{\rm ring}^2/2$, ${\mathcal W}_{\rm sp}\sim \pi G\Sigma_{\rm ring}\Sigma_{\rm sp}$, and ${\mathcal W}_{\rm ext}\sim \Sigma_{\rm ring}\sigma_z(2G\rho_b/3)^{1/2}$. Although ${\mathcal W}_{\rm ext}$ is the largest among the three (except for  the  {\tt S3} model), the contribution from ${\mathcal W}_{\rm sp}$ becomes quite significant as $\Sigma_{\rm ring}$ increases. Because of this (and since ${\Sigma}_{\rm sp}$ increases with $\Sigma_{\rm ring})$, the total weight increases superlinearly with surface density (a fit gives ${\mathcal W}_{\rm tot} \sim \Sigma_{\rm ring}^{1.54}$). Meanwhile, Equation \eqref{eq:SFRP} indicate the total feedback yield $\Upsilon_{\rm tot}\equiv P_{\rm mid}/\Sigma_{\rm SFR}$ decreases with $\Sigma_{\rm SFR}$ as $\Upsilon_{\rm tot} = 277\,{\rm km\,s^{-1}}\left(\Sigma_{\rm SFR}/M_\odot\,{\rm yr}^{-1}\,{\rm kpc}^{-2}\right)^{-0.17}$. These two effects act together to steepen the $\Sigma_{\rm SFR}-\Sigma_{\rm ring}$ relation via $\Sigma_{\rm SFR}\propto \Upsilon_{\rm tot}^{-1} {\cal W}_{\rm tot} \propto \Sigma_{\rm ring}^{1.86}$. Although the weight in our simulations is dominated by ${\mathcal W}_{\rm ext}$ and ${\mathcal W}_{\rm sp}$ rather than ${\mathcal W}_{\rm gas}$, it happens that the dependence of $\Sigma_{\rm sp}$ on $\Sigma_\text{ring}$ makes the $\Sigma_{\rm SFR}$--$\Sigma_\text{ring}$ relation close to the self-gravity dominated case, as in \citet{os11} and \citet{so12}.

\section{Summary and Discussion}\label{s:discussion}

\subsection{Summary}
Nuclear rings are the sites of compact and extremely vigorous star formation, powered by bar-driven inflows from the larger-scale ISM.
Despite many observational and theoretical studies of nuclear rings, it has remained unclear what determines the gas ring's properties and SFR, and how star formation in nuclear rings proceeds with time.
To address these issues, we construct a semi-global model of a nuclear ring where the bar-driven mass inflows are treated by the boundary conditions.  An advantage of our framework over fully global simulations is that  it enables us to directly control the mass inflow rate and ring radius.

We have modified the TIGRESS framework \citep{tigress} for self-consistently simulating the star-forming ISM to make it suitable for galactic center regions with non-periodic boundary conditions.
As before, gravitational collapse leading to star formation and feedback utilizes sink particles that model star clusters, which produce FUV heating and SN explosions.
To account for the shielding of FUV radiation in high density environments expected in nuclear rings, we make the FUV intensity decrease exponentially with local density (Equation \ref{eq:JFUV}). To balance cooling in these heavily shielded region, we include a simplified treatment of CR heating by ionization (Equation \ref{eq:cr}).

We have run two series of models that differ in the specific angular momentum of the inflow and therefore the resulting ring size $R_{\rm ring}$: models in the {\tt S} series have small rings with $R_{\rm ring}=150\,{\rm pc}$ and models in the {\tt L} series have large rings with  $R_{\rm ring}=600\,{\rm pc}$.  While the gravitational potential profile is the same for the two series, the ring gas in the {\tt S} series experiences stronger vertical compression than the {\tt L} series because the vertical gravity is stronger closer to the galactic center. The initial gas distribution is near-vacuum and the subsequent evolution is governed by the gas inflows through two nozzles located at the $y$-boundaries (Figure \ref{fig:galcen_schematic}).
In each series, we consider four models that differ in the mass inflow rate $\dot{M}_{\rm in}$ (Table \ref{tb:models}).
We run all models beyond $t=300\,{\rm Myr}$, long enough for the system to reach a quasi-steady state.

The main results of the present work can be summarized as follows:
\begin{enumerate}
  \item
    \emph{Overall Evolution} --- Inflowing gas streams collide with each other after half an orbit,
    forming strong shocks at the contact points between the streams. As the orbital kinetic energy is lost, the gas streams gradually circularize to make a nuclear ring. Star formation soon becomes widely distributed along the whole length of the ring, and SN feedback produces hot gas that fills most of the volume.
    Within $\sim100\,$Myr, the system reaches a quasi-steady state in which the gas properties and the SFR do not vary much with time.

  \item
    \emph{Star Formation and Feedback} --- In the quasi-steady state, star formation occurs throughout the ring and the SFR exhibits only modest (within a factor of $\sim 2$) temporal fluctuations. For both the {\tt L} and {\tt S} series, the SFR is solely determined by the inflow rate as ${\dot M}_{\rm SF} \approx 0.8\dot{M}_{\rm in}$. The ring gas mass $M_{\rm ring}\propto \dot{M}_{\rm in}^{0.6}$, with the rings in the {\tt S} series about four times less massive than in the {\tt L} series.

    SN from clusters create many holes along the ring as superbubbles break out, but the feedback never destroys the entire ring.
    As the star particles diffuse out of the ring, SN from relatively old clusters can occur outside of the ring, dumping most of their energy in the ambient hot medium rather than the ring gas. Because SN feedback is ``wasted'' outside of the ring gas, a higher SFR is required to maintain equilibrium than would be needed if the gas and SNe were cospatial in a more uniform disk.

  \item
    \emph{Winds} ---  Our models naturally develop biconical, helically outflowing winds due to SN feedback (Figure \ref{fig:wind}).
    The opening up of the gas streamlines allows the winds accelerate from subsonic to supersonic velocities, readily reaching $\sim 600$-$900\,{\rm km\,s^{-1}}$ at $z=1\,{\rm kpc}$, with the largest velocity occurring near the symmetry axis ($R=0$).

  \item
    \emph{Ring Properties} ---
    Most of the mass is in cold-warm gas, with pressure and density close to the thermal equilibrium curve in which radiative cooling is balanced by FUV and CR heating. Dissipation of kinetic energy is also a significant
    heating source for the cold-unstable phase with $T<5050\,$K.
    The mean sound speed of the cold-warm gas is only $\sim3-4\,{\rm km\,s^{-1}}$
    (Table \ref{tb:warmcold}),
    much lower than the turbulent velocity dispersion $\sigma_z\sim (10-25)\,{\rm km\,s^{-1}}$, which mildly increases with $\Sigma_{\rm SFR}$ (Figure \ref{fig:veldisp_scaling}).
    The scale height increases monotonically with $\Sigma_{\rm SFR}$ due to an increase in $\sigma_z$ in the {\tt L} series, while it is roughly constant or decreases for high  $\Sigma_{\rm SFR}$ due to the increased gravity of star particles in the {\tt S} series.

    Rings are more eccentric for models with larger $\dot{M}_{\rm in}$ and/or smaller $R_{\rm ring}$. Gas in these models tends to have shorter $t_{\rm dep}$, and is thus rapidly depleted by star formation before the orbits can be fully circularized.
    In eccentric gas rings, the orbits of young star particles also inherit large eccentricities. Due to the precession of orbits, however, the distribution of old star particles is more or less circular, regardless of the ring shape.

  \item
    \emph{Vertical Dynamical Equilibrium} --- The ISM in the nuclear rings satisfies vertical dynamical equilibrium, in which the total pressure (turbulent exceeding thermal) balances the weight of the gas.  The pressure at the $z$-boundaries is non-negligible in models with strong outflows and small vertical extent (Figure \ref{fig:vertical_equilibrium}). For the parameters of our models, the weight is dominated mostly by the external gravity term, with the gravity of the newly created sink particles making a secondary contribution. Because the gas is consumed rapidly, the self-gravity term does not become large.  Scalings of ${\cal W}_{\rm ext}$, ${\cal W}_{\rm sp}$, and ${\cal W}_{\rm gas}$ with the gas surface density in the ring, $\Sigma_{\rm ring}$, are consistent with expectations.

\item
    \emph{Scaling Relations for Pressure and Star Formation}

    Consistent with expectations for driving by momentum injection from SNe, the turbulent pressure varies nearly linearly with the SFR surface density (Equation \ref{eq:Pturb_bestfit}).  The corresponding turbulent yield $\Upsilon_{\rm turb}$ (Equation \ref{eq:Upsturb}) is a factor $\sim 2$--$3$ smaller than the values in local-box simulations; this reduced efficiency may be due to feedback that is ``lost'' when SNe go off outside of the ring.

    The combination of vertical dynamical equilibrium with the $P_{\rm turb}$--$\Sigma_{\rm SFR}$ relation leads to a nearly linear dependence of $\Sigma_{\rm SFR}$ on $P_{\rm mid}$ (or the total weight ${\cal W}_\text{tot}$), given by Equation~\eqref{eq:SFRP} (or \ref{eq:SFRW}).   Both the {\tt S} and {\tt L} series lie along a single relation. The power is the same as found by \citet{kok13}, while the coefficient is a factor $\sim 1.5-2.3$ higher due to the reduced feedback efficiency.

    For the individual {\tt S} and {\tt L} series, there are power-law relationships between $\Sigma_{\rm SFR}$ and $\Sigma_{\rm ring}$ with slopes between 1 and 2, consistent with expectations for scalings intermediate between $\Sigma_{\rm SFR} \propto {\cal W}_{\rm ext}\propto \Sigma_{\rm ring}$ and $\Sigma_{\rm SFR} \propto {\cal W}_{\rm sp}+{\cal W}_{\rm gas} \propto \Sigma_{\rm ring}^2$.  However, the {\tt S} series is offset to higher $\Sigma_{\rm SFR}$ than the {\tt L} series due to the stronger vertical gravity of the bulge at the locations of the smaller rings.

    Importantly, we conclude that the $\Sigma_{\rm SFR}-P_{\rm mid}$ (or ${\cal W}_\text{tot}$) relation is more general (as well as more physically fundamental) than the $\Sigma_{\rm SFR}-\Sigma_{\rm ring}$ relation.  The pressure relation provides a better predictor for star formation because it explicitly allows for variations in the vertical compression of the ISM by stellar gravity, which differ with environment even within a given galactic center region.

\end{enumerate}

\subsection{Discussion}\label{s:discuss}

\paragraph{Inflows and star formation}
Our results show that the ring SFR, $\dot{M}_\text{SF}$, is controlled primarily by the mass inflow rate $\dot{M}_\text{in}$ rather than the ring mass. This is overall consistent with the numerical
results that the ring SFR is, when averaged over a few 100 Myrs, roughly equal to the mass inflow rate to the ring in global simulations of barred galaxies \citep{sk13,seo14,seo19}, suggesting
that observed star formation histories in nuclear rings \citep[e.g.,][]{allard06,sarzi07,timer} may primarily reflect the time variations in the mass inflow rates.
The strong correlation between $\dot{M}_\text{SF}$ and $\dot{M}_\text{in}$ is a direct (and causal) consequence of the mass conservation: in our models, 80\% of the inflowing gas is consumed by star formation, while the remaining 20\% is ejected as galactic winds. While $\dot{M}_\text{SF}$ and $M_\text{ring}$, or $\Sigma_{\rm SFR}$ and $\Sigma_\text{ring}$,  are also correlated, this relationship is more indirect, depending on environmental parameters such as the external gravity and on the feedback strength (see Section \ref{s:self_regulation}). In fact, given that $\dot{M}_\text{SF}$ appears to be causally determined by $\dot{M}_\text{in}$, $M_\text{ring}$ may be considered as {\it responding} to  the inflow rate. That is, $M_\text{ring}\sim t_{\rm dep} \dot{M}_{\rm in}$, where in some circumstances the depletion time is relatively insensitive to the ring properties and depends primarily on the bulge potential (see below).

In our semi-global simulations, $\dot{M}_\text{in}$ is fixed to a constant value and the resulting $\dot{M}_\text{SF}$ and $t_\text{dep}$ do not change appreciably with time (see discussion of $t_\text{dep}$ below).
Our simulations do not have the boom/bust behavior of other models \citep[e.g.,][]{kruijssen14,krumholz17,torrey17,armillotta19} because star formation is distributed throughout the ring and the associated feedback is never strong enough to disperse the ring or make it quiescent as a whole. The recent global simulations of \citet{sormani20} also found that the depletion time in the CMZ is roughly constant, with the SFR varying linearly with the CMZ mass, which is most likely affected by the mass inflow rate.

Our models employ a steady and symmetric injection of gas streams from two nozzles. The simple, symmetric  inflow we impose is intentionally idealized, since this is our first study of nuclear star formation with a new numerical framework.  The inflow results in the formation of nuclear rings over which gas and star particles are roughly uniformly distributed. A visual inspection of the nuclear rings in 78 barred galaxies presented by \citet{comeron10} shows that rings with symmetric star formation are more common than those with lopsided star formation.\footnote{Color composite images of 16 symmetric rings among the sample are available in  \citet{ma18}.}  Nevertheless, there are notable examples of lopsided nuclear rings, including our own CMZ \citep[e.g.,][]{barnes17,henshaw16} and the ring in M83 \citep[e.g.,][]{harris01}. Asymmetric star formation in such rings may be caused by the mass inflow rates that are highly asymmetric and nonsteady \citep[e.g.,][]{harada19}. Large-scale simulations \citep[e.g.][]{seo19,armillotta19,tress20} with time-varying, asymmetric inflows, as well as observations \citep[e.g.,][]{sb19} motivate further study  of lopsided ring formation at high resolution, and we plan to explore this question in a forthcoming paper.

\paragraph{Feedback yield}
The turbulent yield (Equation \ref{eq:Upsturb}) in our simulations is smaller than the value in the previous simulations with SN feedback of \citet{so12} and \citet{kok13}. Considering the balance between the turbulent driving and dissipation, \citet{os11} showed that the turbulent yield is given by $\Upsilon_{\rm turb} = f_p p_*/(4m_*)$, where $p_*$ is the asymptotic radial momentum injected per SN and $m_*$ is the total mass in stars per SN ($m_* \approx 100\,M_\odot$ for a Kroupa IMF). The parameter $f_p$ encapsulates details of the driving and dissipation and is expected to be $f_p\sim 0.5-2$, with tests showing $f_p\approx 1$ for a range of parameters, and a slight decreasing trend towards higher SFR \citep{os11,so12,kko11,kok13}.
\citet{kok13} and \citet{kim15} (including magnetic fields) adopted  $p_*=3\times 10^5\,M_\odot\,{\rm km\,s^{-1}}$ as a fiducial value (based on isolated SNe in uniform gas), and found $\Upsilon_{\rm turb}\sim 700 -900 \,{\rm km\,s^{-1}}$ for solar-neighborhood conditions (irrespective of magnetic field strengths; see \citealt{kim15}), decreasing to $\Upsilon_{\rm turb}\sim 500 \,{\rm km\,s^{-1}}$ for a factor $\sim 20$ higher SFR. In a local-disk TIGRESS suite \citep[][Ostriker \& Kim 2021 in prep.]{kim20}, exploring more extreme conditions up to $\Sigma_{\rm SFR} \sim 1\,M_\odot\,{\rm yr^{-1}\,kpc^{-2}}$, $\Upsilon_{\rm turb}$ further decreases to $\sim 270\,{\rm km\,s^{-1}}$, which is about $\sim 36\%$ lower than an extrapolation from \citet{kok13} and 70\% larger than what we found in this paper (Equation \ref{eq:Upsturb}).

It should be noted that $\sim80$--$90\%$ of all SNe in our simulations are resolved so that, as in the local-disk TIGRESS suite, $p_*$ is determined self-consistently via non-linear gas interactions. Assuming $f_p=1$, Equation \eqref{eq:Upsturb} translates to $p_*\sim 0.4-0.8\times 10^5\,M_\odot\,{\rm km\,s^{-1}}$. This is a factor of $\sim 5$ lower than what was found for isolated SNe at densities comparable to solar neighborhood ISM, but closer to the momentum injection in high density environments  \citep[e.g.][]{ko15,martizzi15}, and also quite comparable to results from experiments in which multiple SNe explode at short ($\sim 0.01\,{\rm Myr}$) intervals for a range of ambient density \citep{kor17}.
This could explain the generally smaller turbulent feedback yield in simulations using the full TIGRESS framework \citep[][and this work]{kim20,kko20} than the idealized simulations with a fixed $p_*$ \citep{so12, kok13, kim15}.

In the present simulations, an additional effect
comes into play to reduce $p_*$: many SNe explode slightly exterior to the ring or close to the ring boundary, dumping their energy into the ambient hot gas and bulk motions of the ring instead of driving internal turbulence within the ring. This could be the primary reason for the smaller turbulent yield in our simulations compared to TIGRESS simulations where the momentum is more fully captured by the surrounding warm-cold gas.
Additionally, when SNe are very crowded, partial cancellation of the vertical momenta due to interactions of neighboring shells could also contribute to reducing $\Upsilon_{\rm turb}$.
All of these considerations explain why SN feedback is somewhat less efficient in driving turbulence within star-forming nuclear ring regions compared to  outer disk environments.

\paragraph{Depletion time}
The depletion time measured in our simulations, $t_{\rm dep,ring}\sim 10-100\,{\rm Myr}$, is very short compared to the solar neighborhood TIGRESS model ($\sim 2\,{\rm Gyr}$; \citealt{tigress}), although analogous TIGRESS simulations modeling regions with higher gas and stellar density -- closer to those of the present models -- have $t_{\rm dep}\sim 70-400\, {\rm Myr}$ \citep{kim20}.
If star formation is locally regulated by stellar feedback (including SNe), the depletion time is determined by balancing the ISM weight $\mathcal{W} = (\Sigma/2) \left<g_z\right>$ and the midplane pressure $P_{\rm mid} \equiv \Upsilon_{\rm tot}\Sigma_{\rm SFR}\equiv\Upsilon_{\rm tot} \Sigma/t_{\rm dep}$, such that $t_{\rm dep} =2 \Upsilon_{\rm tot}/\left< g_z\right>$.  Here, $\Sigma$ is the local gas surface density (equivalent to $\Sigma_{\rm ring}$ for the present case), $\left<g_z\right>$ denotes the mass-weighted vertical gravity, and $\Upsilon_{\rm tot}$ is the total feedback yield.
The analysis given in Section \ref{s:self_regulation} suggests that the short $t_{\rm dep}$ of the present simulations results from the combined effect of reduced $\Upsilon_{\rm tot}$ (see above) and strong $\left<g_z\right>$, with the latter being more important.  Similar to our results, \citet{sormani20} found from analysis of their simulations that the decrease in the CMZ $t_{\rm dep}\sim 100\, {\rm Myr}$ compared to the outer region  $t_{\rm dep}\sim (1-2){\rm Gyr}$ was consistent with expectations from self-regulation, with $\langle g_z\rangle$ dominated by the stellar potential.
The small depletion time in our galactic center simulations together with the result that the gravitational field is dominated by the stellar component (including newly-formed stars) appears qualitatively consistent with \citet{utomo17}. They found that $\sim 30\%$ of the EDGE galaxy sample have a central drop in $t_{\rm dep}$, and that the drop in $t_{\rm dep}$ is correlated with a central increase in stellar surface density.

Our depletion time is still much shorter than observational values of $10^2-10^3\,{\rm Myr}$ for most galactic centers and starburst galaxies, although observed values can be as short as 3 - 10\, ${\rm Myr}$ in regions of extremely high surface density
\citep{kennicutt98,bigiel08,genzel10,narayanan12,utomo17,Wilson2019}. Also, observations using the DYNAMO sample -- local analogues of clumpy, high redshift ($z\sim 1-2$) galaxies -- have suggested there may be superlinear pressure enhancement in regions with high $\Sigma_{\rm SFR}$ \citep{fisher19},
%as opposed to the decreasing trend shown in our simulations, while
although these regions are not well resolved spatially, and the trend is moderated when molecular gas (instead of ionized gas) velocity dispersions are used to estimate the ISM weight \citep{girard21}. \citet{Molina2020} find an upper limit on pressure slightly above the prediction of Equation~\eqref{eq:Upsturb}.
While higher-resolution observations are essential (and improved estimates of the stellar gravity are needed), current empirical work indicates that our models for star formation and feedback cannot fully explain real galaxies at high SFRs.

Several physical elements not yet included in the present models may explain the lower $t_{\rm dep}$ than in real galaxies.
First, our current models only include FUV and CR heating and Type II SNe, while neglecting magnetic fields, CR pressure, and forms of early feedback (see below). All of these contribute to support and/or dispersal of the large-scale ISM and/or individual clouds, and thus may help to lengthen the depletion time \citep[e.g.,][]{hennebelle14,kim15,girichidis16,kim20,jgkim20,kko20}.
For instance, typical magnetic fields in nuclear rings are of order $\sim 60\,\mu$G  \citep[see][and references therein]{beck15}.
The corresponding magnetic pressure is $P_{\rm mag}/k_B\sim 1\times 10^6\rm\,K\,cm^{-3}$, which can be dynamically significant.

Second, since we do not model the destruction of star clusters, the old, massive sink particles remain concentrated at the midplane, increasing $\mathcal{W}_{\rm sp}$ and $\langle g_z\rangle$ and thus reducing $t_\text{dep}$.
In reality, however, they are expected to be disrupted and dispersed both radially and vertically by cluster-cluster collisions and/or the tidal gravity \citep[e.g.,][]{zwart02, grijs12, vaisanen14}, reducing $\Sigma_{\rm sp}$ and increasing $t_{\rm dep}$.

Third, in this paper we do not model early feedback such as stellar winds, photoionization, and radiation pressure from young stars, which can disperse natal clouds even before the onset of the first SNe, limiting their lifetime star formation efficiency \citep[e.g.,][]{rogers13, rahner17, jgkim18,jgkim20}.  Rather than our simple model of 100\% star formation efficiency when sink particles form, a more realistic treatment would include significant mass return over several Myr.  Since the total momentum injection from early feedback is low compared to the injection from SNe \citep[][L. Lancaster et al 2021, submitted]{jgkim18}, these processes are not likely to alter large-scale SFRs in outer-galaxy environments where dynamical times within clouds are several Myr. However, early feedback is potentially quite important in denser galactic-center environments.

Finally, we remark that inclusion of these physical elements may lead to stronger and/or more localized star formation and ensuing feedback. In this case, the rings may become more prone to local destruction, and the SFR and the depletion time may exhibit large temporal fluctuations, even if the mass inflow rate is kept constant. Assessing the relation between the ring SFR and the mass inflow rate would thus require more realistic treatments of star formation and feedback.

% \acknowledgments

\begin{center}
    {\sc Acknowledgements}
\end{center}

We acknowledge an insightful and constructive report from the referee.
We thank Jeong-Gyu Kim for useful discussions and help in radiation post-processing for the development of the approximate form of FUV heating rate. We thank Munan Gong for highlighting the importance of CR heating and suggesting an approximate form for the CR heating rate. The work of S.M.\ was supported by an NRF (National Research Foundation of Korea) grant funded by the Korean Government (NRF-2017H1A2A1043558-Fostering Core Leaders of the Future Basic Science Program/Global Ph.D. Fellowship Program).
The work of W.-T.K.\ was supported by the grants of National Research Foundation of Korea (2019R1A2C1004857 and 2020R1A4A2002885).  The work of E.C.O and C.-G.K.\ was partly supported by NASA (ATP grant No. NNX17AG26G). Computational resources for this project were provided by Princeton Research Computing, a consortium including PICSciE and OIT at Princeton University,
and by the Supercomputing Center/Korea Institute of Science and Technology Information with supercomputing resources including technical support (KSC-2019-CRE-0052).

\software{{\tt Athena} \citep{athena}, {\tt STARBURST99} \citep{starburst99}}

\appendix

\section{Orbit Integration of Sink Particles with the Coriolis Force}%
\label{s:boris}

Here we describe how we apply the Boris algorithm, which is widely used in plasma simulations, to integrate the equations of motion for sink particles in a rotating frame; we also present a test result. Note that Equation \eqref{eq:sp_eom} is formally equivalent to the equations of motion for charged particles under the Lorentz force if we substitute $-\boldsymbol \nabla \Phi_{\rm tot} \to q{\bf E}/m$ and $2{\bf \Omega}_p \to q{\bf B}/m$.

\begin{figure}[htpb]
  \centering
  \includegraphics[width=0.3\linewidth]{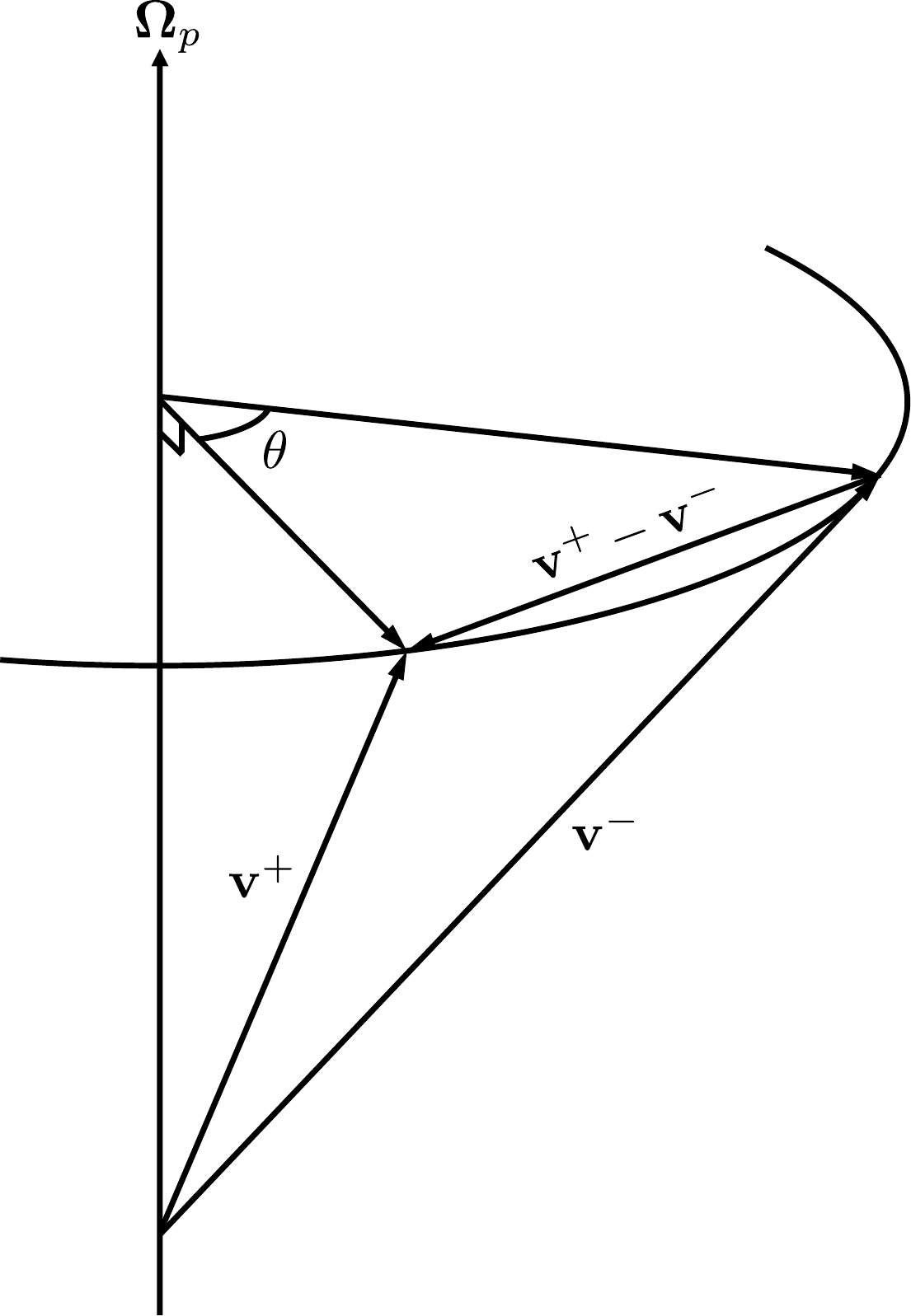}
  \caption{Illustration of the velocity vector rotation described by Equation \eqref{eq:boris_rot}. The rotation angle $\theta$ is defined in the plane perpendicular to ${\bf \Omega}_p$.}%
  \label{fig:boris_schematic}
\end{figure}

As in the original TIGRESS implementation, we adopt a ``Kick-Drift-Kick (KDK)'' leap-frog  integrator, for which a semi-implicit discretization of Equation \eqref{eq:sp_eom} leads to
\begin{equation}\label{eq:discretized_sp_eom}
  \frac{{\bf v}^{n+1/2}-{\bf v}^{n-1/2}}{\Delta t} = {\bf g}^{n} - 2{\bf \Omega}_p\times \left( \frac{{\bf v}^{n+1/2}+{\bf v}^{n-1/2}}{2} \right),
\end{equation}
where ${\bf g} = -\boldsymbol \nabla\Phi_{\rm tot}$ is the gravitational acceleration.  \citet{boris70} noted that the ``electric'' and ``magnetic'' forces in Equation \eqref{eq:discretized_sp_eom} can be separated by changing the variables to
\begin{align}
  {\bf v}^{-} &\equiv {\bf v}^{n-1/2} + \frac{\Delta t}{2} {\bf g}^{n}\label{eq:boris_first},\\
  {\bf v}^{+} &\equiv {\bf v}^{n+1/2} - \frac{\Delta t}{2} {\bf g}^{n}\label{eq:boris_third},
\end{align}
yielding
\begin{equation}\label{eq:boris_rot}
  \frac{{\bf v}^{+}-{\bf v}^{-}}{\Delta t} = ({\bf v}^{+}+{\bf v}^{-})\times {\bf \Omega}_p.
\end{equation}
Taking the inner product of Equation \eqref{eq:boris_rot} with $({\bf v}^{+}+{\bf v}^{-})$ reveals $|{\bf v}^{+}| = |{\bf v}^{-}|$, i.e., Equation \eqref{eq:boris_rot} describes a pure rotation of the vector ${\bf v}^{-}$ into ${\bf v}^{+}$, with ${\bf \Omega}_p$ being the axis of the rotation, as depicted in Figure \ref{fig:boris_schematic}. It can be shown that the rotation angle $\theta$ in Figure \ref{fig:boris_schematic} satisfies $\tan(\theta/2) = \Omega_p\Delta t$. Assuming ${\bf \Omega}_p$ is parallel to the $z$-axis, Equation \eqref{eq:boris_rot} can be decomposed to
\begin{align}
  v^{+}_x &= \frac{1-(\Omega_p\Delta t)^2}{1+(\Omega_p\Delta t)^2} v^{-}_x + \frac{2\Omega_p \Delta t}{1+(\Omega_p\Delta t)^2} v^{-}_y,\label{eq:boris_second1}\\
  v^{+}_y &= -\frac{2\Omega_p \Delta t}{1+(\Omega_p\Delta t)^2}v^{-}_x + \frac{1-(\Omega_p\Delta t)^2}{1+(\Omega_p\Delta t)^2}v^{-}_y,\label{eq:boris_second2}\\
  v^{+}_z &= v^{-}_z.\label{eq:boris_second3}
\end{align}
In the Boris algorithm, therefore, it takes three steps to update ${\bf v}^{n-1/2}$ to ${\bf v}^{n+1/2}$. (1) Apply the gravitational force for a half time step as in Equation \eqref{eq:boris_first}, (2) solve for the epicyclic rotation using Equation \eqref{eq:boris_second1}--\eqref{eq:boris_second3}, and (3) apply the gravitational force for the remaining half time step according to Equation \eqref{eq:boris_third}.

As a test calculation, we consider a particle orbiting solely under a rotating, rigid-body potential $\Phi_{\rm ext} = \tfrac{1}{2}R^2\Omega_0^2$ with $\Omega_0 = 1$, $\Omega_p = 0.5$ and $\Phi_{\rm self}=0$. The orbit is limited to the $z=0$ plane.  The equations of motion for such a  particle are given by
\begin{subequations}\label{eq:parteom}
\begin{align}
  \ddot{x} &= -\left(\Omega_0^2-\Omega_p^2\right)x + 2\Omega_p \dot{y},\label{eq:parteom1}\\
  \ddot{y} &= -\left(\Omega_0^2-\Omega_p^2\right)y - 2\Omega_p \dot{x}.\label{eq:parteom2}
\end{align}
\end{subequations}
Equations \eqref{eq:parteom1}--\eqref{eq:parteom2} are linear, coupled ordinary differential equations with analytic solutions
\begin{subequations}\label{eq:orbit}
\begin{align}
  x(t) &= A_1\cos(\Omega_- t) + A_2\sin(\Omega_- t) + A_3\cos(\Omega_+ t) + A_4\sin(\Omega_+ t),\label{eq:orbitx}\\
  y(t) &= -A_2\cos(\Omega_- t) + A_1\sin(\Omega_- t) + A_4\cos(\Omega_+ t) -A_3\sin(\Omega_+ t),\label{eq:orbity}
\end{align}
\end{subequations}
where $\Omega_\pm \equiv \Omega_0\pm \Omega_p$ and
\begin{align}
  A_1 &= (\Omega_+x(0) + \dot{y}(0))/(2\Omega_0),\\
  A_2 &= (\dot{x}(0) - \Omega_+y(0))/(2\Omega_0),\\
  A_3 &= (\Omega_-x(0) - \dot{y}(0))/(2\Omega_0),\\
  A_4 &= (\dot{x}(0) + \Omega_-y(0))/(2\Omega_0),
\end{align}
with initial position $[x(0), y(0)]$ and velocity  $[\dot{x}(0), \dot{y}(0)]$.

\begin{figure*}[htpb]
  \centering
  \includegraphics[width=\linewidth]{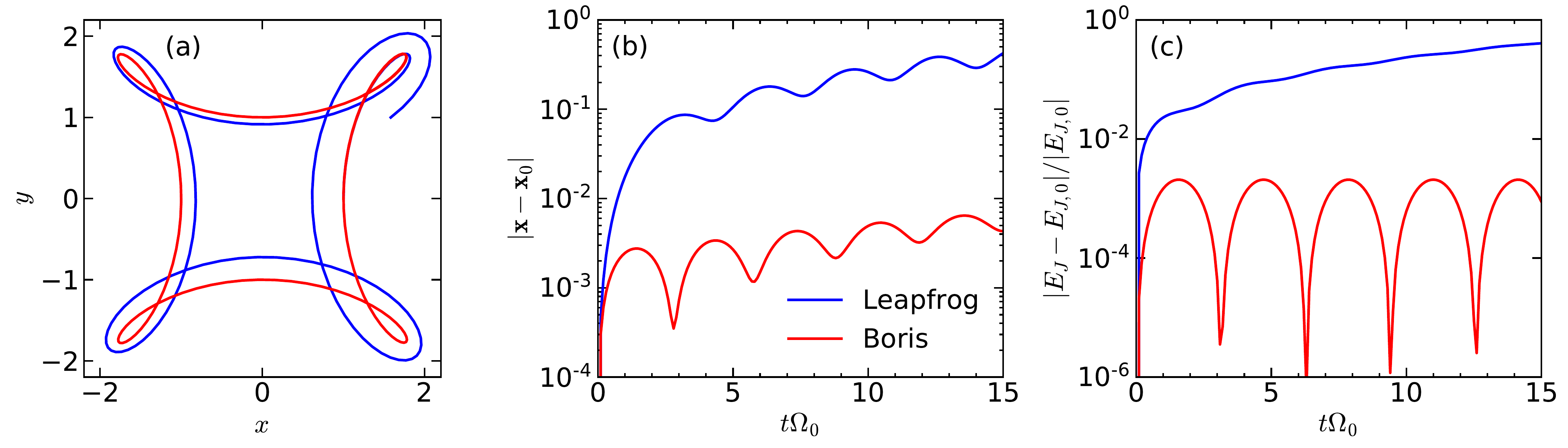}
  \caption{Comparison between the Boris algorithm (red) and the standard leap-frog integrator (blue) for (a) the particle orbits in the $x$-$y$ plane, (b) the position offsets between the numerical and analytic solutions, and (c) the relative errors in the Jacobi integral $E_J$ under the rotating, rigid-body potential $\Phi_{\rm ext} = \frac{1}{2}R^2\Omega_0^2$ with $\Omega_0=1$ and $\Omega_p=0.5$. The initial conditions are $(x,y) = (1,0)$ and $(\dot{x}, \dot{y}) = (0, 2)$ and $\Delta t = 0.1/\Omega_0$ is taken for the timestep.
  }%
  \label{fig:boris_test}
\end{figure*}

As the initial conditions, we take $(x,y)=(1,0)$ and $(\dot{x}, \dot{y})=(0,2)$ at $t=0$ and integrate Equation \eqref{eq:parteom} using the Boris algorithm with a fixed timestep of $\Delta t= 0.1/\Omega_0$.  For comparison, we also integrate Equation \eqref{eq:parteom} using the standard KDK leap-frog integrator.
Figure \ref{fig:boris_test} compares the resulting orbits in the $x$-$y$ plane, the position offsets relative to the analytic predictions (Equation \ref{eq:orbit}), and the errors in the Jacobi integral $E_J\equiv \frac{1}{2} (\dot{x}^2 + \dot{y}^2) + \Phi_\text{tot}$.
The position offsets oscillate and secularly grow with time in both methods, but the growth rate is much lower in the Boris algorithm. The relative errors in the Jacobi integral are bounded below $0.2\%$ in the Boris algorithm.

\bibliography{mybib}

\end{document}